\documentclass[10pt]{article}
\usepackage[a4paper,margin=22mm]{geometry}
\usepackage[T1]{fontenc}
\usepackage{lmodern,microtype,amsmath,amssymb,graphicx,xcolor}
\usepackage[round,authoryear]{natbib}
\usepackage[font=small,labelfont=bf]{caption}
\usepackage[hypertexnames=false,colorlinks=true,urlcolor=blue,citecolor=blue,linkcolor=black,
  pdftitle={QuantileMix: A Python package for quantile g-computation},
  pdfauthor={Nilson Chapagain}]{hyperref}
\usepackage{bookmark}
\usepackage{float,doi,needspace}
\providecommand{\pandocbounded}[1]{#1}
\setkeys{Gin}{width=\linewidth,keepaspectratio}
\let\standardSubsection\subsection
\makeatletter
\renewcommand\subsection{\Needspace{8\baselineskip}\@startsection{subsection}{2}{\z@}
  {-1ex plus -.3ex minus -.2ex}{.3ex plus .1ex}
  {\normalfont\normalsize\bfseries}}
\makeatother
\title{QuantileMix: A Python package for quantile g-computation}
\author{Nilson Chapagain\\[3pt]\small Texas A\&M University, United States\\
  \small\href{mailto:nchapagain@tamu.edu}{nchapagain@tamu.edu}}
\date{}

\usepackage{listings,longtable,booktabs,array,calc,upquote,etoolbox}
\usepackage{footnotehyper}
\makesavenoteenv{longtable}

\newcommand{\passthrough}[1]{#1}
\makeatletter
\patchcmd\longtable{\par}{\if@noskipsec\mbox{}\fi\par}{}{}
\makeatother
\begin{document}
\addtocontents{toc}{\protect\setcounter{tocdepth}{-10}}
\pdfbookmark[-1]{Paper}{paper-start}
\maketitle
\begingroup
\renewcommand{\thefootnote}{}
\footnotetext{Paper: CC BY 4.0. Appended reference manual: GPL-2.0-or-later.}
\addtocounter{footnote}{-1}
\endgroup
\section{Summary}\label{summary}

Health researchers often study how a mixture of pollutants or nutrients
relates to health \citep{Keil2020}. Quantile g-computation addresses this question
by giving each exposure an ordered score and fitting a model for the outcome.
For example, dividing an exposure into four quantile groups gives scores
0, 1, 2, and 3. The model can compare score 1 with score 2 for every exposure
at once. These groups need not contain equal numbers of observations when
measurements are tied. A one-score increase can represent different changes
in the original units for different exposures \citep{Keil2020}.

\href{https://github.com/nilson01/quantilemix}{QuantileMix} provides selected
quantile g-computation methods in Python. It returns estimates, confidence intervals, predictions, and diagnostics for
continuous, binary, count, and survival outcomes.
Other models handle measurement limits, outcomes with several categories,
and counts with many zeros. Fitting and prediction use NumPy and SciPy
\citep{Harris2020, Virtanen2020} and do not require R. The package includes a
reference manual, runnable examples, and numerical comparisons with R.

\section{Motivation}\label{statement-of-need}

When the joint effect is a sum of exposure coefficients, its uncertainty
depends on the variance of each coefficient estimate and how the estimates vary together
(their covariances) \citep{Keil2020}. QuantileMix performs these calculations and
stores the group boundaries, or cutpoints, for use in prediction.

The package is intended for researchers who prepare and analyse data in Python.
Functions for named columns allow categorical adjustment variables and removal
of rows with missing values. The default is to reject missing values.
Removing incomplete rows does not by itself prevent bias \citep{Sterne2009}.

A causal interpretation requires well-defined exposure interventions. The
outcome under a person's observed exposure must equal the observed outcome
(consistency), and one person's exposure must not affect another person's
outcome (no interference). After adjustment, no unmeasured common causes of
exposure and outcome may remain. The compared exposure combinations must have
support at the adjustment values used (positivity), and the outcome model must
be correctly specified \citep{Keil2020}. The software cannot verify these assumptions.

\section{Related software}\label{state-of-the-field}

Alexander Keil's R package qgcomp provides a broader set of quantile
g-computation methods \citep{qgcomp}. We use it as the numerical reference for
selected QuantileMix calculations. QuantileMix is a separate, unofficial
implementation with fewer functions and options.

Other Python tools also address parts of this problem. Statsmodels provides
regression models and linear contrasts \citep{Seabold2010}. DeepMix implements
additive quantile g-computation with fitted cutpoints and prediction \citep{DeepMix}.
At the cited revision, its Python result does not include a coefficient
covariance matrix, standard error, or confidence interval.

We developed QuantileMix to combine joint effect estimates, uncertainty, and
prediction in one Python interface. Each model states what its effect estimate
measures. The contribution is the software and checks that others can repeat;
the statistical methods come from earlier work.

\section{Methods and implementation}\label{software-design}

QuantileMix separates exposure scoring, model fitting, and results. Fitted
cutpoints are saved so that the same scoring rule can be used for new data.
Results store method-specific estimates, uncertainty, diagnostics, and the
values needed for prediction.

\subsection{Joint effect estimation}\label{joint-effect-estimation}

The additive Gaussian model illustrates the joint effect. Following
\citet{Keil2020}, let \(Y_i\) be the outcome for person \(i\), \(X_{ji}^{q}\) the score for
exposure \(j\), and \(d\) the number of exposures. Omitting adjustment variables
for clarity, with \(E(\epsilon_i\mid X_{1i}^{q},\ldots,X_{di}^{q})=0\), gives

\[
Y_i = \beta_0 + \sum_{j=1}^{d}\beta_j X_{ji}^{q} + \epsilon_i,
\qquad
\psi = \sum_{j=1}^{d}\beta_j.
\]

Here \(\beta_0\) is the intercept, \(\beta_j\) an exposure coefficient, and
\(\epsilon_i\) the random error. The parameter \(\psi\) is the change in the mean
outcome when all exposure scores rise by one, with adjustment variables held
fixed. This comparison applies only where both exposure combinations are
supported by the study population \citep{Keil2020}.

We add the fitted coefficients to obtain
\(\widehat\psi=\sum_{j=1}^{d}\widehat\beta_j\). Its estimated variance includes
the variance of each coefficient estimate and the covariances between estimates
\citep{Keil2020}:

\[
\widehat{\operatorname{Var}}(\widehat\psi)
=\sum_{j=1}^{d}\sum_{k=1}^{d}
\widehat{\operatorname{Cov}}(\widehat\beta_j,\widehat\beta_k).
\]

\begin{figure}
\centering
\pandocbounded{\includegraphics[keepaspectratio,alt={Estimated coefficients and joint exposure contrast in an additive Gaussian model. (a) \textbackslash widehat\textbackslash beta\_j and \textbackslash widehat\textbackslash psi=\textbackslash sum\_\{j=1\}\^{}\{3\}\textbackslash widehat\textbackslash beta\_j. (b) Estimated mean difference a\textbackslash widehat\textbackslash psi between X\_j\^{}q=a and X\_j\^{}q=0 for all j, with a\textbackslash in\textbackslash\{0,1,2,3\textbackslash\}. Bars in (a) and shading in (b) denote pointwise 95\% normal confidence intervals. Shading is not a prediction interval for an individual outcome. The synthetic example has n=64, d=3, and two adjustment variables.}]{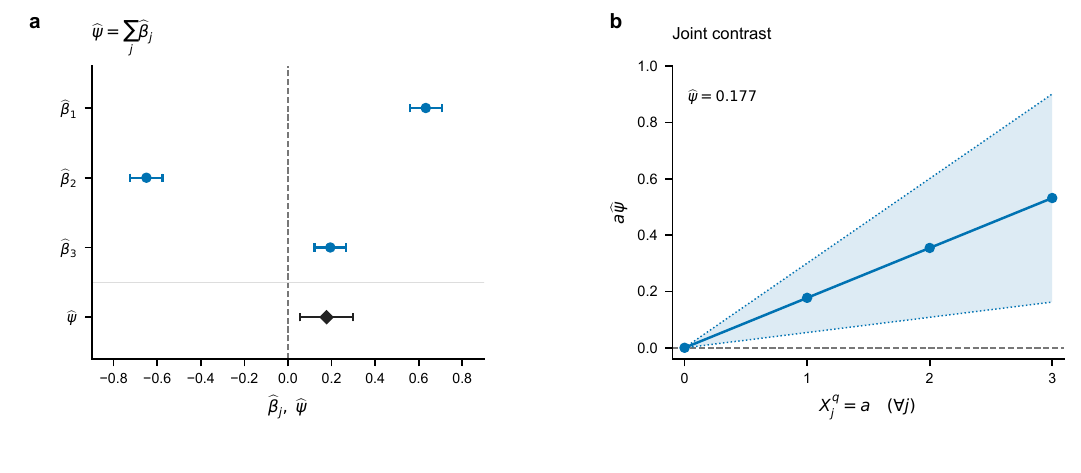}}
\caption{Estimated coefficients and joint exposure contrast in an additive Gaussian model.
(a) \(\widehat\beta_j\) and \(\widehat\psi=\sum_{j=1}^{3}\widehat\beta_j\).
(b) Estimated mean difference \(a\widehat\psi\) between \(X_j^q=a\) and
\(X_j^q=0\) for all \(j\), with \(a\in\{0,1,2,3\}\).
Bars in (a) and shading in (b) denote pointwise 95\% normal confidence
intervals. Shading is not a prediction interval for an individual outcome. The synthetic example has \(n=64\), \(d=3\),
and two adjustment variables.\label{fig:contrast}}
\end{figure}

Coefficients with opposite signs can cancel, giving a small \(\widehat\psi\)
even when individual \(\widehat\beta_j\) are large (Figure \ref{fig:contrast}).
For a nonzero fitted coefficient \(\widehat\beta_k\), its sign weight is
\(w_k=|\widehat\beta_k|/\sum_{j:\,\widehat\beta_j\widehat\beta_k>0}|\widehat\beta_j|\): the denominator sums absolute coefficients with the same sign
\citep{Keil2020, qgcomp}. These weights describe the fitted model, not causal
importance.

\subsection{Supported models and prediction}\label{supported-models-and-prediction}

For binary and count outcomes, separate functions estimate conditional and
marginal effects. Conditional effects hold adjustment variables fixed; marginal
effects average over their distribution. For binary outcomes, a conditional
odds ratio and a marginal odds ratio can differ even without confounding
\citep{Greenland1999}.
Marginal methods predict outcomes under each exposure setting and average over
the observed adjustment values; the reported effect follows the specified
marginal model \citep{Keil2020, qgcomp}. Bootstrap and estimating-equation methods
have separate limits on nonlinear terms and clusters \citep{qgcomp}. A row bootstrap
samples independent observations with replacement \citep{Efron1979}; it does not
preserve within-cluster dependence.

\begin{figure}
\centering
\pandocbounded{\includegraphics[keepaspectratio,alt={Conditional predictions for six outcome models. (a, c) \textbackslash widehat E(Y) under Gaussian and Poisson models; (b) \textbackslash widehat\{\textbackslash Pr\}(Y=1); (d) the latent Gaussian mean before censoring; (e) \textbackslash widehat\{\textbackslash Pr\}(Y=k), k=1,2,3 (low, mid, high); (f) Cox survival estimates \textbackslash widehat S\_i(t) for four observed profiles i. In (a--e), X\_j\^{}q=a for all j and adjustment variables are zero; lines connect the discrete score levels. Shading in (a--d) denotes pointwise 95\% Wald confidence intervals, with binary and Poisson limits transformed from the linear predictor scale. No intervals are shown in (e--f). Each panel uses separate synthetic data; t is in the example's time units.}]{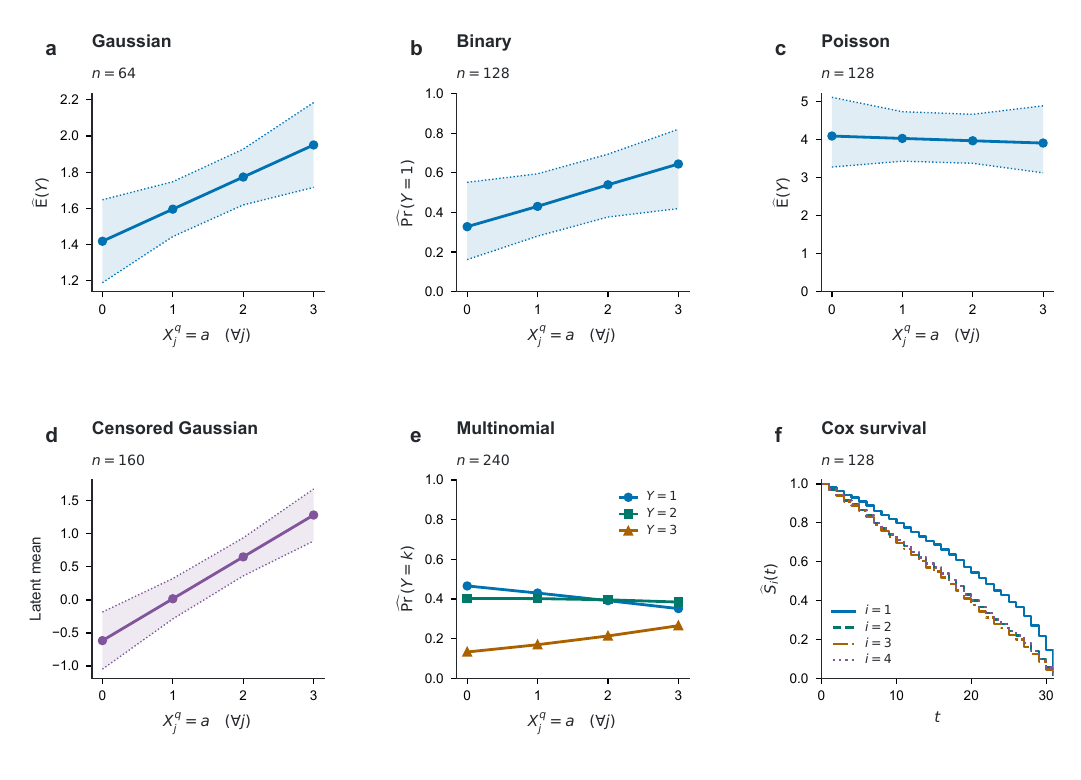}}
\caption{Conditional predictions for six outcome models.
(a, c) \(\widehat E(Y)\) under Gaussian and Poisson models;
(b) \(\widehat{\Pr}(Y=1)\);
(d) the latent Gaussian mean before censoring;
(e) \(\widehat{\Pr}(Y=k)\), \(k=1,2,3\) (low, mid, high);
(f) Cox survival estimates \(\widehat S_i(t)\) for four observed profiles \(i\).
In (a--e), \(X_j^q=a\) for all \(j\) and adjustment variables are zero; lines
connect the discrete score levels. Shading in (a--d) denotes pointwise 95\%
Wald confidence intervals, with binary and Poisson limits transformed from
the linear predictor scale. No intervals are shown in (e--f). Each panel uses
separate synthetic data; \(t\) is in the example's time units.\label{fig:families}}
\end{figure}

\begin{figure}
\centering
\pandocbounded{\includegraphics[keepaspectratio,alt={Fitted probability mass functions for two-part count models. (a) Zero-inflated Poisson; (b) zero-inflated negative binomial; (c) hurdle Poisson; (d) hurdle negative binomial. Points show \textbackslash widehat\{\textbackslash Pr\}(Y=y), y=0,\textbackslash ldots,10, at X\_j\^{}q=0 or X\_j\^{}q=3 for all j, with adjustment variables zero. Probabilities are not renormalized over the displayed counts; \textbackslash widehat\{\textbackslash Pr\}(Y\textgreater10) is retained in the source table. Each panel uses separate synthetic data. No confidence intervals are shown.}]{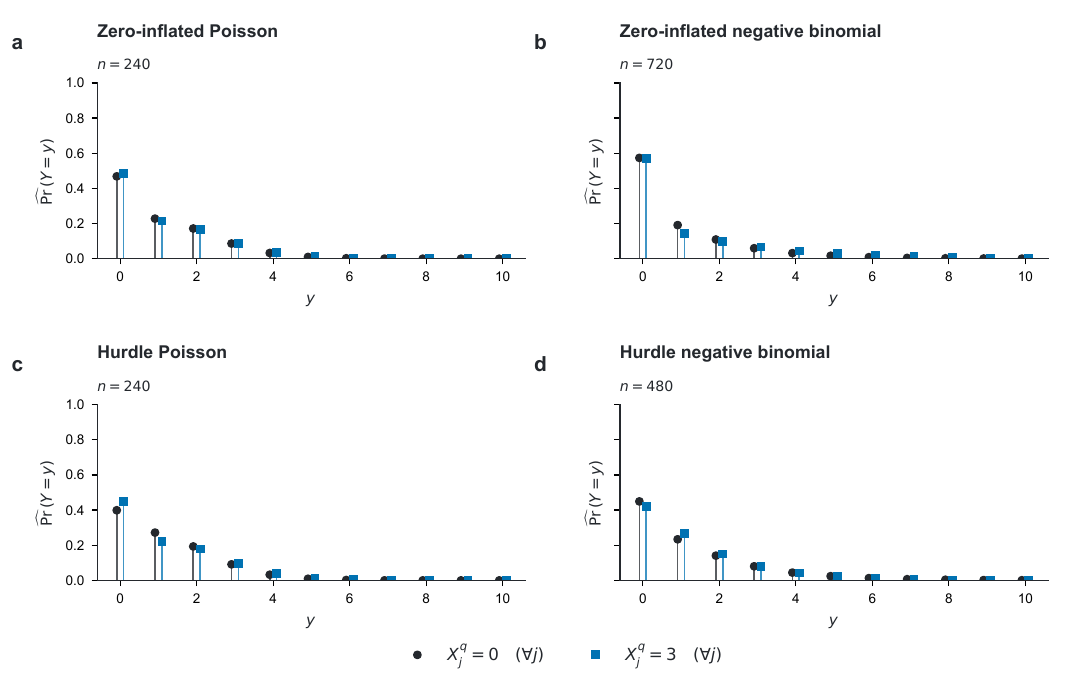}}
\caption{Fitted probability mass functions for two-part count models.
(a) Zero-inflated Poisson; (b) zero-inflated negative binomial;
(c) hurdle Poisson; (d) hurdle negative binomial.
Points show \(\widehat{\Pr}(Y=y)\), \(y=0,\ldots,10\), at \(X_j^q=0\) or
\(X_j^q=3\) for all \(j\), with adjustment variables zero. Probabilities are not
renormalized over the displayed counts; \(\widehat{\Pr}(Y>10)\) is retained
in the source table. Each panel uses separate synthetic data.
No confidence intervals are shown.\label{fig:counts}}
\end{figure}

Figures \ref{fig:families} and \ref{fig:counts} show predictions from synthetic
examples, including Cox survival models \citep{Cox1972} and zero-inflated and
hurdle count models \citep{Zeileis2008}.

Matplotlib supplies the plots \citep{Hunter2007}. The manual lists supported formula
expressions and explains the different roles of case weights, sign weights,
and cluster identifiers.

\section{Validation and limitations}\label{research-impact-statement}

\subsection{Numerical agreement and interval coverage}\label{numerical-agreement-and-interval-coverage}

Our \href{https://github.com/nilson01/quantilemix/tree/main/validation}{validation records}
contain numerical comparisons and repeated-dataset studies. We compared nine
Gaussian cases with installed R qgcomp 2.19.6. In a 24,000-dataset Gaussian
study, we compared Python, installed R,
and an independent calculation using QR factorization. All fits succeeded,
and all paired comparisons met the recorded numerical tolerances. The 12
conditions used \(d=4\) exposures, \(n\in\{100,500\}\), and 2,000 datasets per
condition (\(12\times2{,}000=24{,}000\)).
Across these conditions, 94.55\%--96.10\% of the nominal 95\% Python/R intervals
contained the true effect. These intervals use normal critical values rather
than exact Gaussian Student-\(t\) limits.

\begin{figure}
\centering
\pandocbounded{\includegraphics[keepaspectratio,alt={Empirical coverage of nominal 95 percent Gaussian confidence intervals}]{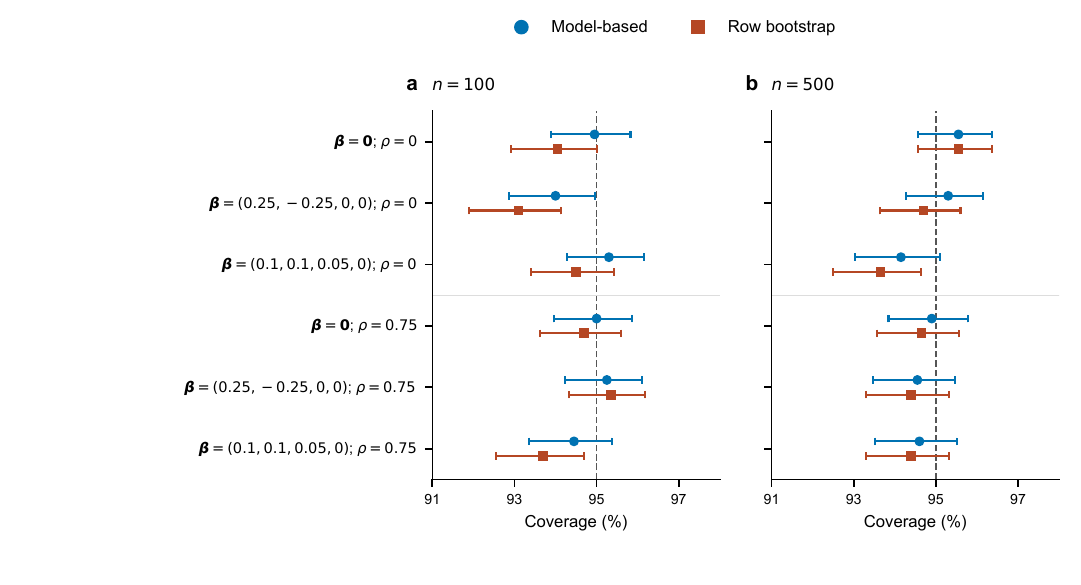}}
\caption{Empirical coverage of nominal 95\% Gaussian confidence intervals.
(a) \(n=100\); (b) \(n=500\). Points show
\(\frac{100}{2{,}000}\sum_{r=1}^{2{,}000}\mathbf{1}\{\psi\in I_r\}\),
where \(I_r\) is the confidence interval from dataset \(r\) and \(\mathbf{1}\) is
the indicator function. Bars denote pointwise 95\% Wilson Monte Carlo intervals
\citep{Wilson1927}; the dashed line marks 95\% coverage.
Rows identify \(\boldsymbol\beta=(\beta_1,\ldots,\beta_4)\) and residual latent
exposure correlation \(\rho\). Model-based and row-bootstrap methods use the
same 2,000 datasets per condition (\(24{,}000\) total), with no fit failures.
Bootstrap intervals use normal critical values and standard errors from 200
resamples with fixed exposure scores. The Wilson intervals are neither
simultaneous intervals across conditions nor intervals for paired coverage
differences.\label{fig:coverage}}
\end{figure}

We also evaluated Gaussian row-bootstrap intervals in a separate
24,000-dataset study (Figure \ref{fig:coverage}). The normal Wald intervals
are approximately \(\widehat\psi\pm1.96\sqrt{\widehat{\operatorname{Var}}_{\mathrm{boot}}(\widehat\psi)}\),
where \(\widehat{\operatorname{Var}}_{\mathrm{boot}}(\widehat\psi)\) is the sample
variance of 200 bootstrap estimates. These are not percentile intervals. All fits succeeded. In three conditions, observed
coverage was below 95\%, with pointwise 95\% Wilson intervals lying entirely
below 95\%. These comparisons were not adjusted for testing multiple conditions.
The studies used known population scores and correctly specified Gaussian
models. They do not establish coverage when cutpoints are estimated, the model
is incorrect, or other outcome families are fitted. Agreement between programs
and interval coverage answer different questions. Neither establishes causality.

\subsection{Reproducible materials and limits}\label{reproducible-materials-and-limits}

The repository includes study settings, scripts, results, reference comparisons,
and worked examples. These materials allow others to inspect and repeat the
calculations. We have not yet documented use by other researchers or a completed
real-data research application. The evidence applies to the tested methods and
settings; it does not establish widespread use or complete R-package coverage.

\section{AI usage disclosure}\label{ai-usage-disclosure}

ChatGPT was used as an AI assistant.
AI-assisted language editing was used to improve wording in parts of the manuscript.
All code and text generated by AI was proofread by the authors. AI-generated
code was also extensively tested.

\section{Acknowledgements}\label{acknowledgements}

Computations were performed on the Arseven Computing Cluster, Department of
Statistics, Texas A\&M University.

This work received no grant funding.

\clearpage
\setlength{\paperwidth}{8.5in}
\setlength{\paperheight}{11in}
\pdfpagewidth=8.5in
\pdfpageheight=11in
\newgeometry{layoutwidth=8.5in,layoutheight=11in,margin=1in}
\setcounter{page}{1}
\setcounter{figure}{0}
\setcounter{table}{0}
\setcounter{lstlisting}{0}
\let\subsection\standardSubsection
\floatplacement{figure}{htbp}
\setlength{\parskip}{6pt plus 2pt minus 1pt}
\setlength{\emergencystretch}{3em}
\pdfbookmark[-1]{Reference manual}{manual-start}
\begin{center}
{\LARGE QuantileMix Reference Manual}
\end{center}
\bigskip
\addtocontents{toc}{\protect\setcounter{tocdepth}{3}}
\setcounter{tocdepth}{3}
\tableofcontents
{\def\LTcaptype{none} % do not increment counter
\begin{longtable}[]{@{}
  >{\raggedright\arraybackslash}p{(\linewidth - 2\tabcolsep) * \real{0.1370}}
  >{\raggedright\arraybackslash}p{(\linewidth - 2\tabcolsep) * \real{0.8630}}@{}}
\toprule\noalign{}
\endhead
\bottomrule\noalign{}
\endlastfoot
Version & \passthrough{\lstinline!0.1.1!} \\
Author & Nilson Chapagain \\
Contact &
\href{mailto:nchapagain@tamu.edu}{\passthrough{\lstinline!nchapagain@tamu.edu!}} \\
Import & \passthrough{\lstinline!quantilemix!} \\
Requires & Python 3.10 or later; NumPy 2.0 or later; SciPy 1.14 or
later \\
License & GNU General Public License, version 2 or later \\
\end{longtable}
}

\section{Description}\label{manual-description}

Quantile g-computation transforms exposures into ordinal quantile scores
and estimates the joint effect of increasing all exposures
simultaneously by one quantile. The package supports selected models for
continuous, binary, count, and survival outcomes, with more limited
support for censored, multicategory, and zero-inflated outcomes. Each
API entry documents the effect scale, required inputs, and
methodological limitations.

Quantile cutpoints estimated during model fitting are reused for
prediction. Sign-specific weights summarize fitted exposure coefficients
rather than causal importance. Successful model fitting does not
establish the validity of model assumptions or identify a causal effect
without additional causal identification assumptions.

The methods follow Keil et al.~(2020),
\url{https://doi.org/10.1289/EHP5838}. Selected results were checked
against R qgcomp 2.19.6. This unofficial Python package has not been
endorsed by the original authors. Fitting and prediction do not require
R.

\section{Documentation source and
adaptations}\label{manual-documentation-source-and-adaptations}

The reference entries follow the R manual's Description, Usage,
Arguments, Details, Value, Examples and See Also structure. Applicable
wording is retained from the GPL-licensed R qgcomp 2.19.6 documentation.
Python signatures, containers, restrictions and examples describe this
package's behavior. The adapted source and license are identified in
\href{https://github.com/nilson01/quantilemix/blob/main/NOTICE.md}{NOTICE.md}.

\section{License and credit}\label{manual-license-and-credit}

Copyright in the adapted R documentation belongs to its original
contributors, including Alexander Keil. The Python manual and
adaptations are distributed under GPL-2.0-or-later. See
\href{https://github.com/nilson01/quantilemix/blob/main/LICENSE}{LICENSE}
for the full license terms and warranty disclaimer, and
\href{https://github.com/nilson01/quantilemix/blob/main/NOTICE.md}{NOTICE.md}
for attribution. DeepMix is acknowledged as prior partial Python
coverage; no DeepMix code was copied.

\section{Using the package}\label{manual-using-the-package}

\subsection{Install and run a Gaussian
fit}\label{manual-install-and-run-a-gaussian-fit}

Install from PyPI:

\begin{lstlisting}[language=bash]
pip install quantilemix
\end{lstlisting}

Or install the latest source from GitHub with Git installed:

\begin{lstlisting}[language=bash]
python -m pip install "git+https://github.com/nilson01/quantilemix.git@main"
\end{lstlisting}

Or, if you have downloaded the source, run this from the folder
containing \passthrough{\lstinline!pyproject.toml!}:

\begin{lstlisting}[language=bash]
python -m pip install .
\end{lstlisting}

In this example, the first two columns are exposure scores and the third
column is a continuous outcome:

\begin{lstlisting}[language=Python]
import numpy as np
from quantilemix import fit

data = np.array([
    [0, 0, 1.0], [0, 1, 0.6],
    [1, 0, 2.1], [1, 1, 1.4],
    [2, 0, 3.0], [2, 1, 2.7],
    [3, 0, 4.2], [3, 1, 3.5],
])
result = fit(data[:, 2], data[:, :2], q=None)
print(result.summary())
\end{lstlisting}

Here \passthrough{\lstinline!q=None!} means the exposure columns already
contain scores. Use \passthrough{\lstinline!quantilemix!} for
installation and imports. For a separate Python environment, see the
\href{https://packaging.python.org/en/latest/guides/installing-using-pip-and-virtual-environments/}{Python
Packaging User Guide}. Contributor setup and tests are in
\href{https://github.com/nilson01/quantilemix/blob/main/CONTRIBUTING.md}{CONTRIBUTING.md}.

\subsection{Method and interpretation}\label{manual-sec:method}

Keil et al.~use \(X_{ji}^{q}\) for the score of exposure \(j\) in person
\(i\), \(Y_i\) for the outcome, \(\beta_j\) for an exposure coefficient,
and \(\psi\) for the joint effect in their quantile g-computation
section. We use those symbols here. We write \(X_i^q\) for the vector of
person \(i\)'s exposure scores and \(Z_{i\ell}\) for adjustment
covariate \(\ell\), which the paper omits from its displayed additive
equation. There are \(n\) observations, \(d\) exposures and \(c\)
covariates. The fitted model has \(k=1+d+c\) coefficients and
conditional mean

\[
E(Y_i\mid X_i^q,Z_i)=\beta_0+\sum_{j=1}^{d}\beta_j X_{ji}^{q}
                         +\sum_{\ell=1}^{c}\gamma_\ell Z_{i\ell}.
\]

The joint effect is \(\psi=\sum_{j=1}^{d}\beta_j\); its estimate is
\(\widehat\psi=\sum_{j=1}^{d}\widehat\beta_j\). Holding covariates
fixed, the difference between two fitted means whose exposure scores all
differ by one is

\[
\widehat m(X_i^q+\mathbf{1},Z_i)-\widehat m(X_i^q,Z_i)
 =\sum_{j=1}^{d}\widehat\beta_j\{(X_{ji}^{q}+1)-X_{ji}^{q}\}
 =\sum_{j=1}^{d}\widehat\beta_j=\widehat\psi.
\]

Here \(\mathbf{1}\) is a vector of \(d\) ones and \(\widehat m\) is the
fitted conditional mean. Both exposure profiles must be meaningful for
the intended interpretation. The contrast in \(\widehat\psi\) raises
each exposure by one score category, not by one unit of its original
measurement. It does not describe an intervention that keeps scores at
the top category when they would otherwise rise above it.

In this manual, a hat marks an estimated quantity. Each other method
entry defines its own \(\psi\) and states its scale. For example, the
Gaussian \(\psi\) is a mean difference, while the additive binary
\(\psi\) is a log odds ratio. The two values have different meanings
even when the software calls both of them \passthrough{\lstinline!psi!}.
For a common exposure score \(a\), this manual uses \(\widehat\mu(a)\)
for a Gaussian standardized mean, \(\widehat R(a)\) for a binary risk,
and \(\widehat M(a)\) for a Poisson count mean. Each entry defines its
own intervention and adjustment rule. Symbols for methods beyond the
paper's examples are defined in their individual entries.

\subsubsection{Confidence interval and p-value for the Gaussian
effect}\label{manual-confidence-interval-and-p-value-for-the-gaussian-effect}

For the variance calculation, let \(\widehat\theta\) collect the fitted
intercept, exposure coefficients and covariate coefficients, in that
order. This vector notation is only for the calculation; the paper names
exposure coefficients separately. The contrast vector \(a\) is zero for
the intercept and covariates and one for each exposure. With full design
matrix \(D\) and residuals \(e_i\), the ordinary least-squares
covariance is

\[
\begin{aligned}
 \widehat\sigma^2 &= \frac{\sum_{i=1}^{n}e_i^2}{n-k}, &
 \widehat V &= \widehat\sigma^2(D^{\mathsf T}D)^{-1},\\
 \widehat{\mathrm{Var}}(\widehat\psi)&=a^{\mathsf T}\widehat V a
 =\sum_j\widehat{\mathrm{Var}}(\widehat\beta_j)
  +2\sum_{j<l}\widehat{\mathrm{Cov}}(\widehat\beta_j,\widehat\beta_l).
\end{aligned}
\]

The software uses singular-value decomposition, a matrix method, to fit
the regression and calculate \(\widehat{\mathrm{Var}}(\widehat\psi)\)
without directly forming the inverse shown above. The covariance terms
account for the fact that the exposure coefficient estimates can vary
together.

The standard error is \(s=\sqrt{\widehat{\mathrm{Var}}(\widehat\psi)}\).
For significance level \(\alpha\), R qgcomp 2.19.6's Gaussian convention
uses

\[
\mathrm{CI}=\widehat\psi\ \pm\ z_{1-\alpha/2}s,\qquad
 p=2\,\Pr\{T_{n-1-d}\geq |\widehat\psi/s|\}.
\]

Here \(z_{1-\alpha/2}\) is a standard-normal quantile and \(T_\nu\) has
a Student \(t\) distribution with \(\nu\) degrees of freedom. The
p-value uses \(n-1-d\), whereas residual variance uses \(n-1-d-c\). Thus
the interval and test need not agree at a decision threshold. The
interval is not an exact finite-sample \(t\) interval. If the variation
is too small to separate from numerical rounding, the software warns
that the test result is undefined. The result entry explains this case.

\subsubsection{Positive case weights}\label{manual-positive-case-weights}

Case weights give observations different influence in the Gaussian
regression. They differ from the descriptive sign weights below. For
strictly positive weights \(w_i\), the fitted coefficients minimize
\(\sum_i w_i(Y_i-D_i\theta)^2\). Let \(W\) be the diagonal matrix of
these weights and \(S\) the weighted residual sum of squares. Then

\[
\widehat\sigma_w^2=\frac{S}{n-k},\qquad
 \widehat V_w=\widehat\sigma_w^2(D^{\mathsf T}WD)^{-1}.
\]

The same joint contrast uses \(a^{\mathsf T}\widehat V_w a\). R qgcomp
computes exposure cutpoints before fitting, so case weights do not
change them. These model-based limits do not account for complex survey
sampling or correlation among observations in the same cluster. The
study of 24,000 datasets used equal weights and does not show how often
weighted intervals cover the true effect.

For Bernoulli-logit and Poisson-log fits, these are GLM prior weights.
Each positive weight multiplies its row's log likelihood and Fisher
information. For integer weights, the coefficients and model-based
covariance agree with repeating that row. Formula fits may use
\passthrough{\lstinline!missing="drop"!} to omit rows with missing
weights; the returned result saves the retained row positions. Weighted
fixed-case agreement does not establish repeated-sample coverage.

\subsubsection{Sign weights}\label{manual-sign-weights}

Following the paper's \(w_j\) notation, the fitted sign weight is

\[
\widehat w_j=
\begin{cases}
\displaystyle\frac{\widehat\beta_j}
 {\sum_{\ell:\widehat\beta_\ell>0}\widehat\beta_\ell},
 & \widehat\beta_j>0,\\[8pt]
\displaystyle\frac{|\widehat\beta_j|}
 {\sum_{\ell:\widehat\beta_\ell<0}|\widehat\beta_\ell|},
 & \widehat\beta_j<0.
\end{cases}
\]

A zero coefficient has no sign weight. The positive weights sum to one,
and the negative weights separately sum to one, when those groups are
present. They describe the fitted coefficients; no confidence intervals
are reported for these weights. Opposite signs can cancel in
\(\widehat\psi\) even when individual coefficients are large. Weights
are not probabilities of selection or evidence of causal importance.

\subsubsection{Example plots: sign weights and fitted
means}\label{manual-example-plots-sign-weights-and-fitted-means}

The two plots use constructed data with three exposure scores. The left
plot shows descriptive sign weights. The right plot shows
\(\widehat\mu(a)\), the standardized fitted mean when all three exposure
scores are set to level \(a\). This example does not show how often
confidence intervals cover the true effect across repeated datasets.

\begin{figure}
\centering
\pandocbounded{\includegraphics[keepaspectratio,alt={Descriptive sign weights}]{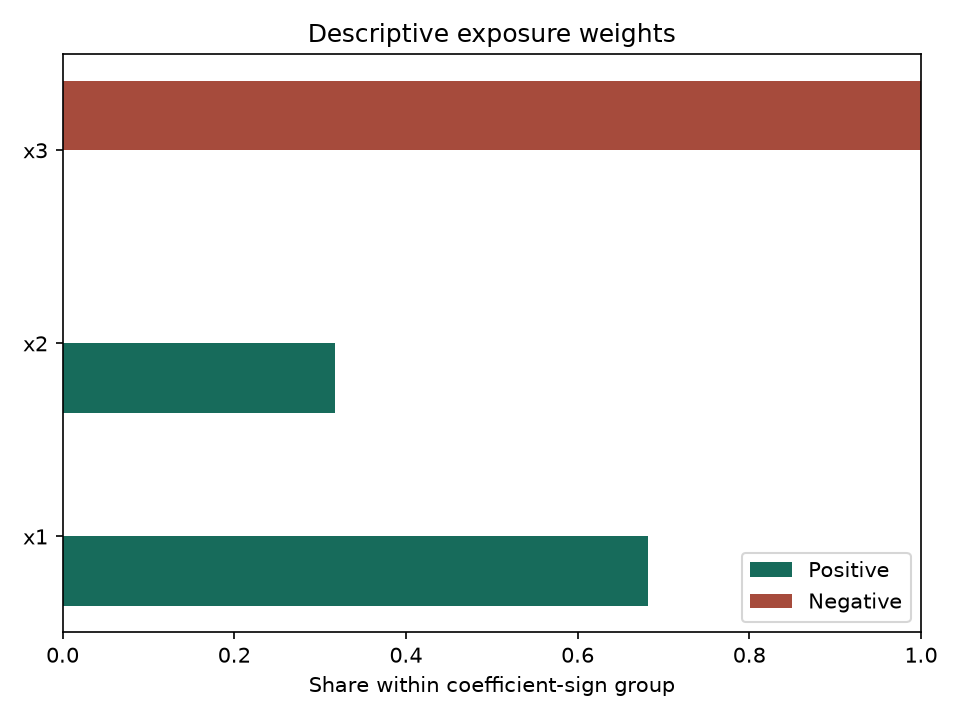}}
\caption{Descriptive sign weights}
\end{figure}

\begin{figure}
\centering
\pandocbounded{\includegraphics[keepaspectratio,alt={Standardized fitted means}]{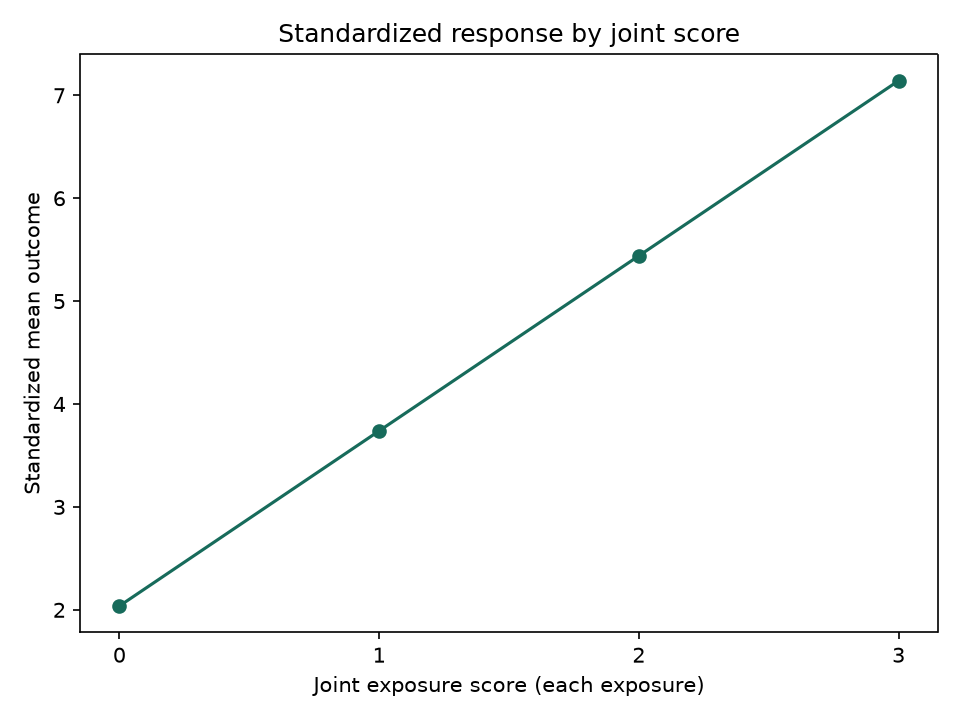}}
\caption{Standardized fitted means}
\end{figure}

Constructed example. Left: sign weights from a conditional Gaussian fit.
\passthrough{\lstinline!x1!} and \passthrough{\lstinline!x2!} share the
positive group; \passthrough{\lstinline!x3!} alone fills the negative
group. The groups have separate denominators, so their bar lengths do
not compare causal importance. Right: standardized means from a separate
additive Gaussian row-bootstrap fit. Each point gives \(\widehat\mu(a)\)
at the shown score \(a\). The line joins fitted means; it is not an
uncertainty band. Twelve seeded resamples make the example repeatable
but do not assess interval coverage.

\subsubsection{Quadratic Gaussian response to joint exposure
changes}\label{manual-quadratic-gaussian-response-to-joint-exposure-changes}

The conditional model includes every exposure score and its square:

\[
E(Y_i\mid X_i^q,Z_i)=\beta_0+\sum_{j=1}^d\beta_jX_{ji}^{q}
                    +\sum_{j=1}^d\beta_{j+d}(X_{ji}^{q})^{2}
                    +\gamma^{\mathsf T}Z_i.
\]

Set every exposure score to the same level \(a\), retain each row's
adjustment values, and average the fitted means. The paper calls the
linear and quadratic mixture coefficients \(\psi_1\) and \(\psi_2\). In
the restricted model, \(\widehat\psi_1=\sum_j\widehat\beta_j\) and
\(\widehat\psi_2=\sum_j\widehat\beta_{j+d}\). When requested, the model
also includes pairwise products \(\beta_{j,k}X_{ji}^qX_{ki}^q\) for
\(j<k\), following the paper's example \(\beta_{1,2}\).
Exposure-covariate products are a package extension; we write them as
\(\eta_{j\ell}X_{ji}^qZ_{i\ell}\). Then \(\widehat\psi_2\) also includes
\(\sum_{j<k}\widehat\beta_{j,k}\), while \(\widehat\psi_1\) also
includes \(\sum_{j,\ell}\widehat\eta_{j\ell}\overline Z_\ell\), where
\(\overline Z_\ell\) is the sample mean of covariate \(\ell\). Each
bootstrap draw uses its own sampled covariate mean. The fitted mean
change from \(a\) to \(a+1\) is \(\widehat\psi_1+(2a+1)\widehat\psi_2\),
so it depends on the starting level. The API's
\passthrough{\lstinline!effect\_at(a)!} evaluates this change.

Exposure cutpoints are fixed before row-bootstrap resampling. The
covariance of the two saved resampled coefficients gives their
uncertainty; normal-based confidence limits and Student \(t\) p-values
follow the restricted R qgcomp 2.19.6 degree-two convention. Two fixed
installed-R comparisons verify the software calculation, not
repeated-sample interval coverage. Separate fixed R cases also verify
pairwise and exposure-covariate products. This method does not include
clusters, case weights or other outcomes.

The separate \passthrough{\lstinline!fit\_quadratic\_gaussian\_ee!} uses
the same intervention curve for repeated observations with known cluster
IDs. Its robust covariance sums both conditional and marginal scores
within each cluster. Positive case weights enter those scores but leave
exposure cuts unweighted. The fitted curve agrees with three fixed
installed-R cases. For nonlinear formulas, R qgcomp 2.19.6 forms cluster
scores from a saved model frame: after it changes exposure columns,
squared and interaction columns can retain their original values. Python
rebuilds these terms at each intervention. A separate fresh-design R
calculation agrees with Python's covariance, while the installed R
interval can differ. These fixed checks do not establish
repeated-dataset interval coverage.

\subsubsection{Assumptions and limits}\label{manual-assumptions-and-limits}

The covariance calculation assumes independent observations and the same
conditional error variance for each observation. Exact small-sample
calculations also require a correct normal-error model and enough
independent model columns to estimate every coefficient. A successful
fit does not verify these assumptions. The Gaussian validation study
uses known scores from fixed population cutpoints. It does not show how
intervals behave when cutpoints are estimated from each dataset.

A causal interpretation additionally needs a well-defined intervention,
consistency between observed and intervention outcomes, adequate joint
exposure support for the exposure combinations being compared, no effect
of one person's exposure on another person's outcome, and sufficient
control of confounding. These conditions depend on the study design and
subject knowledge; software diagnostics cannot establish them. The API
entries describe the supported nonlinear, interaction, clustered and
resampling methods. The
\href{https://github.com/nilson01/quantilemix/blob/main/docs/r-api-parity.md}{coverage
map} lists other methods and limits.

\section{API reference}\label{manual-api-reference}

In a Python function signature, an argument without a default is
required; one with a default may be omitted. After
\passthrough{\lstinline!*!}, supply arguments by name. The examples show
valid calls, not full signatures. Each entry lists its other settings.

\subsection{Exposure scores and
inputs}\label{manual-exposure-scores-and-inputs}

\subsubsection{\texorpdfstring{\texttt{quantize}: Assign scores to
exposures}{quantize: Assign scores to exposures}}\label{manual-api:quantize}

\paragraph{Description}\label{manual-description-1}

Assign each exposure an ordered score. The function can calculate group
boundaries from the data, use boundaries you supply, or accept scores
you have already assigned. It saves the scores and, when boundaries are
used, saves them so new observations can be scored in the same way.

\paragraph{Usage}\label{manual-usage}

\begin{lstlisting}[language=Python]
from quantilemix import quantize

quantization = quantize(exposures)
\end{lstlisting}

\paragraph{Arguments}\label{manual-arguments}

{\def\LTcaptype{none} % do not increment counter
\begin{longtable}[]{@{}
  >{\raggedright\arraybackslash}p{(\linewidth - 2\tabcolsep) * \real{0.5000}}
  >{\raggedright\arraybackslash}p{(\linewidth - 2\tabcolsep) * \real{0.5000}}@{}}
\toprule\noalign{}
\begin{minipage}[b]{\linewidth}\raggedright
Argument
\end{minipage} & \begin{minipage}[b]{\linewidth}\raggedright
Meaning and accepted inputs
\end{minipage} \\
\midrule\noalign{}
\endhead
\bottomrule\noalign{}
\endlastfoot
\passthrough{\lstinline!exposures!} & Nonempty real, finite array of
shape \passthrough{\lstinline!(n, d)!}, or a vector for one exposure. No
missing or actively masked entries. Labels, when provided, must be
unique. \\
\passthrough{\lstinline!q!} & Default \passthrough{\lstinline!4!}.
Requested automatic group count: an integer from 2 to 100,000, excluding
booleans. Ties may reduce the fitted count.
\passthrough{\lstinline!None!} with no breaks means that inputs are
already nonnegative integer-valued scores below \(2^{53}\). \\
\passthrough{\lstinline!breaks!} & Default
\passthrough{\lstinline!None!}. One finite, strictly increasing vector
per exposure, including both outer endpoints; lengths may differ. A
single vector is accepted for one exposure. Supplied breaks take
priority over \passthrough{\lstinline!q!}. This is an alternative to
using \passthrough{\lstinline!q!} to define cutpoints. \\
\end{longtable}
}

\paragraph{Details}\label{manual-details}

With automatic grouping, a continuous exposure is represented by scores
\(0,\ldots,q-1\), subject to reductions caused by ties. For example,
score 1 identifies observations in the second fitted category. These are
ordered numeric scores, not a matrix of separate binary indicators.

Automatic mode uses type-7 empirical quantiles, removes duplicate values
from the full quantile sequence, and replaces its outer endpoints by
\(-10^{64}\) and \(10^{64}\). A value exactly on an interior cut belongs
to the lower category. The outer endpoints are included. Automatic
training magnitudes must be strictly smaller than \(10^{64}\). Constant
exposures are rejected in automatic mode; ties can otherwise reduce the
result to one category, which cannot identify a regression coefficient.

Explicit mode keeps the supplied endpoints. Values outside those
endpoints raise an error; the endpoints are not automatically expanded.
Prequantized mode retains the supplied scores and stores no cutpoints.
Scores need not be consecutive. This mode is narrower than R's general
untransformed \passthrough{\lstinline!q=NULL!} input mode.

The \passthrough{\lstinline!mode!} field is
\passthrough{\lstinline!"quantile"!},
\passthrough{\lstinline!"explicit"!} or
\passthrough{\lstinline!"prequantized"!}. The stored
\passthrough{\lstinline!q!} is the requested integer only in automatic
mode. It is \passthrough{\lstinline!None!} in both other modes, even
when a \passthrough{\lstinline!q!} value was supplied with explicit
breaks. The upper bound of 100,000 automatic groups limits allocation in
this implementation; it is not a restriction of the statistical method.

\paragraph{Errors.}\label{manual-errors.}

\passthrough{\lstinline!ValueError!} is raised for invalid numeric
inputs, masks, shapes, labels, group counts or cut vectors; observations
outside supplied endpoints; invalid prequantized scores; automatic
constant exposures; or automatic training magnitudes at least
\(10^{64}\). The function does not fit a model, check regression rank,
or emit \passthrough{\lstinline!QGCompWarning!}.

\paragraph{Value}\label{manual-value}

A \passthrough{\lstinline!QuantizationResult!} containing read-only
scores, saved boundaries, category counts, feature names, training
ranges and diagnostics. Its \passthrough{\lstinline!transform!} method
applies the saved map to new observations.

\paragraph{Examples}\label{manual-examples}

\begin{lstlisting}[language=Python]
from quantilemix import quantize

tied = quantize([0, 0, 0, 1], q=4)
print(tied.scores.ravel())       # [0. 0. 0. 1.]
print(tied.category_counts)      # (2,)
print(tied.diagnostics["reduced_category_columns"])  # ('x1',)

explicit = quantize([0, 1, 1.5, 2, 3], breaks=[0, 1.5, 3])
print(explicit.scores.ravel())   # [0. 0. 0. 1. 1.]
print(explicit.mode, explicit.q) # explicit None
\end{lstlisting}

In the first example, ties reduce four requested groups to two. In the
second, the exact boundary value 1.5 belongs to the lower category.

\paragraph{See Also}\label{manual-see-also}

\hyperref[manual-api:quantization]{\passthrough{\lstinline!QuantizationResult!}},
\hyperref[manual-api:quantization-transform]{\passthrough{\lstinline!QuantizationResult.transform!}},
\hyperref[manual-api:fit]{\passthrough{\lstinline!fit!}}.

\paragraph{Source note.}\label{manual-source-note.}

The title, category-score explanation and alternative-cutpoint wording
are retained or adapted from Alexander Keil's R qgcomp 2.19.6
\passthrough{\lstinline!man/quantize.Rd!}, printed manual pages 76--77,
GPL (version 2 or later). The Python array interface, complete boundary
vectors, mode restrictions and result object replace the R data-frame
and column-selection interface.

\subsubsection{\texorpdfstring{\texttt{QuantizationResult}: Exposure
scores and their saved
transformation}{QuantizationResult: Exposure scores and their saved transformation}}\label{manual-api:quantization}

\paragraph{Description}\label{manual-description-2}

The object returned by \passthrough{\lstinline!quantize!} and either
\passthrough{\lstinline!transform!} method. It stores scores, the
original exposure map, and diagnostics for the rows represented by those
scores.

\paragraph{Usage}\label{manual-usage-1}

\begin{lstlisting}[language=Python]
quantization = quantize(exposures, q=4)
quantization.scores
quantization.breaks
quantization.transform(new_exposures)
\end{lstlisting}

\paragraph{Arguments}\label{manual-arguments-1}

Obtain the object from \passthrough{\lstinline!quantize!} or
\passthrough{\lstinline!transform!}. The generated dataclass constructor
requires internal label state and does not validate manually supplied
fields.

\paragraph{Details}\label{manual-details-1}

Results made by \passthrough{\lstinline!quantize!} or
\passthrough{\lstinline!transform!} cannot be changed in place: their
stored arrays and mappings are read-only. A transformed result retains
the original map's boundaries, training ranges and fitted category
counts. Its scores and diagnostics describe the new rows. It does not
replace the training map.

The saved category count can differ from the number of categories
present in new data. When boundaries are available, the saved count
includes categories that no new observation occupies. With previously
assigned scores, no full list of possible categories was supplied. The
saved count is then the number of distinct training scores, which need
not equal the largest score plus one.

\paragraph{Value}\label{manual-value-1}

Nine public fields are available:

{\def\LTcaptype{none} % do not increment counter
\begin{longtable}[]{@{}
  >{\raggedright\arraybackslash}p{(\linewidth - 2\tabcolsep) * \real{0.5000}}
  >{\raggedright\arraybackslash}p{(\linewidth - 2\tabcolsep) * \real{0.5000}}@{}}
\toprule\noalign{}
\begin{minipage}[b]{\linewidth}\raggedright
Field
\end{minipage} & \begin{minipage}[b]{\linewidth}\raggedright
Meaning
\end{minipage} \\
\midrule\noalign{}
\endhead
\bottomrule\noalign{}
\endlastfoot
\passthrough{\lstinline!scores!} & Read-only float array
\passthrough{\lstinline!(n, d)!} or \passthrough{\lstinline!(m, d)!};
integer-valued exposure scores for the represented rows. \\
\passthrough{\lstinline!breaks!} & Tuple of read-only boundary vectors,
including outer endpoints; \passthrough{\lstinline!None!} in
prequantized mode. \\
\passthrough{\lstinline!category\_counts!} & Tuple of fitted bin counts,
or distinct training score counts when no boundaries were supplied.
Preserved by transform. \\
\passthrough{\lstinline!mode!} & \passthrough{\lstinline!"quantile"!},
\passthrough{\lstinline!"explicit"!} or
\passthrough{\lstinline!"prequantized"!}. \\
\passthrough{\lstinline!q!} & Requested group count in automatic mode;
\passthrough{\lstinline!None!} in explicit or prequantized mode. \\
\passthrough{\lstinline!feature\_names!} & Tuple of exposure names in
column order. \\
\passthrough{\lstinline!training\_min!} & Read-only vector
\passthrough{\lstinline!(d,)!} of minima of the original training
inputs, before grouping. \\
\passthrough{\lstinline!training\_max!} & Read-only vector
\passthrough{\lstinline!(d,)!} of corresponding training maxima. \\
\passthrough{\lstinline!diagnostics!} & Read-only mapping with the six
keys below. \\
\end{longtable}
}

The public method is \passthrough{\lstinline!transform!}. There is no
\passthrough{\lstinline!summary()!} method. The underscore-prefixed
label field is internal state.

\paragraph{Quantization diagnostics.}\label{manual-quantization-diagnostics.}

All row indices and bin numbers are zero-based. The same mapping appears
under \passthrough{\lstinline!result.diagnostics["quantization"]!} for
training observations.

{\def\LTcaptype{none} % do not increment counter
\begin{longtable}[]{@{}
  >{\raggedright\arraybackslash}p{(\linewidth - 2\tabcolsep) * \real{0.5000}}
  >{\raggedright\arraybackslash}p{(\linewidth - 2\tabcolsep) * \real{0.5000}}@{}}
\toprule\noalign{}
\begin{minipage}[b]{\linewidth}\raggedright
Key
\end{minipage} & \begin{minipage}[b]{\linewidth}\raggedright
Meaning
\end{minipage} \\
\midrule\noalign{}
\endhead
\bottomrule\noalign{}
\endlastfoot
\passthrough{\lstinline!extrapolated\_rows!} & Tuple of row positions
having at least one exposure below its training minimum or above its
training maximum. \\
\passthrough{\lstinline!extrapolated\_columns!} & Tuple of names having
at least one such observation. \\
\passthrough{\lstinline!extrapolated\_counts!} & One count of such
values per exposure. \\
\passthrough{\lstinline!observed\_category\_counts!} & One
distinct-score count per exposure in the currently represented rows. May
differ from saved \passthrough{\lstinline!category\_counts!} after
transformation. \\
\passthrough{\lstinline!empty\_bins!} & With boundaries, one tuple of
currently unoccupied fitted bin numbers per exposure.
\passthrough{\lstinline!None!} in prequantized mode because the complete
category domain was not specified. \\
\passthrough{\lstinline!reduced\_category\_columns!} & Names whose
number of fitted bins is less than requested
\passthrough{\lstinline!q!}. Only automatic mode can populate it. \\
\end{longtable}
}

Initial training extrapolation counts are zero by construction. Neither
these counts nor observed categories establish adequate joint exposure
support or validate the regression model.

\paragraph{Examples}\label{manual-examples-1}

\begin{lstlisting}[language=Python]
from quantilemix import quantize

quantization = quantize([0, 2, 8, 10], q=4)
print(quantization.scores.ravel())   # [0. 1. 2. 3.]
print(quantization.breaks[0])
# [-1.0e+64  1.5e+00  5.0e+00  8.5e+00  1.0e+64]
print(quantization.feature_names)    # ('x1',)
print(quantization.training_min)     # [0.]
print(quantization.training_max)     # [10.]
\end{lstlisting}

\paragraph{See Also}\label{manual-see-also-1}

\hyperref[manual-api:quantize]{\passthrough{\lstinline!quantize!}},
\hyperref[manual-api:quantization-transform]{\passthrough{\lstinline!QuantizationResult.transform!}},
\hyperref[manual-api:result-transform]{\passthrough{\lstinline!QGCompResult.transform!}}.

\paragraph{Source note.}\label{manual-source-note.-1}

R qgcomp 2.19.6 \passthrough{\lstinline!man/quantize.Rd!} describes
returned quantized data and retained cutpoints. This Python class
extends that organization with explicit mode, category-count,
training-range and diagnostic fields. It does not reproduce the R list
interface.

\subsubsection{\texorpdfstring{\texttt{QuantizationResult.transform}:
Score new data with saved
cutpoints}{QuantizationResult.transform: Score new data with saved cutpoints}}\label{manual-api:quantization-transform}

\paragraph{Description}\label{manual-description-3}

Apply existing category boundaries to new exposures without calculating
new cutpoints. In prequantized mode, check and retain the new scores.

\paragraph{Usage}\label{manual-usage-2}

\begin{lstlisting}[language=Python]
transformed = quantization.transform(exposures)
\end{lstlisting}

\paragraph{Arguments}\label{manual-arguments-2}

\passthrough{\lstinline!exposures!}: nonempty real, finite array
\passthrough{\lstinline!(m, d)!}, or a vector for one exposure. It must
have the fitted column count and order. When the original inputs were
labeled, names and order must match exactly, including the presence of
labels.

\paragraph{Details}\label{manual-details-2}

The original boundaries, mode, names, category counts and training
ranges stay the same. Diagnostics identify new observations outside the
original training ranges. Crossing a training range alone does not
trigger a warning.

\passthrough{\lstinline!ValueError!} is raised for invalid data, shapes,
masks or labels; an exposure count different from the fitted map; values
outside saved endpoints; or invalid prequantized scores. For automatic
maps, new values can enter the extreme bins beyond the training range,
provided they lie within the saved \(\pm10^{64}\) endpoints. No
regression is fitted.

\paragraph{Value}\label{manual-value-2}

A new \passthrough{\lstinline!QuantizationResult!}. Its
\passthrough{\lstinline!scores!} and diagnostics refer to the supplied
rows; its map metadata remains unchanged.

\paragraph{Examples}\label{manual-examples-2}

\begin{lstlisting}[language=Python]
from quantilemix import quantize

quantization = quantize([0, 1, 2, 3], q=2)
new = quantization.transform([1.5, 2])
print(new.scores.ravel())        # [0. 1.]
print(new.category_counts)      # (2,)
\end{lstlisting}

The original cut is 1.5. The value equal to that cut belongs to the
lower category; cutpoints are not recomputed from the two new
observations.

\paragraph{See Also}\label{manual-see-also-2}

\hyperref[manual-api:quantize]{\passthrough{\lstinline!quantize!}},
\hyperref[manual-api:quantization]{\passthrough{\lstinline!QuantizationResult!}},
\hyperref[manual-api:result-transform]{\passthrough{\lstinline!QGCompResult.transform!}}.

\subsubsection{\texorpdfstring{\texttt{qgcomp}: Select an additive model
from a
formula}{qgcomp: Select an additive model from a formula}}\label{manual-qgcomp-select-an-additive-model-from-a-formula}

\paragraph{Usage}\label{manual-usage-3}

\begin{lstlisting}[language=Python]
from quantilemix import qgcomp
result = qgcomp("y ~ x1 + x2 + z", data, ("x1", "x2"),
                family="gaussian", q=4)
binary_rr = qgcomp("y ~ x1 + x2", data, ("x1", "x2"),
                   family="binomial", B=200, seed=17)
binary_or = qgcomp("y ~ x1 + x2", data, ("x1", "x2"),
                   family="binomial", rr=False, q=2)
survival = qgcomp("Surv(time,event) ~ x1 + x2", data,
                  ("x1", "x2"), family="cox", q=4)
\end{lstlisting}

\paragraph{Arguments and routing}\label{manual-arguments-and-routing}

\passthrough{\lstinline!formula!} names an outcome, exposures and
adjustment terms; \passthrough{\lstinline!data!} supplies named columns
and \passthrough{\lstinline!exposures!} lists the exposure names.
\passthrough{\lstinline!C(z)!} encodes a categorical adjustment column,
with sorted levels and the first as reference. Gaussian and Poisson
outcomes use the existing conditional additive fitters. For a binary
outcome, the default \passthrough{\lstinline!rr=True!} selects a
marginal risk ratio with fixed-cut row-bootstrap uncertainty. With the
default \passthrough{\lstinline!q=4!}, its one-category ratio is the
exponentiated slope of a degree-one log-risk marginal model. Set
\passthrough{\lstinline!rr=False!} to select the conditional log
odds-ratio fit. The ratio and odds ratio are different measures;
switching between them changes the scientific question. R ignores
\passthrough{\lstinline!rr!} for Gaussian and Poisson outcomes, and so
does this dispatcher. An additive
\passthrough{\lstinline!Surv(time,event)!} formula with
\passthrough{\lstinline!family="cox"!} selects the conditional Cox fit.
It accepts complete rows, positive case weights, exact integer cluster
IDs and known log-hazard offsets; it does not select a marginal Cox
bootstrap.

\passthrough{\lstinline!q!}, \passthrough{\lstinline!breaks!} and
\passthrough{\lstinline!alpha!} pass to the selected fitter.
\passthrough{\lstinline!weights!} and
\passthrough{\lstinline!missing="drop"!} are available for the
conditional fits. \passthrough{\lstinline!B!} and
\passthrough{\lstinline!seed!} are available only for the binary
risk-ratio route. An \passthrough{\lstinline!exposure\_time!} column
name gives a Poisson observation-time offset; its additive route remains
conditional, as in R. The Python formula omits the R term
\passthrough{\lstinline!offset(log(time))!} and instead uses
\passthrough{\lstinline!exposure\_time="time"!}. An explicit R-style
offset term in the Python formula is rejected.

\paragraph{Limits and value}\label{manual-limits-and-value}

The binary marginal method requires complete numeric data without case
weights and may fail on a separated or boundary bootstrap draw. Its
four-level interval was assessed in four known-score conditions (4,000
datasets; see the
\href{https://github.com/nilson01/quantilemix/blob/main/validation/RISK_RATIO_Q4_CALIBRATION_RESULTS.md}{study
report}). Formulas with nonlinear terms or zero-inflated outcomes, and
other unsupported formulas, raise an error. The function returns the
unchanged result type of the selected fitter. Six original routes and
two four-level binary cases agree with isolated installed R qgcomp
2.19.6. The dispatch and four-level comparison reports give the saved
values and limits.

\subsubsection{\texorpdfstring{\texttt{simdata\_quantized}: Generate
example data with exposure
scores}{simdata\_quantized: Generate example data with exposure scores}}\label{manual-simdata_quantized-generate-example-data-with-exposure-scores}

\paragraph{Usage}\label{manual-usage-4}

\begin{lstlisting}[language=Python]
sample = simdata_quantized("survival", n=100, coef=[0.2, -0.1],
                           q=4, scale0=5, seed=17)
\end{lstlisting}

\paragraph{Arguments and Value}\label{manual-arguments-and-value}

\passthrough{\lstinline!outcometype!} is
\passthrough{\lstinline!"continuous"!},
\passthrough{\lstinline!"logistic"!},
\passthrough{\lstinline!"survival"!} or
\passthrough{\lstinline!"multinomial"!}. \passthrough{\lstinline!coef!}
has one entry per exposure, except multinomial outcomes, where it has
one column per nonreference outcome category.
\passthrough{\lstinline!corr!} controls how often later exposures copy
the first exposure's score, or its reverse order for a negative value.
It does not set an exact Pearson correlation.
\passthrough{\lstinline!n!}, \passthrough{\lstinline!q!},
\passthrough{\lstinline!b0!}, \passthrough{\lstinline!yscale!},
\passthrough{\lstinline!shape0!}, \passthrough{\lstinline!scale0!} and
\passthrough{\lstinline!censtime!} control sample size, exposure levels
and outcome generation. \passthrough{\lstinline!seed!} uses a local
NumPy generator. The read-only
\passthrough{\lstinline!QuantizedSimulation!} mapping contains
\passthrough{\lstinline!x1!}, \passthrough{\lstinline!x2!}, and an
outcome \passthrough{\lstinline!y!}, or survival
\passthrough{\lstinline!time!} and event indicator
\passthrough{\lstinline!d!}. Its \passthrough{\lstinline!truecoefs!}
property holds known generation coefficients separately.
\passthrough{\lstinline!to\_dict()!} returns editable column copies. It
can be passed as named \passthrough{\lstinline!data!} to formula
fitters. When \passthrough{\lstinline!ncheck=True!}, Python increases
\passthrough{\lstinline!n!} to the next multiple of
\passthrough{\lstinline!q!}; R 2.19.6 adds the remainder instead. Python
also uses the supplied survival \passthrough{\lstinline!scale0!}, which
that R version does not forward. Identical numeric seeds do not give
identical Python and R draws.

\subsubsection{\texorpdfstring{\texttt{impute\_leftcenslognorm}:
Left-censored lognormal
imputation}{impute\_leftcenslognorm: Left-censored lognormal imputation}}\label{manual-impute_leftcenslognorm-left-censored-lognormal-imputation}

\paragraph{Usage}\label{manual-usage-5}

\begin{lstlisting}[language=Python]
values = impute_leftcenslognorm(y, ry, x, lod=1.0, seed=17)
\end{lstlisting}

\paragraph{Arguments and Value}\label{manual-arguments-and-value-1}

\passthrough{\lstinline!ry!} marks observed positive values in
\passthrough{\lstinline!y!}; \passthrough{\lstinline!wy!} defaults to
their complement and selects the censored rows to impute.
\passthrough{\lstinline!x!} is a complete numeric predictor matrix
without an intercept. \passthrough{\lstinline!lod!} is a positive scalar
or one limit per row; if omitted, the smallest observed value is used.
The function fits a lognormal model using observed values and the fact
that censored values lie below their limits. It draws model parameters
using a normal approximation based on the fitted information matrix,
then draws each replacement value below its limit. It returns positive
imputed values in selected-row order.
\passthrough{\lstinline!impute\_tobit!} is the same function, not
another algorithm. This is a Python helper for an external imputation
workflow, not an R \passthrough{\lstinline!mice!} callback or a pooling
method. Fixed input checks pass; installed-R agreement and
repeated-imputation calibration remain open.

\subsection{Gaussian models and
results}\label{manual-gaussian-models-and-results}

\subsubsection{\texorpdfstring{\texttt{fit}: Additive Gaussian quantile
g-computation}{fit: Additive Gaussian quantile g-computation}}\label{manual-api:fit}

\paragraph{Description}\label{manual-description-4}

With automatic quantile grouping, the following description from R
qgcomp applies:

\begin{quote}
This function estimates a linear dose-response parameter representing a
one quantile increase in a set of exposures of interest. This function
is limited to linear and additive effects of individual components of
the exposure.
\end{quote}

The Python function fits a continuous outcome using an intercept,
exposure scores and optional numeric adjustment covariates. It returns
\(\widehat\psi=\sum_j\widehat\beta_j\), the sum of fitted exposure
coefficients, and its model-based uncertainty. For supplied boundaries
or prequantized scores, \(\widehat\psi\) is the fitted mean change when
every exposure score increases by one category.

\paragraph{Usage}\label{manual-usage-6}

\begin{lstlisting}[language=Python]
from quantilemix import fit

result = fit(y, exposures)
\end{lstlisting}

\paragraph{Arguments}\label{manual-arguments-3}

Let \(n\) be the number of observations, \(d\geq1\) the number of
exposures, and \(c\) the number of adjustment covariates.

{\def\LTcaptype{none} % do not increment counter
\begin{longtable}[]{@{}
  >{\raggedright\arraybackslash}p{(\linewidth - 2\tabcolsep) * \real{0.5000}}
  >{\raggedright\arraybackslash}p{(\linewidth - 2\tabcolsep) * \real{0.5000}}@{}}
\toprule\noalign{}
\begin{minipage}[b]{\linewidth}\raggedright
Argument
\end{minipage} & \begin{minipage}[b]{\linewidth}\raggedright
Meaning and accepted inputs
\end{minipage} \\
\midrule\noalign{}
\endhead
\bottomrule\noalign{}
\endlastfoot
\passthrough{\lstinline!y!} & Nonempty real, finite outcome vector of
shape \passthrough{\lstinline!(n,)!}. A column matrix of shape
\passthrough{\lstinline!(n, 1)!} is rejected. \\
\passthrough{\lstinline!exposures!} & Real, finite exposure array of
shape \passthrough{\lstinline!(n, d)!}. A vector of shape
\passthrough{\lstinline!(n,)!} represents one exposure. All selected
columns contribute to \(\widehat\psi\). \\
\passthrough{\lstinline!covariates!} & \passthrough{\lstinline!None!} by
default, or a real, finite array of shape
\passthrough{\lstinline!(n, c)!}; a vector represents one covariate.
These coefficients adjust the model but do not enter the joint sum. For
this array entry point, encode categorical covariates in advance;
\passthrough{\lstinline!fit\_formula!} handles complete text categories.
Do not supply an intercept column. Use \passthrough{\lstinline!None!},
rather than an empty-column matrix, when there are no covariates. \\
\passthrough{\lstinline!q!} & Requested number of quantile groups;
default \passthrough{\lstinline!4!}. In automatic mode, an integer from
2 through 100,000; booleans are invalid. With
\passthrough{\lstinline!q=None!} and no breaks, inputs must already be
nonnegative integer-valued scores strictly below \(2^{53}\). Supplied
breaks take priority. \\
\passthrough{\lstinline!breaks!} & Default
\passthrough{\lstinline!None!}. When supplied, use one finite, strictly
increasing vector per exposure, including both outer endpoints. Vector
lengths may differ across exposures. One vector is sufficient for a
single exposure. This is an alternative to using
\passthrough{\lstinline!q!} to define cutpoints. \\
\passthrough{\lstinline!alpha!} & Alpha level for confidence limit
calculation; default \passthrough{\lstinline!0.05!}. Must be a finite
real number strictly between 0 and 1; booleans are invalid. The nominal
interval coverage is \(1-\alpha\). \\
\passthrough{\lstinline!weights!} & Default
\passthrough{\lstinline!None!}. When supplied, use a positive, finite
case-weight vector of shape \passthrough{\lstinline!(n,)!}. One value is
required per training row. Weights affect the Gaussian regression, not
the quantile cuts. Zero, negative, missing and masked weights are
rejected. \\
\end{longtable}
}

\paragraph{Details}\label{manual-details-3}

Write \(X_{ji}^{q}\) for exposure \(j\)'s score in observation \(i\),
and \(Z_i\) for the vector of adjustment covariates. The fitted mean
model is

\[
E(Y_i\mid X_i^q,Z_i)
 =\beta_0+\sum_{j=1}^{d}\beta_j X_{ji}^{q}+\gamma^\top Z_i.
\]

Here \(\beta_0\) is the intercept, \(\beta_j\) is an exposure-score
coefficient, and \(\gamma\) contains the covariate coefficients. The
joint contrast is

\[
\psi=\sum_{j=1}^{d}\beta_j.
\]

Holding covariates fixed, increasing each exposure score by one changes
the fitted mean by \(\psi\), where both score profiles are meaningful.
This is a category-scale contrast, not a one-unit increase in each
original measurement. It does not define a policy that clips increases
at the highest category.

For quantile grouping, the R documentation states:

\begin{quote}
For continuous outcomes, under a linear model with no interaction terms,
this is equivalent to g-computation of the effect of increasing every
exposure by 1 quantile.
\end{quote}

A causal interpretation also requires appropriate study assumptions,
including well-defined interventions, consistency, adequate joint
exposure support and no unmeasured confounding. The software does not
establish these assumptions.

Let \(D=[\mathbf1,X^q,Z]\) be the design matrix, with \(k=1+d+c\)
columns. The function requires full column rank and \(n>k\). It fits
ordinary least squares by default. With positive case weights \(w_i\),
it minimizes \(\sum_i w_i(Y_i-D_i\theta)^2\). Both modes use a
singular-value decomposition (SVD). If \(\widehat V\) is the estimated
covariance of all coefficients, and \(a\) is one for exposure
coefficients and zero elsewhere, then

\[
\widehat\psi=a^\top\widehat\theta,\qquad
 \widehat{\mathrm{Var}}(\widehat\psi)=a^\top\widehat V a,\qquad
 \mathrm{SE}(\widehat\psi)=\sqrt{a^\top\widehat V a},
\]

where \(\widehat\theta\) is the vector of all fitted coefficients. Thus
uncertainty includes covariance between exposure coefficients. Adding
their standard errors, or only their individual variances, gives a
different quantity. The software uses singular-value decomposition (SVD)
to reduce loss of precision when calculating the joint variance.

Residual variance uses \(n-1-d-c\) degrees of freedom. The confidence
interval uses a standard-normal critical value:

\[
\widehat\psi\ \mathbin{\pm}\
 z_{1-\alpha/2}\mathrm{SE}(\widehat\psi),
\]

where \(z_u\) is the standard-normal quantile at probability \(u\). The
two-sided p-value instead uses a Student \(t\) distribution with
\(n-1-d\) degrees of freedom and statistic
\(\widehat\psi/\mathrm{SE}(\widehat\psi)\). These are the pinned R
qgcomp 2.19.6 conventions. The interval and test can disagree near a
significance threshold. The normal interval is not an exact
finite-sample Gaussian interval. Changing
\passthrough{\lstinline!alpha!} changes only the interval limits.

Model-based uncertainty assumes independent observations and constant
conditional residual variance. Case weights use model-based Gaussian
covariance, not a complex-survey or cluster-robust variance. The current
interface does not provide bootstrap inference, penalization,
imputation, nonlinear exposure effects, exposure interactions, or other
outcome families.

Inputs must be complete. Missing values, active masks, infinities,
complex values, strings, object arrays and ragged arrays are rejected.
Rows and variables are not silently removed. Numeric inputs are
converted to double precision. An input with column labels must have
unique labels after their conversion to strings. All coefficient names
must be distinct across the intercept, exposures and covariates.
Unlabeled columns receive names \passthrough{\lstinline!x1!},
\passthrough{\lstinline!x2!}, \ldots{} and \passthrough{\lstinline!z1!},
\passthrough{\lstinline!z2!}, \ldots.

\paragraph{Errors and warnings.}\label{manual-errors-and-warnings.}

Invalid inputs, invalid quantization, insufficient residual degrees of
freedom, rank deficiency and nonfinite numerical results raise
\passthrough{\lstinline!ValueError!}. Unsupported keyword arguments
raise Python's \passthrough{\lstinline!TypeError!}. The function emits
\passthrough{\lstinline!QGCompWarning!} when the design condition number
exceeds \(10^8\), or when residual or contrast variation is numerically
zero. In the latter case, the interval is a point and
\passthrough{\lstinline!se!} is zero, but
\passthrough{\lstinline!test\_statistic!} and
\passthrough{\lstinline!p\_value!} are \passthrough{\lstinline!NaN!}.
Undefined inference is not evidence of significance.

\paragraph{Value}\label{manual-value-3}

A \passthrough{\lstinline!QGCompResult!} object containing the joint
estimate \passthrough{\lstinline!psi!} and its variance
\passthrough{\lstinline!var\_psi!}, the fitted model coefficients and
covariance, and the positive and negative within-group sign weights in
\passthrough{\lstinline!pos\_weights!} and
\passthrough{\lstinline!neg\_weights!}. It also stores residuals and the
supplied case weights, confidence limits, p-value, saved exposure
boundaries and diagnostics. The complete fields and properties appear in
the \hyperref[manual-api:result]{\passthrough{\lstinline!QGCompResult!} entry}.

\paragraph{Examples}\label{manual-examples-3}

These 16 constructed observations use two exposures already scored 0--3.
The final term supplies a fixed residual orthogonal to the additive
model columns; it is not an interaction term fitted by the interface.

\begin{lstlisting}[language=Python]
import numpy as np
from quantilemix import fit

x1, x2 = np.meshgrid(np.arange(4), np.arange(4))
X = np.column_stack([x1.ravel(), x2.ravel()])
y = 1 + 2*X[:, 0] - 0.5*X[:, 1]
y += 0.2*(X[:, 0]-1.5)*(X[:, 1]-1.5)
result = fit(y, X, q=None)

print(f"psi = {result.psi:.6f}")
print(f"SE = {result.se:.6f}")
print(f"95% CI = ({result.ci[0]:.6f}, {result.ci[1]:.6f})")
print(f"p = {result.p_value:.6g}")
\end{lstlisting}

\begin{lstlisting}
psi = 1.500000
SE = 0.087706
95% CI = (1.328100, 1.671900)
p = 2.70924e-10
\end{lstlisting}

The fitted exposure coefficients are 2 and \(-0.5\), so their sum is
1.5. The standard error describes uncertainty in that sum. The p-value
tests a zero joint score contrast under the stated model and reference
convention. These constructed values illustrate the calculation; they
are not empirical evidence about an exposure's health effect.

A fixed check with unequal positive case weights uses eight scored rows
with two correlated exposures. Exact rational arithmetic gives
\(\widehat\psi=833/183\) and
\(\widehat{\mathrm{Var}}(\widehat\psi)=198527/334890\). Use
\passthrough{\lstinline!fit(y, X, q=None, weights=w)!} with one weight
per row. The weights affect the fit but do not change exposure scores or
represent a survey variance.

For a raw-measurement example, see
\hyperref[manual-api:predict]{\passthrough{\lstinline!QGCompResult.predict!}}.

\paragraph{See Also}\label{manual-see-also-3}

\hyperref[manual-api:quantize]{\passthrough{\lstinline!quantize!}},
\hyperref[manual-api:result]{\passthrough{\lstinline!QGCompResult!}},
\hyperref[manual-api:predict]{\passthrough{\lstinline!QGCompResult.predict!}}.

\paragraph{Source note.}\label{manual-source-note.-2}

The quoted description and continuous-outcome explanation, the
\passthrough{\lstinline!alpha!} wording, and the sequence of returned
quantities are retained or adapted from Alexander Keil's R qgcomp 2.19.6
\passthrough{\lstinline!man/qgcomp.glm.noboot.Rd!}, printed manual pages
51--53, licensed GPL (version 2 or later). Python signatures,
restrictions, object fields and examples describe this separate,
unofficial implementation. Binary/count outcomes and other R options
have not been imported into this entry's supported scope.

\subsubsection{\texorpdfstring{\texttt{fit\_formula}: Additive Gaussian
fit from named
data}{fit\_formula: Additive Gaussian fit from named data}}\label{manual-api:fit-formula}

\paragraph{Description}\label{manual-description-5}

This restricted Python interface prepares a numeric design from named
columns, then calls the same Gaussian \passthrough{\lstinline!fit!}
estimator documented above. It supports additive main effects and text
adjustment categories. It does not change the joint-effect or
uncertainty formulas.

\paragraph{Usage}\label{manual-usage-7}

\begin{lstlisting}[language=Python]
from quantilemix import fit_formula

result = fit_formula(
    "y ~ x1 + x2 + C(group)", data, ("x1", "x2"),
    q=4, breaks=None, alpha=0.05, weights=None,
    missing="raise"
)
new_mean = result.predict_data(new_data)
new_scores = result.transform_data(new_data)
\end{lstlisting}

\paragraph{Arguments}\label{manual-arguments-4}

\passthrough{\lstinline!formula!} is a string with one outcome name, a
tilde, and additive terms separated by plus signs. Names are simple
letters, digits and underscores, starting with a letter or underscore.
\passthrough{\lstinline!C(name)!} marks a text categorical adjustment
column. \passthrough{\lstinline!data!} is a mapping or table with named,
one-dimensional columns of equal length.
\passthrough{\lstinline!exposures!} is a separate, nonempty sequence of
numeric exposure names in the formula. The remaining numeric terms are
adjustment variables. \passthrough{\lstinline!q!},
\passthrough{\lstinline!breaks!}, \passthrough{\lstinline!alpha!} and
\passthrough{\lstinline!weights!} have the same meaning as in
\passthrough{\lstinline!fit!}. \passthrough{\lstinline!missing!} is
\passthrough{\lstinline!"raise"!} by default or
\passthrough{\lstinline!"drop"!} for explicit complete-case fitting.

\paragraph{Details}\label{manual-details-4}

For a text category, the sorted first observed level is the reference.
Each later level creates one indicator column. The result saves this
level order and uses it again for
\passthrough{\lstinline!predict\_data!}; an unseen level is an error.
\passthrough{\lstinline!transform\_data!} needs only exposure columns.
The fitted coefficient order is intercept, exposures in the supplied
exposure order, then adjustment columns in formula order. R's factor
columns use names such as \passthrough{\lstinline!gB!}; Python uses
\passthrough{\lstinline!g[T.B]!} for the same value and position.

The default rejects missing values. With
\passthrough{\lstinline!missing="drop"!}, the fit omits rows missing an
outcome, exposure, adjustment value, category or case weight. Automatic
exposure cutpoints use all observed exposure values before this
omission, matching the order in R qgcomp 2.19.6. This option does not
impute missing values. Infinities, categorical exposures, interactions,
transformations, offsets and no-intercept formulas remain unsupported.
New prediction data must be complete.

\paragraph{Value}\label{manual-value-4}

A \passthrough{\lstinline!QGCompResult!}. Its
\passthrough{\lstinline!formula!} property records the formula. The
\passthrough{\lstinline!categorical\_levels!} property stores the
observed level order. \passthrough{\lstinline!training\_row\_indices!}
lists retained zero-based original positions;
\passthrough{\lstinline!original\_row\_count!} gives the number of input
rows. The named-data methods use these saved columns and cutpoints.

\paragraph{Examples}\label{manual-examples-4}

\begin{lstlisting}[language=Python]
from itertools import product
import numpy as np
from quantilemix import fit_formula

rows = list(product(range(4), range(4), ("A", "B", "C")))
x1 = np.array([row[0] for row in rows])
x2 = np.array([row[1] for row in rows])
group = [row[2] for row in rows]
y = (2 + 1.5*x1 - .4*x2 + .7*(np.array(group) == "B")
     - .3*(np.array(group) == "C") + .2*(x1-1.5)*(x2-1.5))
result = fit_formula("y ~ x1 + x2 + C(group)",
                     {"y": y, "x1": x1, "x2": x2, "group": group},
                     ("x1", "x2"), q=None)
assert abs(result.psi - 1.1) < 1e-12
\end{lstlisting}

The two exposure coefficients are \(1.5\) and \(-0.4\), so their joint
one-category score contrast is \(1.1\).

For explicit complete-case fitting, copy the input and mark the missing
entry. The returned row positions refer to the original input order:

\begin{lstlisting}[language=Python]
incomplete_y = y.astype(float).copy()
incomplete_y[0] = np.nan
incomplete = fit_formula(
    "y ~ x1 + x2 + C(group)",
    {"y": incomplete_y, "x1": x1, "x2": x2, "group": group},
    ("x1", "x2"), q=None, missing="drop"
)
assert incomplete.training_row_indices == tuple(range(1, 48))
\end{lstlisting}

\paragraph{See Also}\label{manual-see-also-4}

\hyperref[manual-api:fit]{\passthrough{\lstinline!fit!}},
\hyperref[manual-api:result]{\passthrough{\lstinline!QGCompResult!}}.

\subsubsection{\texorpdfstring{\texttt{fit\_bootstrap}: Fixed-cut
Gaussian row
bootstrap}{fit\_bootstrap: Fixed-cut Gaussian row bootstrap}}\label{manual-api:fit-bootstrap}

\paragraph{Description}\label{manual-description-6}

This function estimates the mean change when every exposure score rises
by one category. It keeps the group boundaries calculated from the
original data, then repeatedly draws whole rows with replacement to
estimate uncertainty. It supports the additive Gaussian, linear-curve
case of R \passthrough{\lstinline!qgcomp.glm.boot!}.

\paragraph{Usage}\label{manual-usage-8}

\begin{lstlisting}[language=Python]
from quantilemix import fit_bootstrap
result = fit_bootstrap(y, exposures, covariates=None,
                       q=4, breaks=None, alpha=0.05, B=200, seed=None,
                       weights=None)
\end{lstlisting}

\paragraph{Arguments}\label{manual-arguments-5}

\passthrough{\lstinline!y!}, \passthrough{\lstinline!exposures!},
\passthrough{\lstinline!covariates!}, \passthrough{\lstinline!q!},
\passthrough{\lstinline!breaks!} and \passthrough{\lstinline!alpha!}
follow the numeric-array Gaussian \passthrough{\lstinline!fit!} entry
below. This entry requires complete finite inputs and an intercept. For
\passthrough{\lstinline!q=None!}, pooled observed scores must have 2--9
levels. Supplied boundaries must yield the same category count for every
exposure. \passthrough{\lstinline!B!} is an integer of at least two
resamples. \passthrough{\lstinline!seed!} is an optional nonnegative
integer below \(2^{64}\); it controls a local NumPy generator.
\passthrough{\lstinline!weights!} optionally supplies one strictly
positive finite case weight per row. Exposure cutpoints remain
unweighted. Formulas and missing-row options are not accepted by this
array entry point. Use
\hyperref[manual-api:fit-bootstrap-formula]{\passthrough{\lstinline!fit\_bootstrap\_formula!}}
for a restricted named-data formula.

\paragraph{Details}\label{manual-details-5}

For intervention level \(a\), define the standardized fitted mean

\[
\widehat\mu(a)=\frac{\sum_{i=1}^n w_i\widehat E\{Y_i\mid
 X_{1i}^{q}=\cdots=X_{di}^{q}=a,Z_i\}}{\sum_{i=1}^n w_i}.
\]

Here \(n\) is the row count, \(w_i>0\) is row \(i\)'s case weight (one
if weights are omitted), \(d\) is the exposure count, and \(X^q\)
contains exposure scores and \(Z_i\) contains observation \(i\)'s
covariates. The slope of \(\widehat\mu(a)\) is the joint effect
\(\widehat\psi\). For this linear additive Gaussian model it equals the
sum of exposure coefficients; that equality does not generally hold with
nonlinear terms or interactions.

Each resample draws \(n\) whole rows with replacement, keeps each
selected row's weight and refits the model using the original score
boundaries. This calculation does not account for complex survey
sampling or dependence within clusters. Let \(s_B\) be the sample
standard deviation of the \(B\) fitted slopes. The interval is
\(\widehat\psi\pm z_{1-\alpha/2}s_B\), where \(z_u\) is a
standard-normal quantile. It is not a percentile interval. The two-sided
test uses a Student \(t\) distribution with \(n-d-c-2\) degrees of
freedom, where \(c\) is the number of numeric adjustment columns. A
failed replicate raises an error identifying its position; no
replacement draw is silently substituted.

\paragraph{Value}\label{manual-value-5}

\passthrough{\lstinline!GaussianBootstrapResult!} stores
\passthrough{\lstinline!psi!}, \passthrough{\lstinline!var\_psi!},
\passthrough{\lstinline!se!}, \passthrough{\lstinline!ci!},
\passthrough{\lstinline!test\_statistic!},
\passthrough{\lstinline!p\_value!},
\passthrough{\lstinline!p\_value\_df!}, \passthrough{\lstinline!alpha!},
\passthrough{\lstinline!B!}, \passthrough{\lstinline!seed!}, read-only
\passthrough{\lstinline!bootstrap\_effects!},
\passthrough{\lstinline!intervention\_levels!} and
\passthrough{\lstinline!intervention\_means!}. Its
\passthrough{\lstinline!point\_fit!} is the original
\passthrough{\lstinline!QGCompResult!}; its model-based uncertainty is
separate from the bootstrap values.
\passthrough{\lstinline!summary(digits=6)!} displays saved values and
method limits without refitting.

\paragraph{Examples}\label{manual-examples-5}

\begin{lstlisting}[language=Python]
boot = fit_bootstrap(y, exposures, q=None, B=200, seed=17)
print(boot.psi, boot.se, boot.ci)
print(boot.summary())
import numpy as np
case_weights = np.ones(len(y))
case_weights[::3] = 2.0
weighted_boot = fit_bootstrap(y, exposures, q=None,
                              weights=case_weights, B=200, seed=17)
print(weighted_boot.summary())
\end{lstlisting}

The installed-package worked example uses a smaller, fixed resample
count and checks the returned values. To compare individual resamples
between Python and R, both must draw the same rows in the same order.
The same integer seed does not ensure that. Two unweighted and two
weighted fixed installed-R cases pass. The weighted cases and optional
replay command are in
\passthrough{\lstinline!validation/WEIGHTED\_BOOTSTRAP\_RESULTS.md!}.
The 24,000-dataset study covers the earlier unweighted method; weighted
interval calibration remains open.

\paragraph{See Also}\label{manual-see-also-5}

\passthrough{\lstinline!fit!} for model-based Gaussian uncertainty, and
\passthrough{\lstinline!docs/reference.md!} for the optional installed-R
comparison.

\subsubsection{\texorpdfstring{\texttt{fit\_bootstrap\_formula}:
Gaussian row bootstrap from named
data}{fit\_bootstrap\_formula: Gaussian row bootstrap from named data}}\label{manual-api:fit-bootstrap-formula}

\paragraph{Description}\label{manual-description-7}

This function prepares data from named columns, then runs the Gaussian
row-resampling method described above with the original exposure
cutpoints. It supports additive numeric exposures and numeric or
text-category adjustment terms. It does not change the estimator or
interval method.

\paragraph{Usage}\label{manual-usage-9}

\begin{lstlisting}[language=Python]
from quantilemix import fit_bootstrap_formula

result = fit_bootstrap_formula(
    "y ~ x1 + x2 + C(group)", data, ("x1", "x2"),
    q=4, breaks=None, alpha=0.05, B=200, seed=None,
    missing="raise", weights=None
)
\end{lstlisting}

\paragraph{Arguments}\label{manual-arguments-6}

\passthrough{\lstinline!formula!}, \passthrough{\lstinline!data!},
\passthrough{\lstinline!exposures!} and
\passthrough{\lstinline!missing!} follow the restricted
\hyperref[manual-api:fit-formula]{\passthrough{\lstinline!fit\_formula!}}
interface. \passthrough{\lstinline!q!},
\passthrough{\lstinline!breaks!}, \passthrough{\lstinline!alpha!},
\passthrough{\lstinline!B!} and \passthrough{\lstinline!seed!} follow
\passthrough{\lstinline!fit\_bootstrap!}. The default requires complete
rows. With \passthrough{\lstinline!missing="drop"!}, rows missing a
model value or weight are removed before resampling. Automatic exposure
cutpoints still use every observed exposure value before that removal.
Positive case weights follow sampled rows. Interactions,
transformations, offsets and imputation are not supported.

\paragraph{Value}\label{manual-value-6}

A \passthrough{\lstinline!GaussianBootstrapResult!}. Its
\passthrough{\lstinline!point\_fit!} stores the formula, category
levels, original row count and retained zero-based row positions.
\passthrough{\lstinline!point\_fit.predict\_data!} uses the saved
category levels and cutpoints. The returned result stores the resampling
standard error and normal interval. Its
\passthrough{\lstinline!point\_fit!} field separately stores model-based
uncertainty from the fit to the original rows.

\paragraph{Examples}\label{manual-examples-6}

\begin{lstlisting}[language=Python]
boot_named = fit_bootstrap_formula(
    "y ~ x1 + x2 + C(group)",
    {"y": y, "x1": x1, "x2": x2, "group": group},
    ("x1", "x2"), q=None, B=12, seed=19
)
print(boot_named.psi, boot_named.se, boot_named.ci)
\end{lstlisting}

The installed-package example checks this route against the array fit on
the same named columns and seed. A separate weight-omission check uses a
manually encoded array fit. These check input preparation; the fixed
installed-R comparisons test the underlying weighted and unweighted
bootstrap estimator.

\paragraph{See Also}\label{manual-see-also-6}

\hyperref[manual-api:fit-bootstrap]{\passthrough{\lstinline!fit\_bootstrap!}}
and \hyperref[manual-api:fit-formula]{\passthrough{\lstinline!fit\_formula!}}.

\subsubsection{\texorpdfstring{\texttt{fit\_quadratic\_bootstrap}:
Quadratic Gaussian response
curve}{fit\_quadratic\_bootstrap: Quadratic Gaussian response curve}}\label{manual-api:fit-quadratic-bootstrap}

\paragraph{Description}\label{manual-description-8}

Fit every exposure score and its square in a conditional Gaussian model.
Optionally add all distinct pairwise exposure-score products and every
exposure-score by numeric-covariate product. Set all scores to each
joint intervention level, average fitted means over observed adjustment
values, and fit a degree-two marginal response. A fixed-cut row
bootstrap estimates uncertainty for both curve coefficients.

\paragraph{Usage}\label{manual-usage-10}

\begin{lstlisting}[language=Python]
from quantilemix import fit_quadratic_bootstrap
result = fit_quadratic_bootstrap(
    y, exposures, covariates=None,
    q=4, breaks=None, alpha=0.05, B=200, seed=None,
    pairwise_interactions=False, covariate_interactions=False
)
print(result.psi, result.covariance)
print(result.effect_at(1))
\end{lstlisting}

\paragraph{Arguments}\label{manual-arguments-7}

\passthrough{\lstinline!y!} is a complete continuous outcome.
\passthrough{\lstinline!exposures!} contain complete numeric exposure
columns; \passthrough{\lstinline!covariates!} are optional complete
numeric adjustment columns. \passthrough{\lstinline!q!},
\passthrough{\lstinline!breaks!} and \passthrough{\lstinline!alpha!} use
the same quantization and interval conventions as
\passthrough{\lstinline!fit!}. At least three consecutive joint levels
starting at zero are required. \passthrough{\lstinline!B!} is an integer
of at least two row resamples. \passthrough{\lstinline!seed!} controls
NumPy's local random generator. Set
\passthrough{\lstinline!pairwise\_interactions=True!} to add every
distinct exposure-score product; this requires at least two exposures.
Set \passthrough{\lstinline!covariate\_interactions=True!} to add every
exposure-score by numeric-covariate product; this requires at least one
covariate.

\paragraph{Details}\label{manual-details-6}

The conditional model includes \(X_j^q\) and \((X_j^q)^2\) for each
exposure score. When requested, it also includes \(X_j^qX_k^q\) for
every \(j<k\). Each pairwise coefficient adds to the quadratic
coefficient of the joint marginal curve. Each requested
exposure-covariate product adds its coefficient multiplied by the
observed covariate mean to the linear coefficient. In a bootstrap
replicate, the covariate mean comes from that replicate's sampled rows.
The fitted marginal mean at joint level \(a\) is quadratic. Following
the paper, call its linear and quadratic coefficients \(\widehat\psi_1\)
and \(\widehat\psi_2\). The saved \passthrough{\lstinline!psi!} vector
contains \((\widehat\psi_1,\widehat\psi_2)\), not one constant effect.
The fitted mean difference from \(a\) to \(a+1\) is
\(\widehat\psi_1+(2a+1)\widehat\psi_2\). The method fixes exposure cuts
before independently resampling rows. Bootstrap coefficients give a
two-by-two covariance, normal Wald intervals and two-sided Student \(t\)
p-values. The reference degrees of freedom are \(n-2d-h-g-c-3\), with
\(n\) rows, \(d\) exposures and \(c\) numeric adjustment columns. Here
\(h=d(d-1)/2\) when pairwise products are fitted, and zero otherwise;
\(g=dc\) when exposure-covariate products are fitted, and zero
otherwise. R qgcomp 2.19.6 calls this a degree-two marginal structural
model. The method does not include clusters, weights or other families.

\paragraph{Value}\label{manual-value-7}

\passthrough{\lstinline!QuadraticGaussianBootstrapResult!} saves
conditional coefficients and covariance, intervention levels and means,
three marginal polynomial coefficients, \passthrough{\lstinline!psi!},
its bootstrap covariance, standard errors, intervals, statistics,
p-values, reference degrees of freedom and every resampled coefficient
pair. The Booleans \passthrough{\lstinline!pairwise\_interactions!} and
\passthrough{\lstinline!covariate\_interactions!} record the fitted
design. \passthrough{\lstinline!effect\_at(level)!} returns the fitted
one-category mean difference from that starting level.
\passthrough{\lstinline!transform!} reuses saved cuts;
\passthrough{\lstinline!summary!} displays the two coefficients and
uncertainty. Prequantized Python inputs retain
\passthrough{\lstinline!breaks=None!}, whereas R saves synthetic
endpoints.

\paragraph{Examples}\label{manual-examples-7}

\begin{lstlisting}[language=Python]
import numpy as np
from quantilemix import fit_quadratic_bootstrap

x = np.array([(a, b) for a in range(4) for b in range(4)] * 5)
z = np.tile([0, 1, 2, 1, 0], 16)
y = (1 + .4*x[:, 0] - .2*x[:, 1]
     + .3*x[:, 0]**2 - .1*x[:, 1]**2 + .5*z
     + .05*((7*np.arange(len(x))) % 9 - 4))
curve = fit_quadratic_bootstrap(y, x, z, q=None, B=12, seed=19)
print(curve.effect_at(1))  # fitted mean change from 1 to 2
interaction_curve = fit_quadratic_bootstrap(
    y + .12*x[:, 0]*x[:, 1], x, z, q=None, B=12, seed=23,
    pairwise_interactions=True)
print(interaction_curve.psi)  # includes the pairwise quadratic term
covariate_curve = fit_quadratic_bootstrap(
    y + .17*x[:, 0]*z - .13*x[:, 1]*z,
    x, z, q=None, B=12, seed=31, covariate_interactions=True)
print(covariate_curve.psi)  # averages products over observed z
\end{lstlisting}

Fixed checks compare both curve coefficients with an independent
least-squares calculation. Two square-only, two pairwise and two
exposure-covariate cases also check exact shared bootstrap row
schedules. Neither check establishes repeated-sample interval coverage.

\paragraph{See Also}\label{manual-see-also-7}

\hyperref[manual-api:fit-bootstrap]{\passthrough{\lstinline!fit\_bootstrap!}}
for the additive Gaussian degree-one case, and
\passthrough{\lstinline!docs/reference.md!} for the installed-R
comparison.

\subsubsection{\texorpdfstring{\texttt{fit\_gaussian\_ee}: Clustered
Gaussian estimating
equations}{fit\_gaussian\_ee: Clustered Gaussian estimating equations}}\label{manual-api:fit-gaussian-ee}

\paragraph{Description}\label{manual-description-9}

Estimate an additive Gaussian joint mean difference while accounting for
repeated observations within identified clusters. Sum each cluster's
conditional and marginal estimating-function contributions before
computing the robust covariance. This method has no small-cluster
correction.

\paragraph{Usage}\label{manual-usage-11}

\begin{lstlisting}[language=Python]
from quantilemix import fit_gaussian_ee
result = fit_gaussian_ee(
    y, exposures, cluster_ids, covariates=None,
    q=4, breaks=None, alpha=0.05, weights=None
)
print(result.psi, result.se, result.ci)
\end{lstlisting}

\paragraph{Arguments}\label{manual-arguments-8}

\passthrough{\lstinline!y!} is a complete continuous outcome;
\passthrough{\lstinline!exposures!} contain complete numeric columns.
\passthrough{\lstinline!cluster\_ids!} supplies one finite exact integer
ID per row. At least two distinct clusters are needed.
\passthrough{\lstinline!covariates!} are optional complete numeric
adjustment columns. \passthrough{\lstinline!q!} and
\passthrough{\lstinline!breaks!} set the saved exposure scores as in
\passthrough{\lstinline!fit!}; \passthrough{\lstinline!alpha!} is the
interval error level. Joint intervention scores must include at least
two consecutive levels starting at zero. The conditional design must
have full column rank. Optional positive
\passthrough{\lstinline!weights!} multiply each row's conditional and
marginal estimating functions. They do not change the exposure cuts.

\paragraph{Details}\label{manual-details-7}

The conditional equation fits the additive Gaussian mean described in
\hyperref[manual-sec:method]{Method and interpretation}. For each common
exposure level \(a\), the marginal equation fits a line to the
standardized mean \(\widehat\mu(a)\). Its one-category mean difference
\(\widehat\psi\) equals the sum of fitted exposure coefficients under
this additive model. Cluster totals enter the sandwich covariance
\(A^{-1}BA^{-\top}\), where \(A\) is the derivative of both estimating
equations and \(B\) sums cluster-score outer products. R qgcomp 2.19.6
uses a numerical derivative; Python uses the exact derivative. Weighted
fits average each intervention mean using the same case weights.
Confidence limits use a normal reference and the two-sided p-value uses
a Student \(t\) reference with \(n-d-c-2\) degrees of freedom, where
\(n\) is the row count, \(d\) the exposure count and \(c\) the
adjustment-column count. A small cluster count can make these
approximations unreliable.

\paragraph{Value}\label{manual-value-8}

\passthrough{\lstinline!GaussianEEResult!} saves conditional and
marginal coefficients and robust covariances, the full stacked
covariance, derivative bread, cluster meat, cluster IDs and count,
optional case weights, intervention levels and means, joint estimate,
variance, standard error, interval, test statistic, p-value and
reference degrees of freedom. \passthrough{\lstinline!transform!} reuses
saved exposure cuts; \passthrough{\lstinline!summary!} explains the
cluster-robust quantities. Prequantized Python inputs retain
\passthrough{\lstinline!breaks=None!}, whereas R saves synthetic
endpoints.

\paragraph{Examples}\label{manual-examples-8}

\begin{lstlisting}[language=Python]
import numpy as np
from quantilemix import fit_gaussian_ee

x = np.array([(a, b) for a in range(4) for b in range(4)] * 5)
ids = np.arange(len(x)) // 2
z = np.arange(len(x)) % 3
y = (1 + .4*x[:, 0] - .2*x[:, 1] + .3*z
     + .07*((ids*3) % 5 - 2))
clustered = fit_gaussian_ee(y, x, ids, z, q=None)
print(clustered.psi, clustered.se, clustered.cluster_count)
\end{lstlisting}

Fixed checks compare the cluster-robust contrast with an independent
least-squares and cluster-score calculation. Two fixed installed-R cases
check the full stacked covariance; neither check establishes interval
coverage across datasets. One further weighted installed-R case and an
integer-weight row-replication check cover this option; see
\passthrough{\lstinline!validation/WEIGHTED\_EE\_RESULTS.md!}.

\paragraph{See Also}\label{manual-see-also-8}

\hyperref[manual-api:fit-bootstrap]{\passthrough{\lstinline!fit\_bootstrap!}}
for independent-row resampling and
\passthrough{\lstinline!docs/reference.md!} for the optional R
comparison.

\subsubsection{\texorpdfstring{\texttt{fit\_quadratic\_gaussian\_ee}:
Quadratic Gaussian curve with clustered
data}{fit\_quadratic\_gaussian\_ee: Quadratic Gaussian curve with clustered data}}\label{manual-fit_quadratic_gaussian_ee-quadratic-gaussian-curve-with-clustered-data}

\paragraph{Description}\label{manual-description-10}

Estimate a degree-two population-average Gaussian response when rows may
share a cluster. This method adds each exposure score and its square to
the conditional mean. It can also add every pairwise exposure product
and every exposure-by-numeric-covariate product. Its uncertainty sums
estimating functions within each cluster; it does not resample rows.

\paragraph{Usage}\label{manual-usage-12}

\begin{lstlisting}[language=Python]
from quantilemix import fit_quadratic_gaussian_ee
result = fit_quadratic_gaussian_ee(
    y, exposures, cluster_ids, covariates=None,
    q=4, breaks=None, alpha=0.05, weights=None,
    pairwise_interactions=False,
    covariate_interactions=False
)
print(result.psi, result.covariance, result.effect_at(0))
\end{lstlisting}

\paragraph{Arguments}\label{manual-arguments-9}

\passthrough{\lstinline!y!} is a complete continuous outcome.
\passthrough{\lstinline!exposures!} and optional
\passthrough{\lstinline!covariates!} are complete finite numeric
columns. \passthrough{\lstinline!cluster\_ids!} has one exact integer ID
per row and at least two distinct IDs. The joint intervention levels
must be three or more consecutive values starting at zero.
\passthrough{\lstinline!q!} or \passthrough{\lstinline!breaks!} defines
exposure scores; \passthrough{\lstinline!q=None!} accepts already scored
columns. \passthrough{\lstinline!alpha!} sets the interval error level.
Optional \passthrough{\lstinline!weights!} are positive case weights.
They affect both estimating equations and their cluster scores, but
leave exposure cuts unweighted. The interaction flags are Boolean.
Exposure-covariate products require at least one numeric covariate.

\paragraph{Details}\label{manual-details-8}

The fitted marginal mean \(\widehat\mu(a)\) is quadratic in the common
score \(a\). The two entries of \passthrough{\lstinline!psi!} are the
paper's linear and quadratic mixture coefficients,
\((\widehat\psi_1,\widehat\psi_2)\). The fitted change from level \(a\)
to \(a+1\) is \(\widehat\psi_1+(2a+1)\widehat\psi_2\); a single constant
joint effect is generally not available. The sandwich covariance is
\(A^{-1}BA^{-\mathsf T}\), where \(A\) differentiates the conditional
and marginal estimating equations and \(B\) sums the outer products of
their cluster totals. Confidence limits use a normal reference.
Two-sided tests use a Student \(t\) reference with \(n-p-2\) degrees of
freedom; \(n\) is the row count and \(p\) is the number of conditional
coefficients, including the intercept. No small-cluster correction or
repeated-dataset coverage claim is made.

The three fixed installed-R cases agree on conditional and marginal
point coefficients and intervention means. The installed R 2.19.6
cluster-meat calculation retains precomputed squared and interaction
columns after changing exposure scores within a saved model frame.
Python recomputes those columns. A separate fresh-design R cluster
calculation agrees with Python's covariance; R's reported intervals can
differ. The companion validation report gives the numerical results.

\paragraph{Value}\label{manual-value-9}

\passthrough{\lstinline!QuadraticGaussianEEResult!} stores the
conditional coefficients, three marginal coefficients, two joint curve
coefficients, their robust covariance, standard errors, intervals, test
statistics and p-values. It also saves the full stacked covariance,
derivative bread, cluster meat, intervention means, scores, cuts, IDs
and optional case weights. Arrays are read-only.
\passthrough{\lstinline!effect\_at(level)!} calculates the fitted
adjacent-level change. \passthrough{\lstinline!summary()!} displays the
two curve terms and their limits.

\paragraph{Examples}\label{manual-examples-9}

\begin{lstlisting}[language=Python]
curve = fit_quadratic_gaussian_ee(
    y, x, ids, z, q=None, pairwise_interactions=True
)
print(curve.psi, curve.effect_at(0), curve.effect_at(2))
\end{lstlisting}

Fixed checks compare the two curve coefficients with an independent
polynomial contrast. These checks do not show that the interval covers a
population value at its stated rate across datasets.

\paragraph{See Also}\label{manual-see-also-9}

\hyperref[manual-api:fit-gaussian-ee]{\passthrough{\lstinline!fit\_gaussian\_ee!}}
gives an additive clustered mean.
\hyperref[manual-api:fit-quadratic-bootstrap]{\passthrough{\lstinline!fit\_quadratic\_bootstrap!}}
uses independent-row bootstrap uncertainty.

\subsubsection{\texorpdfstring{\texttt{QGCompResult}: Fitted model and
joint exposure
summaries}{QGCompResult: Fitted model and joint exposure summaries}}\label{manual-api:result}

\paragraph{Description}\label{manual-description-11}

The object returned by \passthrough{\lstinline!fit!}. It contains the
full fitted regression, the joint exposure contrast and uncertainty,
descriptive sign summaries, the saved exposure map and numerical
diagnostics.

\paragraph{Usage}\label{manual-usage-13}

\begin{lstlisting}[language=Python]
result = fit(y, exposures, covariates=None, q=4)
result.psi
result.ci
result.predict(new_exposures, covariates=new_covariates)
result.transform(new_exposures)
\end{lstlisting}

\paragraph{Arguments}\label{manual-arguments-10}

Obtain this object from \passthrough{\lstinline!fit!}, whose arguments
are documented above. The exported class has a generated dataclass
constructor, but that constructor requires internal transformation state
and does not validate manually supplied fields. Use
\passthrough{\lstinline!fit!} to construct a consistent result.

\paragraph{Details}\label{manual-details-9}

Factory-produced results are frozen dataclasses. Their arrays and
mappings are read-only, and changes to the original input arrays do not
change the stored fit. Coefficients and their full covariance are
ordered as intercept, exposures, then covariates. This covariance is not
the two-by-two intercept/mixture covariance returned by the R
mixture-level \passthrough{\lstinline!vcov!} generic.

For positive exposure coefficients, \passthrough{\lstinline!pos\_psi!}
is their sum and each positive weight is its coefficient divided by that
sum. For negative coefficients, \passthrough{\lstinline!neg\_psi!} is
their signed sum and each negative weight is its absolute coefficient
divided by the absolute sum. Each nonempty group's weights sum to one.
The mappings are sorted by decreasing weight; exact ties retain exposure
order. An absent group gives an empty mapping. Exactly zero coefficients
are listed separately, without rounding near-zero values into that
group.

These are descriptive fitted proportions. They have no confidence
intervals and are not selection probabilities or causal-importance
measures. The relation \(\widehat\psi=\widehat\psi_++\widehat\psi_-\)
holds, but weights are not divided by the net estimate, which can be
close to zero.

Use \passthrough{\lstinline!print(result.summary())!} for a statistical
summary table and diagnostic explanations. The full-precision fields
remain available directly. Printing the object itself gives a dataclass
representation. There are no \passthrough{\lstinline!plot!},
\passthrough{\lstinline!coef!}, \passthrough{\lstinline!confint!},
\passthrough{\lstinline!tidy!} or \passthrough{\lstinline!glance!}
methods.

\paragraph{Value}\label{manual-value-10}

There are 27 public stored fields and two formula properties. Here \(n\)
is the training row count, \(d\) the exposure count, \(c\) the covariate
count and \(k=1+d+c\).

{\def\LTcaptype{none} % do not increment counter
\begin{longtable}[]{@{}
  >{\raggedright\arraybackslash}p{(\linewidth - 2\tabcolsep) * \real{0.5000}}
  >{\raggedright\arraybackslash}p{(\linewidth - 2\tabcolsep) * \real{0.5000}}@{}}
\toprule\noalign{}
\begin{minipage}[b]{\linewidth}\raggedright
Field
\end{minipage} & \begin{minipage}[b]{\linewidth}\raggedright
Meaning
\end{minipage} \\
\midrule\noalign{}
\endhead
\bottomrule\noalign{}
\endlastfoot
\passthrough{\lstinline!coefficients!} & Read-only float array
\passthrough{\lstinline!(k,)!} in intercept, exposure, covariate
order. \\
\passthrough{\lstinline!coefficient\_names!} & Tuple of \(k\) names in
that order. \\
\passthrough{\lstinline!covariance!} & Read-only float array
\passthrough{\lstinline!(k, k)!}; covariance of all coefficient
estimates in the same order. \\
\passthrough{\lstinline!exposure\_names!} & Tuple of \(d\) exposure
names. \\
\passthrough{\lstinline!covariate\_names!} & Tuple of \(c\) covariate
names; empty when absent. \\
\passthrough{\lstinline!fitted\_values!} & Read-only float array
\passthrough{\lstinline!(n,)!} of fitted training conditional means. \\
\passthrough{\lstinline!residuals!} & Read-only float array
\passthrough{\lstinline!(n,)!} of outcome minus fitted mean, computed
before restoring the outcome mean for numerical accuracy. \\
\passthrough{\lstinline!case\_weights!} & Read-only float array
\passthrough{\lstinline!(n,)!} when positive case weights were supplied;
otherwise \passthrough{\lstinline!None!}. These are not descriptive sign
weights. \\
\passthrough{\lstinline!residual\_df!} & Integer \(n-1-d-c\), used for
residual variance. \\
\passthrough{\lstinline!residual\_variance!} & Estimated residual
variance, normally residual sum of squares divided by
\passthrough{\lstinline!residual\_df!}; zero when residual variation is
numerically zero. \\
\passthrough{\lstinline!psi!} & Float; sum of exposure coefficients. \\
\passthrough{\lstinline!var\_psi!} & Float; estimated variance of that
sum, including exposure covariance terms. \\
\passthrough{\lstinline!se!} & Float; square root of
\passthrough{\lstinline!var\_psi!}. \\
\passthrough{\lstinline!ci!} & Tuple
\passthrough{\lstinline!(lower, upper)!}; normal reference interval at
nominal level \(1-\alpha\). \\
\passthrough{\lstinline!alpha!} & Float significance level used for the
interval. \\
\passthrough{\lstinline!test\_statistic!} & Float
\passthrough{\lstinline!psi / se!}; \passthrough{\lstinline!NaN!} when
inference is degenerate. \\
\passthrough{\lstinline!p\_value!} & Two-sided Student \(t\) reference
p-value; \passthrough{\lstinline!NaN!} when inference is degenerate. \\
\passthrough{\lstinline!p\_value\_df!} & Integer \(n-1-d\), as in the
pinned R convention. \\
\passthrough{\lstinline!pos\_psi!} & Sum of strictly positive exposure
coefficients; zero if none. \\
\passthrough{\lstinline!neg\_psi!} & Signed sum of strictly negative
exposure coefficients; zero if none. \\
\passthrough{\lstinline!pos\_weights!} & Read-only mapping of positive
exposure names to within-positive-group proportions; empty if absent. \\
\passthrough{\lstinline!neg\_weights!} & Read-only mapping of negative
exposure names to within-negative-group proportions; empty if absent. \\
\passthrough{\lstinline!zero\_exposures!} & Tuple of exposure names with
exactly zero fitted coefficients. \\
\passthrough{\lstinline!diagnostics!} & Read-only mapping with the 13
keys described below. \\
\passthrough{\lstinline!library\_versions!} & Read-only mapping
containing the keys \passthrough{\lstinline!"numpy"!} and
\passthrough{\lstinline!"scipy"!} and their version strings. It does not
contain the Python or package version. \\
\passthrough{\lstinline!reference\_version!} & String
\passthrough{\lstinline!"2.19.6"!}. \\
\passthrough{\lstinline!inference!} & String
\passthrough{\lstinline!"qgcomp\_2.19.6"!}. \\
\end{longtable}
}

Eight properties provide convenient access to existing stored
quantities:

{\def\LTcaptype{none} % do not increment counter
\begin{longtable}[]{@{}
  >{\raggedright\arraybackslash}p{(\linewidth - 2\tabcolsep) * \real{0.5000}}
  >{\raggedright\arraybackslash}p{(\linewidth - 2\tabcolsep) * \real{0.5000}}@{}}
\toprule\noalign{}
\begin{minipage}[b]{\linewidth}\raggedright
Property
\end{minipage} & \begin{minipage}[b]{\linewidth}\raggedright
Meaning
\end{minipage} \\
\midrule\noalign{}
\endhead
\bottomrule\noalign{}
\endlastfoot
\passthrough{\lstinline!intercept!} & Float equal to
\passthrough{\lstinline!coefficients[0]!}. \\
\passthrough{\lstinline!beta\_exposures!} & Read-only exposure
coefficient slice \passthrough{\lstinline!(d,)!}. \\
\passthrough{\lstinline!beta\_covariates!} & Read-only covariate
coefficient slice \passthrough{\lstinline!(c,)!}; empty when absent. \\
\passthrough{\lstinline!breaks!} & Tuple of saved boundary vectors, or
\passthrough{\lstinline!None!} in prequantized mode. \\
\passthrough{\lstinline!category\_counts!} & Tuple of fitted bin counts,
or distinct training score counts in prequantized mode. \\
\passthrough{\lstinline!quantization\_mode!} &
\passthrough{\lstinline!"quantile"!},
\passthrough{\lstinline!"explicit"!} or
\passthrough{\lstinline!"prequantized"!}. \\
\passthrough{\lstinline!q!} & Requested integer only in automatic mode;
\passthrough{\lstinline!None!} in explicit and prequantized modes. \\
\passthrough{\lstinline!scores!} & Read-only training-score array
\passthrough{\lstinline!(n, d)!}. \\
\passthrough{\lstinline!formula!} & Restricted formula string for a
\passthrough{\lstinline!fit\_formula!} result; otherwise
\passthrough{\lstinline!None!}. \\
\passthrough{\lstinline!categorical\_levels!} & Read-only mapping from
adjustment category name to the saved ordered text levels; empty for an
array fit. \\
\passthrough{\lstinline!training\_row\_indices!} & Retained zero-based
input-row positions for a formula fit; \passthrough{\lstinline!None!}
for an array fit. \\
\passthrough{\lstinline!original\_row\_count!} & Input-row count before
formula row omission; \passthrough{\lstinline!None!} for an array
fit. \\
\end{longtable}
}

The public methods are \passthrough{\lstinline!summary!},
\passthrough{\lstinline!predict!}, \passthrough{\lstinline!transform!},
\passthrough{\lstinline!predict\_data!} and
\passthrough{\lstinline!transform\_data!}. Underscore-prefixed fields
hold internal state.

\paragraph{Fit diagnostics.}\label{manual-fit-diagnostics.}

The mapping \passthrough{\lstinline!result.diagnostics!} contains:

{\def\LTcaptype{none} % do not increment counter
\begin{longtable}[]{@{}
  >{\raggedright\arraybackslash}p{(\linewidth - 2\tabcolsep) * \real{0.5000}}
  >{\raggedright\arraybackslash}p{(\linewidth - 2\tabcolsep) * \real{0.5000}}@{}}
\toprule\noalign{}
\begin{minipage}[b]{\linewidth}\raggedright
Key
\end{minipage} & \begin{minipage}[b]{\linewidth}\raggedright
Meaning
\end{minipage} \\
\midrule\noalign{}
\endhead
\bottomrule\noalign{}
\endlastfoot
\passthrough{\lstinline!rank!} & Numerical design rank; \(k\) in
accepted fits. \\
\passthrough{\lstinline!rank\_tolerance!} & Rank threshold:
\(\max(n,k)\,\epsilon\,s_{\max}\), where \(\epsilon\) is
double-precision machine epsilon and \(s_{\max}\) is the largest design
singular value. \\
\passthrough{\lstinline!condition\_number!} & Largest divided by
smallest design singular value; a measure of numerical sensitivity that
depends on column scaling. \\
\passthrough{\lstinline!condition\_warning\_threshold!} & \(10^8\); an
implementation warning threshold, not a general scientific validity
cutoff. \\
\passthrough{\lstinline!ill\_conditioned!} & Whether the condition
number is strictly above that threshold. \\
\passthrough{\lstinline!inference\_degenerate!} & Whether residual or
contrast variation is numerically zero and the test is undefined. \\
\passthrough{\lstinline!residual\_roundoff\_tolerance!} & Threshold used
to identify numerical-zero residual variation on the centered outcome
scale. \\
\passthrough{\lstinline!variance\_roundoff\_bound!} & \(100\epsilon\)
times the sum of absolute entries of the exposure covariance block; a
numerical roundoff scale. \\
\passthrough{\lstinline!variance\_clamped!} & Always
\passthrough{\lstinline!False!} in this implementation; the reported
joint variance is not produced by clipping a negative value. \\
\passthrough{\lstinline!contrast\_variance\_method!} & String
\passthrough{\lstinline!"svd\_factor"!}. \\
\passthrough{\lstinline!case\_weighted!} & Whether positive case weights
were supplied to the Gaussian regression. \\
\passthrough{\lstinline!covariance\_contrast\_variance!} & Literal
exposure-sum contraction of the covariance matrix; numerical
cancellation can make it differ from the more stable
\passthrough{\lstinline!var\_psi!}. \\
\passthrough{\lstinline!quantization!} & The six training quantization
diagnostics listed in the
\hyperref[manual-api:quantization]{\passthrough{\lstinline!QuantizationResult!}
entry}. \\
\end{longtable}
}

These numerical checks do not verify the additive mean model,
independent observations, constant residual variance, joint exposure
support or causal identification. For new-row quantization diagnostics,
call \passthrough{\lstinline!result.transform(new\_exposures)!}.

\paragraph{Examples}\label{manual-examples-10}

Using \passthrough{\lstinline!result!} from the
\hyperref[manual-api:fit]{\passthrough{\lstinline!fit!} example}:

\begin{lstlisting}[language=Python]
print(result.coefficient_names)
print(result.coefficients.round(6))
print(dict(result.pos_weights))
print(dict(result.neg_weights))
print(round(result.pos_psi, 6), round(result.neg_psi, 6))
print(result.diagnostics["inference_degenerate"])
\end{lstlisting}

\begin{lstlisting}
('(Intercept)', 'x1', 'x2')
[ 1.   2.  -0.5]
{'x1': 1.0}
{'x2': 1.0}
2.0 -0.5
False
\end{lstlisting}

Each sign group has one exposure, so each within-group weight is one.
The partial sums show that their contributions differ in size.

\paragraph{See Also}\label{manual-see-also-10}

\hyperref[manual-api:fit]{\passthrough{\lstinline!fit!}},
\hyperref[manual-api:predict]{\passthrough{\lstinline!QGCompResult.predict!}},
\hyperref[manual-api:result-transform]{\passthrough{\lstinline!QGCompResult.transform!}},
\hyperref[manual-api:warning]{\passthrough{\lstinline!QGCompWarning!}}.

\paragraph{Source note.}\label{manual-source-note.-3}

The ordering of joint-effect, variance, fit and sign-weight information
follows the Value section of R qgcomp 2.19.6
\passthrough{\lstinline!man/qgcomp.glm.noboot.Rd!}. The class, public
field names, immutability, properties and diagnostics are Python
interface documentation. No R list or S3 class compatibility is implied.

\subsubsection{\texorpdfstring{\texttt{QGCompResult.summary}: Display a
saved Gaussian
fit}{QGCompResult.summary: Display a saved Gaussian fit}}\label{manual-qgcompresult.summary-display-a-saved-gaussian-fit}

\paragraph{Description}\label{manual-description-12}

Return readable text containing \(\widehat\psi\), its uncertainty,
descriptive sign weights, coefficient estimates and diagnostic
explanations. The method uses the saved result without fitting another
model or changing any value.

\paragraph{Usage}\label{manual-usage-14}

\begin{lstlisting}[language=Python]
result.summary(*, digits=6)
print(result.summary())
\end{lstlisting}

\paragraph{Arguments}\label{manual-arguments-11}

\passthrough{\lstinline!digits!} is the number of significant digits
used for displayed estimates, uncertainty and weights. It must be an
integer from 1 through 17, inclusive. Boolean values are rejected. An
invalid value raises \passthrough{\lstinline!ValueError!}.

\paragraph{Details}\label{manual-details-10}

The joint-effect row contains \passthrough{\lstinline!psi!}, standard
error (SE), lower and upper confidence limits (CI), the stored \(t\)
statistic and its two-sided \(p\)-value. The display states the fitted
\passthrough{\lstinline!alpha!} and reference degrees of freedom. Normal
confidence limits and the reference \(t\) test can give different
decisions near a significance threshold; neither is recalculated by the
summary.

The opening lines say whether case weights were supplied. These change
the fit and differ from the descriptive sign weights below. Positive and
negative sign weights are shown separately, in stored decreasing order.
The corresponding signed coefficient sums are also shown. These
proportions are descriptive and do not have uncertainty estimates. The
full-model coefficient table contains estimates only, in the fitted
order; it does not add inference for the intercept or individual
exposures.

Scientific notation preserves small nonzero values, including very small
\(p\)-values. R's subtraction-based tail calculation can round such
values to zero; Python retains its existing stable tail calculation. An
undefined statistic or \(p\)-value appears as
\passthrough{\lstinline!NA!}. A degenerate-fit explanation warns that a
point interval does not establish certainty. Ill-conditioning, reduced
categories from ties and empty fitted bins are explained when present.
These diagnostics do not test the model assumptions or establish
causality.

The return value is a Python string, not R's summary list or coefficient
matrix. Full-precision attributes remain the interface for later
calculations. Printed rounding can make weights sum slightly differently
from one or conceal a small difference between estimates; do not parse
the text for further analysis. Column labels are quoted and control
characters are escaped. Long names remain complete and can make the text
wider than a terminal window.

\paragraph{Value}\label{manual-value-11}

A \passthrough{\lstinline!str!}. The method does not print to standard
output or emit warnings. Use \passthrough{\lstinline!print!} to display
it. The fit's original warnings and stored diagnostics remain unchanged.
Prediction warnings concern new data and are handled by
\passthrough{\lstinline!predict!}; \passthrough{\lstinline!transform!}
returns the corresponding row and column diagnostics.

\paragraph{Examples}\label{manual-examples-11}

Using \passthrough{\lstinline!result!} from the
\hyperref[manual-api:fit]{\passthrough{\lstinline!fit!} example}:

\begin{lstlisting}[language=Python]
print(result.summary())
# Joint-effect row (default six significant digits):
# psi  1.5  0.0877058  1.3281  1.6719  17.1026  2.70924e-10
print(result.summary(digits=10))
\end{lstlisting}

Here the columns are estimate, SE, lower CI, upper CI, \(t\) and
\(p\)-value. The interval uses \passthrough{\lstinline!alpha=0.05!}.
Each sign group contains one exposure, so both within-group weights are
one.

\paragraph{See Also}\label{manual-see-also-11}

\passthrough{\lstinline!QGCompResult!},
\passthrough{\lstinline!QGCompResult.predict!}, and
\href{https://github.com/nilson01/quantilemix/blob/main/docs/reference.md}{docs/reference.md}
for differences from R qgcomp's registered summary method.

\subsubsection{\texorpdfstring{\texttt{QGCompResult.predict}: Predict
conditional mean
outcomes}{QGCompResult.predict: Predict conditional mean outcomes}}\label{manual-api:predict}

\paragraph{Description}\label{manual-description-13}

Predict mean outcomes from a fitted
\passthrough{\lstinline!QGCompResult!}. Measurements are converted to
scores using the cutpoints from the original fit. Prequantized fits
accept scores directly.

\paragraph{Usage}\label{manual-usage-15}

\begin{lstlisting}[language=Python]
result.predict(exposures, covariates=None)
\end{lstlisting}

\paragraph{Arguments}\label{manual-arguments-12}

{\def\LTcaptype{none} % do not increment counter
\begin{longtable}[]{@{}
  >{\raggedright\arraybackslash}p{(\linewidth - 2\tabcolsep) * \real{0.5000}}
  >{\raggedright\arraybackslash}p{(\linewidth - 2\tabcolsep) * \real{0.5000}}@{}}
\toprule\noalign{}
\begin{minipage}[b]{\linewidth}\raggedright
Argument
\end{minipage} & \begin{minipage}[b]{\linewidth}\raggedright
Meaning and accepted inputs
\end{minipage} \\
\midrule\noalign{}
\endhead
\bottomrule\noalign{}
\endlastfoot
\passthrough{\lstinline!exposures!} & Real, finite array
\passthrough{\lstinline!(m, d)!}, with at least one row and the fitted
exposure count and order. A vector means \(m\) rows of one exposure.
Supply \passthrough{\lstinline![[x1, x2]]!} for one row with two
exposures. \\
\passthrough{\lstinline!covariates!} & Required when the model was
fitted with covariates: an array \passthrough{\lstinline!(m, c)!}, or a
vector for one covariate. Must match the fitted count and order.
Supplying covariates to a model fitted without them is an error. \\
\end{longtable}
}

Here \(m\geq1\) is the number of new rows; \(d\) and \(c\) are the
fitted exposure and covariate counts.

\paragraph{Details}\label{manual-details-11}

The prediction is \(\widehat\beta_0+X_{\mathrm{new}}^q\widehat\beta+
Z_{\mathrm{new}}\widehat\gamma\), where \(X_{\mathrm{new}}^q\) uses the
saved map. Cutpoints are never estimated again from a new batch. If
training inputs had column labels, the same labels in the same order are
required; dropping those labels is rejected. Unlabeled training arrays
use positional order.

Values outside observed training exposure ranges cause
\passthrough{\lstinline!QGCompWarning!}, provided they remain within
stored cut endpoints. Use \passthrough{\lstinline!transform!} to inspect
row and column details. Values outside stored endpoints raise
\passthrough{\lstinline!ValueError!}. In prequantized mode, new scores
must remain nonnegative integers below \(2^{53}\). The implementation
does not check covariate-range extrapolation.

Invalid shapes, labels, data, missing or unexpected covariates, and
nonfinite predictions raise \passthrough{\lstinline!ValueError!}.
Prediction returns conditional means. It does not return standard
errors, individual-outcome intervals or causal policy values. There is
no \passthrough{\lstinline!type!} argument, marginal-prediction option
or option to omit exposures. Training means are stored in
\passthrough{\lstinline!fitted\_values!}.

\paragraph{Value}\label{manual-value-12}

A read-only NumPy float array of shape \passthrough{\lstinline!(m,)!}.

\paragraph{Examples}\label{manual-examples-12}

\begin{lstlisting}[language=Python]
import numpy as np
from quantilemix import fit

raw = np.arange(8, dtype=float)
y = 1 + 2*np.repeat(np.arange(4), 2) + np.tile([-0.5, 0.5], 4)
raw_fit = fit(y, raw)  # q=4 by default
new_raw = [-1, 1.75, 1.8, 7, 9]
print(raw_fit.transform(new_raw).scores.ravel())
print(raw_fit.predict(new_raw).round(6))
\end{lstlisting}

\begin{lstlisting}
[0. 0. 1. 3. 3.]
[1. 1. 3. 7. 7.]
\end{lstlisting}

The second call also emits \passthrough{\lstinline!QGCompWarning!}:
values at rows 0 and 4 are outside the observed training range
\([0,7]\). The saved boundaries are
\passthrough{\lstinline![-1e64, 1.75, 3.5, 5.25, 1e64]!}. The value 1.75
remains in the lower category; new batches do not change these
boundaries.

\paragraph{See Also}\label{manual-see-also-12}

\hyperref[manual-api:result-transform]{\passthrough{\lstinline!QGCompResult.transform!}},
\hyperref[manual-api:warning]{\passthrough{\lstinline!QGCompWarning!}},
\hyperref[manual-api:result]{\passthrough{\lstinline!QGCompResult!}}.

\paragraph{Source note.}\label{manual-source-note.-4}

The saved-cut prediction explanation is adapted from Alexander Keil's R
qgcomp 2.19.6 \passthrough{\lstinline!man/predict.qgcompfit.Rd!},
printed manual pages 28--29, GPL (version 2 or later). Python returns
Gaussian conditional means and has its own required inputs, checks and
warning behavior. The R documentation's broader outcome, bootstrap and
marginal-model options are outside this interface.

\subsubsection{\texorpdfstring{\texttt{QGCompResult.transform}: Apply
the fitted exposure
map}{QGCompResult.transform: Apply the fitted exposure map}}\label{manual-api:result-transform}

\paragraph{Description}\label{manual-description-14}

Convert new exposures to scores using the fitted model's saved
transformation, and return the scores with quantization diagnostics.

\paragraph{Usage}\label{manual-usage-16}

\begin{lstlisting}[language=Python]
transformed = result.transform(exposures)
\end{lstlisting}

\paragraph{Arguments}\label{manual-arguments-13}

\passthrough{\lstinline!exposures!}: nonempty real, finite new data of
shape \passthrough{\lstinline!(m, d)!}, or a vector for one exposure.
The fitted exposure count and order are required. Labels must match when
training inputs were labeled. No outcome or covariates are needed.

\paragraph{Details}\label{manual-details-12}

This method delegates to the saved
\passthrough{\lstinline!QuantizationResult.transform!} method. It does
not fit regression coefficients or calculate new cuts. Quantile and
explicit modes apply stored boundaries; prequantized mode checks and
retains scores. Values beyond observed training ranges are reported in
diagnostics without a warning from this method. Values outside stored
endpoints, invalid scores or invalid input shapes, values and labels
raise \passthrough{\lstinline!ValueError!}.

\paragraph{Value}\label{manual-value-13}

A new \passthrough{\lstinline!QuantizationResult!} whose
\passthrough{\lstinline!scores!} and diagnostics describe the new rows.
Boundaries, fitted category counts, mode, names and training ranges
remain those of the original fit.

\paragraph{Examples}\label{manual-examples-13}

Using \passthrough{\lstinline!raw\_fit!} from the prediction example:

\begin{lstlisting}[language=Python]
transformed = raw_fit.transform([-1, 1.75, 1.8, 7, 9])
print(transformed.diagnostics["extrapolated_rows"])        # (0, 4)
print(transformed.diagnostics["extrapolated_columns"])     # ('x1',)
print(transformed.diagnostics["extrapolated_counts"])      # (2,)
print(transformed.category_counts)                        # (4,)
print(transformed.diagnostics["observed_category_counts"]) # (3,)
print(transformed.diagnostics["empty_bins"])               # ((2,),)
\end{lstlisting}

Four bins were fitted. This new batch occupies three of them, with bin 2
empty. These are different counts and are kept separately.

\paragraph{See Also}\label{manual-see-also-13}

\hyperref[manual-api:predict]{\passthrough{\lstinline!QGCompResult.predict!}},
\hyperref[manual-api:quantization-transform]{\passthrough{\lstinline!QuantizationResult.transform!}},
\hyperref[manual-api:quantization]{\passthrough{\lstinline!QuantizationResult!}}.

\subsubsection{\texorpdfstring{\texttt{QGCompResult.predict\_data\ and\ transform\_data}:
Named-data
prediction}{QGCompResult.predict\_data and transform\_data: Named-data prediction}}\label{manual-qgcompresult.predict_data-and-transform_data-named-data-prediction}

\paragraph{Description}\label{manual-description-15}

These methods are available on \passthrough{\lstinline!fit\_formula!}
results. They read named columns and apply the saved categorical design
and exposure boundaries. \passthrough{\lstinline!predict\_data!} returns
conditional Gaussian mean predictions.
\passthrough{\lstinline!transform\_data!} returns exposure scores and
diagnostics; it needs no outcome or adjustment columns.

\paragraph{Usage}\label{manual-usage-17}

\begin{lstlisting}[language=Python]
means = result.predict_data({"x1": [1], "x2": [2], "group": ["B"]})
scores = result.transform_data({"x1": [1], "x2": [2]})
\end{lstlisting}

\paragraph{Arguments and details}\label{manual-arguments-and-details}

\passthrough{\lstinline!data!} is a mapping or named table. Prediction
requires every fitted exposure and adjustment column with equal row
counts. Text categories must use one of the fitted levels; new levels
are rejected. The original formula and reference level are retained.
Array-based fits do not have a formula schema and reject these methods.

\paragraph{Value}\label{manual-value-14}

The \passthrough{\lstinline!predict\_data!} method returns a read-only
vector of conditional means. The
\passthrough{\lstinline!transform\_data!} method returns exposure scores
and diagnostics in a \passthrough{\lstinline!QuantizationResult!}. Both
use the existing \passthrough{\lstinline!predict!} and
\passthrough{\lstinline!transform!} calculations.

\paragraph{See Also}\label{manual-see-also-14}

\hyperref[manual-api:fit-formula]{\passthrough{\lstinline!fit\_formula!}},
\hyperref[manual-api:result]{\passthrough{\lstinline!QGCompResult!}}.

\subsection{Binary outcomes}\label{manual-binary-outcomes}

\subsubsection{\texorpdfstring{\texttt{fit\_binomial}: Additive
binary-outcome
fit}{fit\_binomial: Additive binary-outcome fit}}\label{manual-api:fit-binomial}

\paragraph{Description}\label{manual-description-16}

Fit a complete Bernoulli outcome, coded 0 or 1, with an additive logit
model. The estimate \(\widehat\psi=\sum_j\widehat\beta_j\) is a
conditional log odds ratio. Its exponent \(\exp(\widehat\psi)\) is an
odds ratio, not a risk ratio.

\paragraph{Usage}\label{manual-usage-18}

\begin{lstlisting}[language=Python]
from quantilemix import fit_binomial

result = fit_binomial(
    y, exposures, covariates=None,
    q=4, breaks=None, alpha=0.05, weights=None
)
\end{lstlisting}

\paragraph{Arguments}\label{manual-arguments-14}

\passthrough{\lstinline!y!} is a nonempty vector containing both 0 and
1. \passthrough{\lstinline!exposures!} are real, finite columns, with
one row per outcome. \passthrough{\lstinline!covariates!} are optional
complete numeric adjustment columns. \passthrough{\lstinline!q!},
\passthrough{\lstinline!breaks!} and \passthrough{\lstinline!alpha!}
follow \passthrough{\lstinline!fit!}. Positive
\passthrough{\lstinline!weights!} are optional GLM prior weights;
exposure cuts do not use them.

\paragraph{Details}\label{manual-details-13}

The fitted model is

\[
\mathrm{logit}\Pr(Y=1\mid X^q,Z)
=\beta_0+\sum_{j=1}^{d}\beta_j X_j^q+\gamma^\top Z.
\]

The joint log odds ratio is \(\psi=\sum_j\beta_j\) for one
score-category increase in every exposure with adjustment values fixed.
The covariance of the full coefficient vector gives the joint standard
error, including cross terms. Normal critical values form the confidence
interval, and a normal z statistic gives the two-sided p-value. This
matches the binomial branch of R qgcomp 2.19.6; the Gaussian branch uses
a different p-value rule. Predictions return conditional event
probabilities using saved cuts. The fit rejects constant outcomes, rank
deficiency, nonconvergence and numerical separation. Grouped trials,
incomplete rows, offsets, interactions and other links are outside this
entry.

\paragraph{Value}\label{manual-value-15}

A \passthrough{\lstinline!BinomialResult!} with the joint log odds
ratio, model covariance, normal interval, p-value, odds ratio and
conditional predictions.

\paragraph{Examples}\label{manual-examples-14}

\begin{lstlisting}[language=Python]
import numpy as np
from quantilemix import fit_binomial

cells = ((0, 0, 5), (1, 0, 10), (0, 1, 10), (1, 1, 15))
x = np.vstack([np.tile([a, b], (20, 1)) for a, b, _ in cells])
y = np.concatenate([np.r_[np.ones(s), np.zeros(20-s)]
                    for _, _, s in cells])
result = fit_binomial(y, x, q=None)
assert abs(result.psi - np.log(9)) < 1e-12
assert abs(result.odds_ratio - 9) < 1e-12
\end{lstlisting}

The four event fractions are 0.25, 0.5, 0.5 and 0.75. The joint
conditional odds ratio is exactly 9 in this construction.

\paragraph{See Also}\label{manual-see-also-15}

\hyperref[manual-api:fit-binomial-formula]{\passthrough{\lstinline!fit\_binomial\_formula!}},
\hyperref[manual-api:binomial-result]{\passthrough{\lstinline!BinomialResult!}},
\hyperref[manual-api:fit]{\passthrough{\lstinline!fit!}}.

\subsubsection{\texorpdfstring{\texttt{fit\_binomial\_formula}: Binary
fit from named
data}{fit\_binomial\_formula: Binary fit from named data}}\label{manual-api:fit-binomial-formula}

\paragraph{Description}\label{manual-description-17}

Prepare named columns, including optional text adjustment categories,
then call the same additive Bernoulli-logit estimator.

\paragraph{Usage}\label{manual-usage-19}

\begin{lstlisting}[language=Python]
from quantilemix import fit_binomial_formula

result = fit_binomial_formula(
    "y ~ x1 + x2 + C(group)", data, ("x1", "x2"),
    q=4, breaks=None, alpha=0.05, missing="raise", weights=None
)
new_probability = result.predict_data(new_data)
\end{lstlisting}

\paragraph{Arguments}\label{manual-arguments-15}

\passthrough{\lstinline!formula!}, \passthrough{\lstinline!data!} and
\passthrough{\lstinline!exposures!} follow the restricted main-effects
grammar of \passthrough{\lstinline!fit\_formula!}. The outcome must
contain only 0 and 1. \passthrough{\lstinline!q!},
\passthrough{\lstinline!breaks!} and \passthrough{\lstinline!alpha!}
follow \passthrough{\lstinline!fit\_binomial!}.
\passthrough{\lstinline!missing="drop"!} explicitly removes rows with a
missing outcome, exposure or adjustment value. The default rejects them.
Missing case weights are also omitted with
\passthrough{\lstinline!missing="drop"!}. Positive case weights are
accepted.

\paragraph{Details}\label{manual-details-14}

\passthrough{\lstinline!C(name)!} uses the sorted first observed text
level as its reference. The saved levels and cutpoints are reused for
named-data prediction. An unseen category or incomplete new row is an
error. The estimand remains a conditional log odds ratio; category
coefficients are adjustment terms and do not enter its exposure sum.
With row removal, automatic cutpoints use all observed exposure values
before omitting incomplete model rows. This matches the order in the
installed R reference. No imputation occurs.

\paragraph{Value}\label{manual-value-16}

A \passthrough{\lstinline!BinomialResult!} with saved formula and
category levels. \passthrough{\lstinline!training\_row\_indices!}
contains retained zero-based original positions;
\passthrough{\lstinline!original\_row\_count!} records the input length.

\paragraph{Examples}\label{manual-examples-15}

Use the four-cell construction above with a complete text adjustment
column.

\paragraph{See Also}\label{manual-see-also-16}

\hyperref[manual-api:fit-binomial]{\passthrough{\lstinline!fit\_binomial!}},
\hyperref[manual-api:fit-formula]{\passthrough{\lstinline!fit\_formula!}}.

\subsubsection{\texorpdfstring{\texttt{fit\_binomial\_risk\_ratio}:
Marginal binary risk
ratio}{fit\_binomial\_risk\_ratio: Marginal binary risk ratio}}\label{manual-api:fit-binomial-risk-ratio}

\paragraph{Description}\label{manual-description-18}

Summarize how the average event probability changes when every exposure
score rises by one category, using a risk ratio. This supports selected
settings of R \passthrough{\lstinline!qgcomp.glm.boot(..., rr=TRUE)!}.
The separate \passthrough{\lstinline!fit\_binomial!} function reports an
odds ratio at fixed adjustment values, which answers a different
question.

\paragraph{Usage}\label{manual-usage-20}

\begin{lstlisting}[language=Python]
from quantilemix import fit_binomial_risk_ratio
result = fit_binomial_risk_ratio(
    y, exposures, covariates=None,
    q=2, breaks=None, alpha=0.05, B=200, seed=None
)
\end{lstlisting}

\paragraph{Arguments}\label{manual-arguments-16}

\passthrough{\lstinline!y!} contains only zero and one.
\passthrough{\lstinline!exposures!} and optional
\passthrough{\lstinline!covariates!} are complete finite numeric
columns. \passthrough{\lstinline!q=2!} groups exposures into two
categories by default; \passthrough{\lstinline!q=4!} groups them into
four. \passthrough{\lstinline!q=None!} accepts already scored
consecutive levels starting at zero. Supplied boundaries define the
intervention levels. \passthrough{\lstinline!alpha!} is the interval
error level. \passthrough{\lstinline!B!} is at least two independent
row-bootstrap draws; \passthrough{\lstinline!seed!} controls a local
NumPy random generator. Case weights, formulas, missing-row options and
cluster resampling are not accepted by this method.

\paragraph{Details}\label{manual-details-15}

At intervention level \(a\in\{0,\ldots,q-1\}\), replace every exposure
score by \(a\) while retaining each person's adjustment covariates
\(Z_i\). Let \(\widehat p_i(a)\) be that person's predicted event
probability from the conditional logit fit. The standardized risk is

\[
\widehat R(a)=n^{-1}\sum_{i=1}^n\widehat p_i(a).
\]

Here \(n\) is the number of rows. With two levels, the effect is
\(\widehat\psi=\log\{\widehat R(1)/\widehat R(0)\}\). With more levels,
the effect is the slope \(\widehat\psi\) from a degree-one log-binomial
marginal model of all the standardized risks \(\widehat R(a)\) against
\(a\). The reported one-category risk ratio is \(\exp(\widehat\psi)\);
it need not equal \(\widehat R(1)/\widehat R(0)\) when there are more
than two levels. Each bootstrap draw resamples whole rows, refits the
conditional logit and marginal model, and saves its slope. Their sample
standard deviation is the bootstrap standard error. Normal limits on the
log scale are exponentiated for the risk-ratio interval. The two-sided
test uses a normal \(z\) statistic. A failed replicate raises an error
and is never silently replaced.

\paragraph{Value}\label{manual-value-17}

\passthrough{\lstinline!BinomialRiskRatioResult!} saves
\passthrough{\lstinline!psi!} (log risk ratio),
\passthrough{\lstinline!risk\_ratio!}, \passthrough{\lstinline!se!},
both log-scale and ratio-scale intervals,
\passthrough{\lstinline!p\_value!}, and standardized
\passthrough{\lstinline!intervention\_risks!}. It also saves read-only
marginal-model coefficients
(\passthrough{\lstinline!msm\_coefficients!}), fitted risks
(\passthrough{\lstinline!msm\_fitted\_risks!}), bootstrap slopes
(\passthrough{\lstinline!bootstrap\_effects!}), and the conditional
\passthrough{\lstinline!point\_fit!}.
\passthrough{\lstinline!summary(digits=6)!} displays the saved values
without refitting.

\paragraph{Examples}\label{manual-examples-16}

\begin{lstlisting}[language=Python]
rr = fit_binomial_risk_ratio(y, exposures, q=None, B=200, seed=17)
print(rr.risk_ratio, rr.risk_ratio_ci)
\end{lstlisting}

The two-level worked example has an independently known risk ratio of
three and a conditional odds ratio of nine. A second example fits four
levels. Two fixed cases at each level count agree with installed R
qgcomp 2.19.6 under identical saved resampling schedules. A separate
four-level study recorded 4,000 datasets and one bootstrap separation
failure. See the
\href{https://github.com/nilson01/quantilemix/blob/main/validation/RISK_RATIO_Q4_CALIBRATION_RESULTS.md}{study
report} for coverage and limits.

\paragraph{See Also}\label{manual-see-also-17}

\hyperref[manual-api:fit-binomial]{\passthrough{\lstinline!fit\_binomial!}} for
conditional odds ratios.

\subsubsection{\texorpdfstring{\texttt{fit\_binomial\_odds\_ratio}:
Marginal binary odds
ratio}{fit\_binomial\_odds\_ratio: Marginal binary odds ratio}}\label{manual-fit_binomial_odds_ratio-marginal-binary-odds-ratio}

\paragraph{Description}\label{manual-description-19}

Estimate a fitted population-average odds-ratio summary across joint
exposure-score levels. It follows the degree-one logistic marginal model
in R \passthrough{\lstinline!qgcomp.glm.boot(..., rr=FALSE)!}. It can
differ from the conditional odds ratio even without confounding, because
odds ratios are not generally collapsible over adjustment variables.

\paragraph{Usage}\label{manual-usage-21}

\begin{lstlisting}[language=Python]
from quantilemix import fit_binomial_odds_ratio
result = fit_binomial_odds_ratio(
    y, exposures, covariates=None,
    q=2, breaks=None, alpha=0.05, B=200, seed=None
)
\end{lstlisting}

\paragraph{Arguments}\label{manual-arguments-17}

The complete numeric inputs and score controls match
\hyperref[manual-api:fit-binomial-risk-ratio]{\passthrough{\lstinline!fit\_binomial\_risk\_ratio!}}.
\passthrough{\lstinline!B!} is the number of independent-row bootstrap
draws, at least two. \passthrough{\lstinline!seed!} controls the local
generator. Case weights, formulas, missing rows and cluster resampling
are not supported by this method.

\paragraph{Details}\label{manual-details-16}

Let \(\widehat R(a)\) be the mean predicted event probability when every
exposure score is set to \(a\), keeping observed adjustment covariates.
With two levels, the effect is
\(\widehat\psi=\mathrm{logit}\{\widehat R(1)\}-
\mathrm{logit}\{\widehat R(0)\}\), where
\(\mathrm{logit}(p)=\log\{p/(1-p)\}\). With more levels, a degree-one
logistic marginal model fits all \(\widehat R(a)\) against \(a\) with
equal weight per score. The reported fitted odds-ratio summary is
\(\exp(\widehat\psi)\). It need not equal the marginal odds ratio
between any one pair of adjacent score levels. Each fixed-cut bootstrap
draw refits the conditional and marginal models. Their slope standard
deviation supplies normal intervals on the log scale. A failed replicate
is named and not replaced.

\paragraph{Value}\label{manual-value-18}

\passthrough{\lstinline!BinomialOddsRatioResult!} saves the log effect
\passthrough{\lstinline!psi!}, \passthrough{\lstinline!odds\_ratio!},
bootstrap \passthrough{\lstinline!se!}, log and ratio confidence limits,
\passthrough{\lstinline!p\_value!}, standardized
\passthrough{\lstinline!intervention\_risks!}, marginal-model
coefficients and fitted risks, bootstrap slopes, and the conditional
\passthrough{\lstinline!point\_fit!}.
\passthrough{\lstinline!summary(digits=6)!} displays their meanings.

\paragraph{Examples}\label{manual-examples-17}

\begin{lstlisting}[language=Python]
or_fit = fit_binomial_odds_ratio(y, exposures, q=None, B=200, seed=17)
print(or_fit.odds_ratio, or_fit.odds_ratio_ci)
\end{lstlisting}

The installed-wheel example checks a two-level adjusted fit. Two fixed
four-level cases, one adjusted, agree with installed R qgcomp 2.19.6
under the same 24 bootstrap row sets. Independent arithmetic and seeded
checks also pass. Repeated-data interval coverage remains untested.

\paragraph{See Also}\label{manual-see-also-18}

\hyperref[manual-api:fit-binomial-risk-ratio]{\passthrough{\lstinline!fit\_binomial\_risk\_ratio!}}
for a marginal risk ratio, and
\hyperref[manual-api:fit-binomial]{\passthrough{\lstinline!fit\_binomial!}} for
a conditional odds ratio.

\subsubsection{\texorpdfstring{\texttt{fit\_binomial\_ee}: Clustered
binary estimating
equations}{fit\_binomial\_ee: Clustered binary estimating equations}}\label{manual-fit_binomial_ee-clustered-binary-estimating-equations}

\paragraph{Description}\label{manual-description-20}

Estimate a marginal binary-outcome association for a joint increase in
all exposure scores, with observations grouped by cluster. Choose a
marginal risk ratio (\passthrough{\lstinline!rr=True!}) or odds ratio
(\passthrough{\lstinline!rr=False!}).

\paragraph{Usage}\label{manual-usage-22}

\begin{lstlisting}[language=Python]
from quantilemix import fit_binomial_ee
result = fit_binomial_ee(
    outcomes, exposures, cluster_ids, covariates=None,
    q=4, breaks=None, alpha=0.05, rr=True, weights=None
)
print(result.effect_ratio, result.se, result.ci)
\end{lstlisting}

\paragraph{Arguments}\label{manual-arguments-18}

\passthrough{\lstinline!outcomes!} contains zeros and ones.
\passthrough{\lstinline!exposures!} and optional
\passthrough{\lstinline!covariates!} contain complete numeric rows.
\passthrough{\lstinline!cluster\_ids!} has one finite exact integer per
row and at least two distinct values. \passthrough{\lstinline!q!} or
\passthrough{\lstinline!breaks!} sets exposure scores. The resulting
joint intervention levels must run consecutively from zero.
\passthrough{\lstinline!alpha!} sets the interval error level. Optional
positive \passthrough{\lstinline!weights!} apply to both estimating
equations and cluster totals but not exposure cuts. Offsets, nonlinear
terms and time-lag structure are not supported.

\paragraph{Method and interpretation}\label{manual-method-and-interpretation}

A conditional logistic model predicts each person's event risk at every
joint score. Averaging these risks across people, using case weights
when supplied, gives standardized risks. The marginal model then fits a
straight line to their log risks when \passthrough{\lstinline!rr=True!},
or to their log odds when \passthrough{\lstinline!rr=False!}. Its slope
describes one score increase on that fitted marginal scale. With more
than two scores, the exponentiated slope need not equal the ratio of two
adjacent standardized risks or odds. The covariance uses both
conditional and marginal estimating equations, sums their scores within
clusters, and forms a sandwich estimate. The normal interval may be
unreliable with few clusters. Neither fixed-case agreement nor the
exact-cell check establishes interval coverage.

\paragraph{Value}\label{manual-value-19}

\passthrough{\lstinline!BinomialEEResult!} saves conditional and
marginal coefficients, robust covariances, the full stacked covariance,
derivative bread, cluster meat, cluster count, optional case weights,
intervention levels, standardized and marginal fitted risks, joint log
effect, standard error, interval, test and p-value.
\passthrough{\lstinline!effect\_ratio!} and
\passthrough{\lstinline!effect\_ratio\_ci!} exponentiate the log-scale
effect and interval. \passthrough{\lstinline!transform!} reuses the
fitted cuts; \passthrough{\lstinline!summary!} explains the scale and
uncertainty. Four fixed installed-R cases and an independent exact-cell
calculation pass; see
\passthrough{\lstinline!validation/BINOMIAL\_EE\_RESULTS.md!}. Two
additional weighted installed-R cases cover the odds and risk links;
integer-weight row replication checks both sandwiches independently. See
\passthrough{\lstinline!validation/WEIGHTED\_EE\_RESULTS.md!}.

\subsubsection{\texorpdfstring{\texttt{BinomialResult}: Binary model
result and conditional odds
contrast}{BinomialResult: Binary model result and conditional odds contrast}}\label{manual-api:binomial-result}

\paragraph{Description}\label{manual-description-21}

This immutable result stores a Bernoulli-logit fit and the joint
exposure coefficient on the log-odds scale. It does not reuse the
Gaussian residual variance or Student t diagnostics.

\paragraph{Usage}\label{manual-usage-23}

\begin{lstlisting}[language=Python]
result = fit_binomial(y, exposures, q=None)
print(result.psi, result.odds_ratio)
print(result.summary())
probability = result.predict(new_exposures)
\end{lstlisting}

\paragraph{Value}\label{manual-value-20}

{\def\LTcaptype{none} % do not increment counter
\begin{longtable}[]{@{}
  >{\raggedright\arraybackslash}p{(\linewidth - 2\tabcolsep) * \real{0.5000}}
  >{\raggedright\arraybackslash}p{(\linewidth - 2\tabcolsep) * \real{0.5000}}@{}}
\toprule\noalign{}
\begin{minipage}[b]{\linewidth}\raggedright
Field
\end{minipage} & \begin{minipage}[b]{\linewidth}\raggedright
Meaning
\end{minipage} \\
\midrule\noalign{}
\endhead
\bottomrule\noalign{}
\endlastfoot
\passthrough{\lstinline!coefficients!},
\passthrough{\lstinline!coefficient\_names!} & Read-only full-model
estimates in intercept, exposure, then adjustment order. \\
\passthrough{\lstinline!covariance!} & Read-only full coefficient
covariance from final Fisher information. \\
\passthrough{\lstinline!psi!}, \passthrough{\lstinline!var\_psi!},
\passthrough{\lstinline!se!} & Conditional joint log odds ratio, its
variance and standard error. \\
\passthrough{\lstinline!ci!}, \passthrough{\lstinline!p\_value!},
\passthrough{\lstinline!test\_statistic!} & Normal confidence limits,
two-sided z p-value and z statistic. \\
\passthrough{\lstinline!odds\_ratio!},
\passthrough{\lstinline!odds\_ratio\_ci!} & Exponentiated joint effect
and interval; these are not risk ratios. \\
\passthrough{\lstinline!fitted\_values!} & Training event
probabilities. \\
\passthrough{\lstinline!case\_weights!} & Read-only GLM prior weights
when supplied; otherwise \passthrough{\lstinline!None!}. \\
\passthrough{\lstinline!pos\_weights!},
\passthrough{\lstinline!neg\_weights!} & Descriptive coefficient shares
within each sign group. \\
\passthrough{\lstinline!scores!}, \passthrough{\lstinline!breaks!},
\passthrough{\lstinline!q!} & Saved exposure scores and cutpoint
information. \\
\passthrough{\lstinline!family!}, \passthrough{\lstinline!link!} & The
strings \passthrough{\lstinline!"binomial"!} and
\passthrough{\lstinline!"logit"!}. \\
\passthrough{\lstinline!diagnostics!} & Read-only rank, condition,
iteration and grouping information. \\
\end{longtable}
}

\paragraph{Methods}\label{manual-methods}

\passthrough{\lstinline!predict!} returns conditional event
probabilities. \passthrough{\lstinline!transform!} reuses saved exposure
cuts. Formula results also support
\passthrough{\lstinline!predict\_data!} and
\passthrough{\lstinline!transform\_data!}.
\passthrough{\lstinline!summary!} returns a plain-text display of the
saved joint effect, interval and p-value; it never refits the model.

\paragraph{See Also}\label{manual-see-also-19}

\hyperref[manual-api:fit-binomial]{\passthrough{\lstinline!fit\_binomial!}},
\hyperref[manual-api:fit-binomial-formula]{\passthrough{\lstinline!fit\_binomial\_formula!}},
\hyperref[manual-api:result]{\passthrough{\lstinline!QGCompResult!}}.

\subsection{Count outcomes}\label{manual-count-outcomes}

\subsubsection{\texorpdfstring{\texttt{fit\_poisson}: Additive
count-outcome
fit}{fit\_poisson: Additive count-outcome fit}}\label{manual-api:fit-poisson}

\paragraph{Description}\label{manual-description-22}

Fit complete nonnegative integer counts with an additive Poisson
log-link model. The estimate \(\widehat\psi=\sum_j\widehat\beta_j\) is a
conditional log rate ratio. An optional observation-time offset accounts
for unequal periods.

\paragraph{Usage}\label{manual-usage-24}

\begin{lstlisting}[language=Python]
from quantilemix import fit_poisson

result = fit_poisson(
    y, exposures, covariates=None,
    q=4, breaks=None, alpha=0.05, weights=None,
    exposure_time=None
)
\end{lstlisting}

\paragraph{Arguments}\label{manual-arguments-19}

\passthrough{\lstinline!y!} is a nonempty vector of nonnegative integer
counts with a positive total. \passthrough{\lstinline!exposures!} are
real, finite columns with one row per outcome.
\passthrough{\lstinline!covariates!} are optional complete numeric
adjustment columns. \passthrough{\lstinline!q!},
\passthrough{\lstinline!breaks!} and \passthrough{\lstinline!alpha!}
follow \passthrough{\lstinline!fit!}. Positive
\passthrough{\lstinline!weights!} act as GLM prior weights.
\passthrough{\lstinline!exposure\_time!} is an optional positive, finite
vector with one observation period per row.

\paragraph{Details}\label{manual-details-17}

The fitted conditional mean model is

\[
\log E(Y\mid X^q,Z,t)=\log t+\beta_0+
\sum_{j=1}^{d}\beta_jX_j^q+\gamma^\top Z,
\]

where \(t>0\) is the observation time; \(t=1\) when no offset is given.
The joint log rate ratio is \(\psi=\sum_j\beta_j\) for a one-category
increase in every exposure at the same time, holding adjustment values
fixed. The full Fisher covariance includes cross terms in its joint
standard error. Normal critical values form the confidence interval, and
a normal z statistic gives the two-sided p-value, following the Poisson
branch of R qgcomp 2.19.6. Predictions return conditional expected
counts from saved cutpoints. An offset fit needs new times for
prediction. It does not handle overdispersion, interactions or
alternative links.

\paragraph{Value}\label{manual-value-21}

A \passthrough{\lstinline!PoissonResult!} with the joint log rate ratio,
model covariance, normal interval, p-value, exponentiated ratio and
expected-count prediction.

\paragraph{Examples}\label{manual-examples-18}

\begin{lstlisting}[language=Python]
import numpy as np
from quantilemix import fit_poisson

cells = ((0, 0, 1), (1, 0, 2), (0, 1, 3), (1, 1, 6))
x = np.vstack([np.tile([a, b], (10, 1)) for a, b, _ in cells])
y = np.concatenate([np.repeat(m, 10) for _, _, m in cells])
result = fit_poisson(y, x, q=None)
assert abs(result.psi - np.log(6)) < 1e-12
assert abs(result.rate_ratio - 6) < 1e-12
\end{lstlisting}

The cell means are exactly 1, 2, 3 and 6. Their joint conditional rate
ratio is 6 in this construction. For unequal observation periods:

\begin{lstlisting}[language=Python]
time = np.tile(np.arange(1, 6), 4)
x = np.vstack([np.tile([a, b], (5, 1)) for a, b, _ in cells])
y = np.concatenate([rate * time[:5] for _, _, rate in cells])
offset_fit = fit_poisson(y, x, q=None, exposure_time=time)
assert abs(offset_fit.rate_ratio - 6) < 1e-12
assert abs(offset_fit.predict([[1, 1]], exposure_time=[3])[0] - 18) < 1e-12
\end{lstlisting}

\paragraph{See Also}\label{manual-see-also-20}

\hyperref[manual-api:fit-poisson-formula]{\passthrough{\lstinline!fit\_poisson\_formula!}},
\hyperref[manual-api:poisson-result]{\passthrough{\lstinline!PoissonResult!}},
\hyperref[manual-api:fit]{\passthrough{\lstinline!fit!}}.

\subsubsection{\texorpdfstring{\texttt{fit\_poisson\_formula}: Count fit
from named
data}{fit\_poisson\_formula: Count fit from named data}}\label{manual-api:fit-poisson-formula}

\paragraph{Description}\label{manual-description-23}

Prepare named columns, including optional text adjustment categories,
then call the additive Poisson-log estimator.

\paragraph{Usage}\label{manual-usage-25}

\begin{lstlisting}[language=Python]
from quantilemix import fit_poisson_formula

result = fit_poisson_formula(
    "y ~ x1 + x2 + C(group)", data, ("x1", "x2"),
    q=4, breaks=None, alpha=0.05, missing="raise", weights=None,
    exposure_time=None
)
new_mean = result.predict_data(new_data)
\end{lstlisting}

\paragraph{Arguments}\label{manual-arguments-20}

\passthrough{\lstinline!formula!}, \passthrough{\lstinline!data!} and
\passthrough{\lstinline!exposures!} follow the restricted main-effects
grammar of \passthrough{\lstinline!fit\_formula!}. The outcome must
contain only nonnegative integer counts.
\passthrough{\lstinline!missing="drop"!} explicitly removes rows with a
missing outcome, exposure or adjustment value. The default rejects them.
Missing case weights are also omitted with
\passthrough{\lstinline!missing="drop"!}. Positive case weights are
accepted. Set \passthrough{\lstinline!exposure\_time="time"!} to name a
positive observation-time column. Missing times are omitted with
\passthrough{\lstinline!missing="drop"!}; the default rejects them.

\paragraph{Details}\label{manual-details-18}

\passthrough{\lstinline!C(name)!} uses the sorted first observed text
level as its reference. The saved levels and cutpoints are reused for
named-data prediction. Category coefficients adjust the model but do not
enter the joint exposure sum. An unseen category or incomplete new row
is an error. With row removal, automatic cutpoints use all observed
exposure values before omitting incomplete model rows. The restricted
formula parser does not accept inline
\passthrough{\lstinline!offset(log(time))!}; pass the column name
separately. This option does not impute.

\paragraph{Value}\label{manual-value-22}

A \passthrough{\lstinline!PoissonResult!} with saved formula and
category levels. \passthrough{\lstinline!training\_row\_indices!}
contains retained zero-based original positions;
\passthrough{\lstinline!original\_row\_count!} records the input length.

\paragraph{Examples}\label{manual-examples-19}

Use the four-cell construction above with a complete text adjustment
column.

\paragraph{See Also}\label{manual-see-also-21}

\hyperref[manual-api:fit-poisson]{\passthrough{\lstinline!fit\_poisson!}},
\hyperref[manual-api:fit-formula]{\passthrough{\lstinline!fit\_formula!}}.

\subsubsection{\texorpdfstring{\texttt{fit\_poisson\_bootstrap}:
Marginal Poisson row
bootstrap}{fit\_poisson\_bootstrap: Marginal Poisson row bootstrap}}\label{manual-fit_poisson_bootstrap-marginal-poisson-row-bootstrap}

\paragraph{Description}\label{manual-description-24}

Estimate the ratio of average counts when every exposure score rises by
one category. This supports the additive Poisson case of R
\passthrough{\lstinline!qgcomp.glm.boot!}. It keeps the original
exposure boundaries and resamples whole, independent rows. Unlike
\passthrough{\lstinline!fit\_poisson!}, it uses the variation across
those resamples to estimate uncertainty.

\paragraph{Usage}\label{manual-usage-26}

\begin{lstlisting}[language=Python]
from quantilemix import fit_poisson_bootstrap
result = fit_poisson_bootstrap(
    y, exposures, covariates=None,
    q=4, breaks=None, alpha=0.05, B=200, seed=None,
    exposure_time=None
)
\end{lstlisting}

\paragraph{Arguments}\label{manual-arguments-21}

\passthrough{\lstinline!y!} contains nonnegative integer counts.
\passthrough{\lstinline!exposures!} and optional
\passthrough{\lstinline!covariates!} are complete finite numeric
columns. \passthrough{\lstinline!q!} requests exposure groups;
\passthrough{\lstinline!q=None!} accepts already scored consecutive
levels starting at zero. Supplied boundaries must yield the same
category count for each exposure. \passthrough{\lstinline!alpha!} is the
interval error level, \passthrough{\lstinline!B!} is at least two
row-bootstrap draws, and \passthrough{\lstinline!seed!} controls a local
NumPy random generator. \passthrough{\lstinline!exposure\_time!}
contains one positive observation period per row when periods differ. If
omitted, periods are treated as equal. Weights, formulas, missing-row
options and cluster resampling are not accepted here.

\paragraph{Details}\label{manual-details-19}

For intervention level \(a\), set all exposure scores to \(a\) while
leaving each observed covariate row \(Z_i\) and its observation time in
place. If \(\widehat\mu_i(a)\) is that row's fitted count mean over its
original time, the standardized mean and joint log mean ratio are

\[
\widehat M(a)=n^{-1}\sum_{i=1}^n\widehat\mu_i(a),\qquad
 \widehat\psi=\log\{\widehat M(a+1)/\widehat M(a)\}.
\]

Here \(n\) is the row count. Both intervention means use the same
distribution of observation times. With an additive log link and an
offset equal to the log of observation time, their one-step log ratio is
constant across levels and equals the sum of conditional exposure
coefficients. Thus, the count mean ratio is also a rate ratio for this
model. Each bootstrap draw refits the conditional Poisson model after
resampling whole rows, including their observation times. The sample
standard deviation of its saved log ratios is the bootstrap standard
error. Normal limits and a two-sided normal \(z\) test follow R qgcomp
2.19.6's Poisson convention. The mean-ratio interval exponentiates the
log-scale limits. A failed draw raises an error.

\paragraph{Value}\label{manual-value-23}

\passthrough{\lstinline!PoissonBootstrapResult!} stores the joint log
mean ratio \passthrough{\lstinline!psi!},
\passthrough{\lstinline!rate\_ratio!}, bootstrap
\passthrough{\lstinline!se!}, both intervals,
\passthrough{\lstinline!p\_value!}, read-only
\passthrough{\lstinline!intervention\_means!} and
\passthrough{\lstinline!bootstrap\_effects!}, and the conditional
\passthrough{\lstinline!point\_fit!}.
\passthrough{\lstinline!summary(digits=6)!} displays saved values
without refitting.

\paragraph{Examples}\label{manual-examples-20}

\begin{lstlisting}[language=Python]
count_boot = fit_poisson_bootstrap(y, exposures, q=None, B=200, seed=29)
print(count_boot.rate_ratio, count_boot.rate_ratio_ci)

unequal_time = fit_poisson_bootstrap(
    y, exposures, q=None, exposure_time=periods, B=200, seed=29
)
print(unequal_time.intervention_means, unequal_time.rate_ratio)
\end{lstlisting}

The equal-period installed-package example has an independently known
mean ratio of six. Two equal-period and two unequal-time cases agree
with isolated installed R qgcomp 2.19.6 under identical row schedules.
Separate 4,000-dataset studies assess
\href{https://github.com/nilson01/quantilemix/blob/main/validation/POISSON_BOOTSTRAP_CALIBRATION_RESULTS.md}{equal-period}
and
\href{https://github.com/nilson01/quantilemix/blob/main/validation/POISSON_BOOTSTRAP_OFFSET_CALIBRATION_RESULTS.md}{unequal-time}
intervals.

\paragraph{See Also}\label{manual-see-also-22}

\hyperref[manual-api:fit-poisson]{\passthrough{\lstinline!fit\_poisson!}} for
model-based uncertainty.

\subsubsection{\texorpdfstring{\texttt{fit\_poisson\_ee}: Clustered
Poisson estimating
equations}{fit\_poisson\_ee: Clustered Poisson estimating equations}}\label{manual-fit_poisson_ee-clustered-poisson-estimating-equations}

\paragraph{Description}\label{manual-description-25}

Estimate the marginal log mean or rate ratio for one joint score
increase, allowing repeated observations within identified clusters. The
model is additive on the conditional log scale. Positive observation
times can enter as a log offset.

\paragraph{Usage}\label{manual-usage-27}

\begin{lstlisting}[language=Python]
from quantilemix import fit_poisson_ee
result = fit_poisson_ee(
    counts, exposures, cluster_ids, covariates=None,
    q=4, breaks=None, alpha=0.05, weights=None,
    exposure_time=None
)
print(result.mean_ratio, result.se, result.ci)
\end{lstlisting}

\paragraph{Arguments}\label{manual-arguments-22}

\passthrough{\lstinline!counts!} contains nonnegative integer outcomes.
\passthrough{\lstinline!exposures!} and optional
\passthrough{\lstinline!covariates!} contain complete numeric rows.
\passthrough{\lstinline!cluster\_ids!} contains one finite exact integer
ID per row and at least two distinct IDs. \passthrough{\lstinline!q!}
and \passthrough{\lstinline!breaks!} determine saved exposure scores;
joint intervention levels must be consecutive from zero.
\passthrough{\lstinline!alpha!} is the interval error level. Optional
positive \passthrough{\lstinline!weights!} apply to both estimating
equations and cluster totals but not exposure cuts.
\passthrough{\lstinline!exposure\_time!} is a positive time vector with
one value per row. The times are held fixed when the joint exposure
score changes. Nonlinear terms and time-lag structure are not supported
in this subset.

\paragraph{Method and interpretation}\label{manual-method-and-interpretation-1}

Let \(t_i\) be row \(i\)'s observation time, or one if none is supplied.
The conditional equation fits a Poisson model with log mean
\(\log t_i+\beta_0+\sum_j\beta_jX_{ji}^q+\gamma^\top Z_i\). For each
joint score \(a\), the method predicts an expected count after assigning
that score to every exposure while keeping \(t_i\) fixed. A marginal
log-linear model describes these predictions across the common score
\(a\). Case weights, when supplied, enter both estimating equations.
Under this additive conditional model, the fitted one-category log rate
ratio \(\widehat\psi\) equals the sum of fitted exposure coefficients.
The cluster-robust covariance sums both estimating functions within each
cluster before forming the sandwich. Python calculates the derivative
exactly; R qgcomp 2.19.6 uses a numerical derivative. The normal
interval and test use the cluster-robust standard error. They can be
unreliable with few clusters. With unequal times, \(\exp(\widehat\psi)\)
is a rate ratio for the observed time distribution. This does not model
time-varying effects.

\paragraph{Value}\label{manual-value-24}

\passthrough{\lstinline!PoissonEEResult!} saves conditional and marginal
coefficients, their robust covariances, the full stacked covariance,
derivative bread, cluster meat, cluster count, optional case weights and
observation times, fitted counts, intervention levels and means, the
joint log effect, standard error, interval, test and p-value.
\passthrough{\lstinline!mean\_ratio!} and
\passthrough{\lstinline!rate\_ratio!} and
\passthrough{\lstinline!mean\_ratio\_ci!} exponentiate the log-scale
values. \passthrough{\lstinline!transform!} reuses fitted cuts;
\passthrough{\lstinline!summary!} describes the uncertainty. Two fixed
equal-time installed-R cases and an independent exact-cell calculation
pass; see \passthrough{\lstinline!validation/POISSON\_EE\_RESULTS.md!}.
Installed R 2.19.6 ignores its \passthrough{\lstinline!offset!} argument
in this EE method. Python's unequal-time extension instead passes an
independent sandwich calculation and a fixed installed-R difference
check; see
\passthrough{\lstinline!validation/POISSON\_EE\_OFFSET\_RESULTS.md!}.
Repeated-data coverage, overdispersion and small-cluster behavior remain
untested. One weighted installed-R case and an independent
row-replication check cover case weights; see
\passthrough{\lstinline!validation/WEIGHTED\_EE\_RESULTS.md!}.

\subsubsection{\texorpdfstring{\texttt{PoissonResult}: Count model
result and conditional rate
ratio}{PoissonResult: Count model result and conditional rate ratio}}\label{manual-api:poisson-result}

\paragraph{Description}\label{manual-description-26}

This immutable result stores a Poisson-log fit and its joint exposure
coefficient on the log-rate scale. It has no Gaussian residual variance
or Student t diagnostics.

\paragraph{Usage}\label{manual-usage-28}

\begin{lstlisting}[language=Python]
result = fit_poisson(y, exposures, q=None)
print(result.psi, result.rate_ratio)
print(result.summary())
expected_count = result.predict(new_exposures)
\end{lstlisting}

\paragraph{Value}\label{manual-value-25}

{\def\LTcaptype{none} % do not increment counter
\begin{longtable}[]{@{}
  >{\raggedright\arraybackslash}p{(\linewidth - 2\tabcolsep) * \real{0.5000}}
  >{\raggedright\arraybackslash}p{(\linewidth - 2\tabcolsep) * \real{0.5000}}@{}}
\toprule\noalign{}
\begin{minipage}[b]{\linewidth}\raggedright
Field
\end{minipage} & \begin{minipage}[b]{\linewidth}\raggedright
Meaning
\end{minipage} \\
\midrule\noalign{}
\endhead
\bottomrule\noalign{}
\endlastfoot
\passthrough{\lstinline!coefficients!},
\passthrough{\lstinline!coefficient\_names!} & Read-only full-model
estimates in intercept, exposure, then adjustment order. \\
\passthrough{\lstinline!covariance!} & Read-only full coefficient
covariance from final Fisher information. \\
\passthrough{\lstinline!psi!}, \passthrough{\lstinline!var\_psi!},
\passthrough{\lstinline!se!} & Conditional joint log rate ratio, its
variance and standard error. \\
\passthrough{\lstinline!ci!}, \passthrough{\lstinline!p\_value!},
\passthrough{\lstinline!test\_statistic!} & Normal confidence limits,
two-sided z p-value and z statistic. \\
\passthrough{\lstinline!rate\_ratio!},
\passthrough{\lstinline!rate\_ratio\_ci!} & Exponentiated joint rate
effect and interval at equal observation time. \\
\passthrough{\lstinline!fitted\_values!} & Training expected counts. \\
\passthrough{\lstinline!case\_weights!} & Read-only GLM prior weights
when supplied; otherwise \passthrough{\lstinline!None!}. \\
\passthrough{\lstinline!exposure\_time!} & Read-only training times when
an offset was supplied; otherwise \passthrough{\lstinline!None!}. \\
\passthrough{\lstinline!pos\_weights!},
\passthrough{\lstinline!neg\_weights!} & Descriptive coefficient shares
within each sign group. \\
\passthrough{\lstinline!scores!}, \passthrough{\lstinline!breaks!},
\passthrough{\lstinline!q!} & Saved exposure scores and cutpoint
information. \\
\passthrough{\lstinline!family!}, \passthrough{\lstinline!link!} & The
strings \passthrough{\lstinline!"poisson"!} and
\passthrough{\lstinline!"log"!}. \\
\passthrough{\lstinline!diagnostics!} & Read-only rank, condition,
iteration and grouping information. \\
\end{longtable}
}

\paragraph{Methods}\label{manual-methods-1}

\passthrough{\lstinline!predict!} returns conditional expected counts.
Offset fits require new \passthrough{\lstinline!exposure\_time!} values.
\passthrough{\lstinline!transform!} reuses saved exposure cuts. Formula
results also support \passthrough{\lstinline!predict\_data!} and
\passthrough{\lstinline!transform\_data!}.
\passthrough{\lstinline!summary!} returns a plain-text display of the
saved joint effect and inference without refitting.

\paragraph{See Also}\label{manual-see-also-23}

\hyperref[manual-api:fit-poisson]{\passthrough{\lstinline!fit\_poisson!}},
\hyperref[manual-api:fit-poisson-formula]{\passthrough{\lstinline!fit\_poisson\_formula!}},
\hyperref[manual-api:binomial-result]{\passthrough{\lstinline!BinomialResult!}}.

\subsection{Survival, censored and multicategory
outcomes}\label{manual-survival-censored-and-multicategory-outcomes}

\subsubsection{\texorpdfstring{\texttt{fit\_cox}: Conditional
right-censored survival
fit}{fit\_cox: Conditional right-censored survival fit}}\label{manual-api:fit-cox}

\paragraph{Description}\label{manual-description-27}

Fit a restricted Cox proportional-hazards model for observations with a
follow-up time and an event indicator. The estimate
\(\widehat\psi=\sum_j\widehat\beta_j\) is a conditional log hazard ratio
at the same adjustment values. It is not a survival probability or a
marginal effect.

\paragraph{Usage}\label{manual-usage-29}

\begin{lstlisting}[language=Python]
fit_cox(durations, events, exposures, covariates=None,
        *, q=4, breaks=None, alpha=0.05, weights=None,
        offset=None, cluster_ids=None)
\end{lstlisting}

\paragraph{Arguments}\label{manual-arguments-23}

\passthrough{\lstinline!durations!} is a one-dimensional vector of
positive follow-up times. \passthrough{\lstinline!events!} has the same
length and contains 1 for an observed event or 0 for right censoring.
\passthrough{\lstinline!exposures!} has one or more complete numeric
columns. \passthrough{\lstinline!covariates!} contains optional complete
numeric adjustment columns. \passthrough{\lstinline!q!} and
\passthrough{\lstinline!breaks!} use the same exposure grouping rules as
\passthrough{\lstinline!quantize!}; use \passthrough{\lstinline!q=None!}
only for nonnegative integer scores. \passthrough{\lstinline!alpha!}
sets the normal confidence level to \(1-\alpha\).
\passthrough{\lstinline!weights!} supplies one positive finite case
weight per row. Case weights affect risk sets and event contributions
but not exposure cutpoints. \passthrough{\lstinline!offset!} supplies
one known, finite log-hazard contribution per row. It is not an
observation-time offset and has no estimated coefficient.
\passthrough{\lstinline!cluster\_ids!} supplies one finite, exact
integer ID per row, with at least two distinct IDs. Rows sharing an ID
contribute together to robust uncertainty. IDs do not change the fitted
hazard ratio or exposure cuts.

\paragraph{Details}\label{manual-details-20}

For exposure scores \(X_{ji}^q\) and adjustment values \(Z_i\), the
fitted Cox hazard is \(h(t\mid X_i^q,Z_i,o_i)=
h_0(t)\exp(\sum_j\beta_jX_{ji}^q+\gamma^TZ_i+o_i)\), where \(o_i\) is
the optional known log offset (zero when omitted). Here \(h_0(t)\) is an
unspecified baseline hazard. The reported joint log hazard ratio is
\(\psi=\sum_j\beta_j\): every exposure score rises by one, while
adjustment values stay fixed. With no weights or only integer case
weights, the standard error uses the fitted partial-likelihood
information matrix. Fractional case weights use a robust variance built
from each row's weighted score contribution, matching the default R
\passthrough{\lstinline!coxph!} behavior. With cluster IDs, weighted row
scores are first summed within each cluster; the sums form the robust
covariance. No small-cluster correction is applied. Normal limits and a
two-sided normal \(z\) test follow the conditional R qgcomp 2.19.6
convention. When events share a follow-up time, Efron's method
approximates their order within the event group. Records censored at
that time remain at risk for its events. This matches the installed R
reference's default tie method. The array method has no formula or
population-average survival estimate. Its uncentered baseline cumulative
hazard supports conditional survival predictions at the observed
follow-up times. Independent censoring and proportional hazards are
substantive assumptions; a successful fit does not establish either
assumption or a causal interpretation.

\paragraph{Value}\label{manual-value-26}

\passthrough{\lstinline!CoxResult!} stores read-only coefficients, their
covariance, exposure scores and cutpoints, joint
\passthrough{\lstinline!psi!}, variance, standard error, confidence
limits, test statistic, p-value, event count, number of shared-time
groups, maximized partial log likelihood, weighted baseline hazard and
iteration count. \passthrough{\lstinline!case\_weights!} stores the
supplied read-only weights, or \passthrough{\lstinline!None!}.
\passthrough{\lstinline!robust\_variance!} indicates a fractional-weight
or cluster-score sandwich. \passthrough{\lstinline!offset!} stores the
supplied read-only log offsets, or \passthrough{\lstinline!None!}.
\passthrough{\lstinline!cluster\_ids!} stores read-only IDs, and
\passthrough{\lstinline!cluster\_count!} gives their count; both are
\passthrough{\lstinline!None!} when omitted.
\passthrough{\lstinline!hazard\_ratio!} and
\passthrough{\lstinline!hazard\_ratio\_ci!} exponentiate the log-scale
values.
\passthrough{\lstinline!survival\_curve(exposures, covariates=None, offset=...)!}
returns observed follow-up times and a matrix of conditional survival
probabilities with one column per new profile. A fit that used offsets
requires one offset for each new profile; it does not reuse the training
offsets. It reuses saved exposure cuts; censor-only times retain the
previous survival value. \passthrough{\lstinline!summary()!} displays
the saved values without refitting.

\paragraph{See Also}\label{manual-see-also-24}

\hyperref[manual-api:quantize]{\passthrough{\lstinline!quantize!}}. The two
fixed reference cases and their comparison procedure are recorded in the
Cox validation report. The two additional shared-time cases are recorded
in the \passthrough{\lstinline!COX\_TIES\_RESULTS.md!} report. Three
fixed weighted cases are recorded in
\passthrough{\lstinline!COX\_WEIGHTS\_RESULTS.md!}. Three additional
offset cases and the baseline-centering convention are recorded in
\passthrough{\lstinline!COX\_OFFSET\_RESULTS.md!}. Three clustered cases
and the weighted-score convention are recorded in
\passthrough{\lstinline!COX\_CLUSTER\_RESULTS.md!}.

\subsubsection{\texorpdfstring{\texttt{fit\_cox\_formula}: Conditional
Cox formula
fit}{fit\_cox\_formula: Conditional Cox formula fit}}\label{manual-fit_cox_formula-conditional-cox-formula-fit}

\paragraph{Description}\label{manual-description-28}

Fit the same conditional Cox method from named complete-data columns.

\paragraph{Usage}\label{manual-usage-30}

\begin{lstlisting}[language=Python]
fit_cox_formula("Surv(time,event) ~ x1 + x2 + C(group)",
                data, exposures=["x1", "x2"], q=4,
                weights=None, cluster_ids=None, offset=None)
\end{lstlisting}

\paragraph{Arguments and Value}\label{manual-arguments-and-value-2}

\passthrough{\lstinline!Surv(time,event)!} names positive follow-up time
and zero-one event columns. The right side accepts distinct additive
numeric exposure terms, numeric adjustment terms and
\passthrough{\lstinline!C(name)!} for text adjustments. The first sorted
observed text level is the reference. \passthrough{\lstinline!weights!},
\passthrough{\lstinline!cluster\_ids!} and
\passthrough{\lstinline!offset!} are row-aligned vectors. Missing rows,
interactions and transformed terms are not supported. The returned
\passthrough{\lstinline!CoxResult!} has the same effect and uncertainty
interpretation as
\hyperref[manual-api:fit-cox]{\passthrough{\lstinline!fit\_cox!}}. The public
\passthrough{\lstinline!qgcomp(..., family="cox")!} route selects this
additive conditional fit.

\subsubsection{\texorpdfstring{\texttt{fit\_cox\_bootstrap}: Marginal
Cox survival
bootstrap}{fit\_cox\_bootstrap: Marginal Cox survival bootstrap}}\label{manual-api:fit-cox-bootstrap}

\paragraph{Description}\label{manual-description-29}

Fit a marginal log hazard curve to simulated event outcomes under joint
exposure-score interventions. Fixed-cut row resampling estimates
uncertainty. This is a restricted Python counterpart to R
\passthrough{\lstinline!qgcomp.cox.boot!}, not an end-to-end numerical
translation of its event-time generator.

\paragraph{Usage}\label{manual-usage-31}

\begin{lstlisting}[language=Python]
from quantilemix import fit_cox_bootstrap
result = fit_cox_bootstrap(
    durations, events, exposures, covariates=None,
    q=4, breaks=None, alpha=0.05, degree=1,
    B=200, MCsize=None, seed=None)
\end{lstlisting}

\paragraph{Arguments}\label{manual-arguments-24}

\passthrough{\lstinline!durations!} are positive follow-up times;
\passthrough{\lstinline!events!} are zero or one.
\passthrough{\lstinline!exposures!} and optional
\passthrough{\lstinline!covariates!} are complete numeric rows.
\passthrough{\lstinline!q!}, \passthrough{\lstinline!breaks!} and
\passthrough{\lstinline!alpha!} follow
\hyperref[manual-api:fit-cox]{\passthrough{\lstinline!fit\_cox!}}.
\passthrough{\lstinline!degree!} is one or two; degree two requires at
least three distinct intervention levels. \passthrough{\lstinline!B!} is
the number of row resamples and must be at least two.
\passthrough{\lstinline!MCsize!} is the simulated number of people per
intervention level; it must be at least the input row count and defaults
to the larger of that count and 10,000. \passthrough{\lstinline!seed!}
controls a local NumPy random generator.

\paragraph{Details}\label{manual-details-21}

The conditional Cox model gives an event hazard for each exposure-score
level and sampled adjustment profile. Python draws an exponential
threshold and finds its first crossing of the fitted cumulative hazard.
If no crossing occurs, follow-up is censored at the last observed time.
A second Cox model fits the simulated outcomes at all joint levels. Its
coefficients describe the fitted marginal log hazard curve. Degree two
has linear and squared score terms. The fitted change from score \(a\)
to \(a+1\) is \(\widehat\psi_1+(2a+1)\widehat\psi_2\).

Each row resample keeps the original exposure cutpoints, refits both
stages, and yields a marginal curve coefficient. The sample covariance
of these coefficients gives normal confidence limits. A failed replicate
raises an error. The current failure object retains its draw number,
original effect, intervention levels and effects from completed draws;
no interval is returned for that dataset. Case weights, clusters,
offsets and time-varying exposures are not supported. A four-condition
known-score interval study is reported in
\href{https://github.com/nilson01/quantilemix/blob/main/validation/COX_BOOTSTRAP_STUDY_RESULTS.md}{validation/COX\_BOOTSTRAP\_STUDY\_RESULTS.md};
it does not establish coverage for other populations. Installed R 2.19.6
compares hazard increments directly with uniform draws, which can leave
simulated times missing. The same-outcome second-stage and bootstrap
arithmetic agree in two fixed R checks; the full event generators are
intentionally different. See
\href{https://github.com/nilson01/quantilemix/blob/main/validation/COX_BOOTSTRAP_RESULTS.md}{validation/COX\_BOOTSTRAP\_RESULTS.md}.

\paragraph{Value}\label{manual-value-27}

\passthrough{\lstinline!CoxBootstrapResult!} stores
\passthrough{\lstinline!conditional\_fit!}, the read-only marginal
coefficient vector \passthrough{\lstinline!psi!}, full bootstrap
\passthrough{\lstinline!covariance!}, \passthrough{\lstinline!se!},
\passthrough{\lstinline!ci!}, and two-sided normal
\passthrough{\lstinline!p\_value!}. It also stores
\passthrough{\lstinline!intervention\_levels!}, synthetic outcomes and
their scores, original \passthrough{\lstinline!training\_covariates!},
\passthrough{\lstinline!bootstrap\_effects!}, the marginal baseline
cumulative hazard and its timeline, \passthrough{\lstinline!B!},
\passthrough{\lstinline!MCsize!}, \passthrough{\lstinline!degree!},
\passthrough{\lstinline!alpha!} and \passthrough{\lstinline!seed!}.
\passthrough{\lstinline!effect\_at(a)!} gives the one-level log hazard
change starting at \(a\). \passthrough{\lstinline!summary()!} displays
the saved effects and uncertainty. The separate
\passthrough{\lstinline!qgcomp\_survcurve\_boot!} helper returns fitted
curves, without bands.

\paragraph{See Also}\label{manual-see-also-25}

\hyperref[manual-api:fit-cox]{\passthrough{\lstinline!fit\_cox!}} and
\hyperref[manual-api:cox-result]{\passthrough{\lstinline!CoxResult!}}.

\subsubsection{\texorpdfstring{\texttt{qgcomp\_survcurve\_boot}:
Marginal and conditional survival
curves}{qgcomp\_survcurve\_boot: Marginal and conditional survival curves}}\label{manual-qgcomp_survcurve_boot-marginal-and-conditional-survival-curves}

\paragraph{Description}\label{manual-description-30}

Return four curve tables from a
\hyperref[manual-api:fit-cox-bootstrap]{\passthrough{\lstinline!fit\_cox\_bootstrap!}}
result. These display how its marginal Cox curve compares with the
underlying conditional Cox fit. No uncertainty bands are provided.

\paragraph{Usage}\label{manual-usage-32}

\begin{lstlisting}[language=Python]
from quantilemix import qgcomp_survcurve_boot
curves = qgcomp_survcurve_boot(result)
\end{lstlisting}

\paragraph{Arguments}\label{manual-arguments-25}

\passthrough{\lstinline!result!} must be a
\passthrough{\lstinline!CoxBootstrapResult!} from the restricted
marginal survival bootstrap. The helper does not fit or resample again.

\paragraph{Details}\label{manual-details-22}

\passthrough{\lstinline!mdfq!} contains one marginal-model curve at each
fitted joint score. \passthrough{\lstinline!mdfpop!} gives their
equal-weight mean. The conditional-model
\passthrough{\lstinline!cdfpop!} averages over every original adjustment
profile and score. Following R qgcomp 2.19.6,
\passthrough{\lstinline!cdfq!} gives each score's conditional curve at
the \emph{first} original adjustment profile. Thus
\passthrough{\lstinline!cdfq!} is not a population-average curve within
each score. Marginal times come from the simulated outcomes; conditional
times come from the original observations. For prequantized
\passthrough{\lstinline!q=None!}, Python uses the actual fitted
pooled-quartile intervention levels. R's helper instead reconstructs
zero-based integer levels from its saved breaks; displays can differ in
that setting.

\paragraph{Value}\label{manual-value-28}

An immutable \passthrough{\lstinline!CoxSurvivalCurves!} object has
read-only, R-named tables \passthrough{\lstinline!mdfpop!},
\passthrough{\lstinline!cdfpop!}, \passthrough{\lstinline!mdfq!} and
\passthrough{\lstinline!cdfq!}. Each table maps
\passthrough{\lstinline!"time"!} and \passthrough{\lstinline!"surv"!} to
equal-length arrays. The by-score tables also have a one-based
\passthrough{\lstinline!"q"!} column, with rows ordered first by score
and then by increasing time. \passthrough{\lstinline!levels!} contains
the actual joint scores. For example,
\passthrough{\lstinline!curves.mdfq["surv"]!} contains the marginal
probabilities.

\subsubsection{\texorpdfstring{\texttt{CoxResult}: Conditional hazard
ratio and
uncertainty}{CoxResult: Conditional hazard ratio and uncertainty}}\label{manual-api:cox-result}

\paragraph{Description}\label{manual-description-31}

This immutable object stores the restricted conditional Cox fit returned
by \passthrough{\lstinline!fit\_cox!}. Its baseline hazard supports
conditional survival prediction.

\paragraph{Value}\label{manual-value-29}

\passthrough{\lstinline!coefficients!} and
\passthrough{\lstinline!covariance!} are read-only arrays for the full
conditional model, in exposure-then-adjustment order.
\passthrough{\lstinline!coefficient\_names!},
\passthrough{\lstinline!exposure\_names!} and
\passthrough{\lstinline!covariate\_names!} give that order.
\passthrough{\lstinline!scores!} and \passthrough{\lstinline!breaks!}
retain the fitted exposure grouping. \passthrough{\lstinline!psi!} sums
the exposure coefficients; \passthrough{\lstinline!var\_psi!} and
\passthrough{\lstinline!se!} give its variance and standard error. They
are model-based without fractional case weights or clusters and
row-score robust with them. \passthrough{\lstinline!ci!},
\passthrough{\lstinline!test\_statistic!},
\passthrough{\lstinline!p\_value!} and \passthrough{\lstinline!alpha!}
describe normal uncertainty. \passthrough{\lstinline!event\_count!},
\passthrough{\lstinline!tied\_time\_groups!},
\passthrough{\lstinline!log\_likelihood!} and
\passthrough{\lstinline!iterations!} describe the fit.
\passthrough{\lstinline!timeline!} and
\passthrough{\lstinline!baseline\_cumulative\_hazard!} are read-only
arrays at all distinct observed follow-up times, including times with
censoring only. The baseline is uncentered; R
\passthrough{\lstinline!basehaz(centered=FALSE)!} retains the mean
fitted offset in the checked cases. \passthrough{\lstinline!offset!}
stores any fitted log offsets and is required for new profiles if
supplied at fitting. \passthrough{\lstinline!hazard\_ratio!} and
\passthrough{\lstinline!hazard\_ratio\_ci!} exponentiate the joint log
effect and its limits. \passthrough{\lstinline!family!} is
\passthrough{\lstinline!"cox"!};
\passthrough{\lstinline!reference\_version!} records the R qgcomp
package version used for comparison.

\paragraph{Methods}\label{manual-methods-2}

\passthrough{\lstinline!summary(digits=6)!} returns a text display of
the saved values. \passthrough{\lstinline!survival\_curve!} accepts new
exposures and optional adjustment values. It returns a pair: increasing
follow-up times and a time-by-profile probability matrix. New rows must
use the fitted exposure and adjustment column order. The model is not
refitted. These are conditional probabilities, not population-average
curves or uncertainty bands; values after the last observed time are not
extrapolated.

\paragraph{See Also}\label{manual-see-also-26}

\hyperref[manual-api:fit-cox]{\passthrough{\lstinline!fit\_cox!}}.

\subsubsection{\texorpdfstring{\texttt{fit\_tobit}: Gaussian
censored-outcome
fit}{fit\_tobit: Gaussian censored-outcome fit}}\label{manual-api:fit-tobit}

\paragraph{Description}\label{manual-description-32}

Fit a Gaussian Tobit model for a continuous outcome censored at known
limits. The estimate \(\widehat\psi=\sum_j\widehat\beta_j\) changes the
model's \emph{latent mean}. It is not a change in the mean of observed
censored values.

\paragraph{Usage}\label{manual-usage-33}

\begin{lstlisting}[language=Python]
fit_tobit(y, exposures, covariates=None, *, left=0.0,
          right=None, q=4, breaks=None, alpha=0.05)
\end{lstlisting}

\paragraph{Arguments}\label{manual-arguments-26}

\passthrough{\lstinline!y!} is a complete numeric vector that has
already been clipped to \passthrough{\lstinline!left!} and/or
\passthrough{\lstinline!right!}. At least one limit is required;
\passthrough{\lstinline!None!} means no limit on that side.
\passthrough{\lstinline!exposures!} contains one or more complete
numeric columns. \passthrough{\lstinline!covariates!} contains optional
complete numeric adjustment columns. \passthrough{\lstinline!q!} and
\passthrough{\lstinline!breaks!} follow the same grouping rules as
\passthrough{\lstinline!quantize!}; \passthrough{\lstinline!q=None!}
requires nonnegative integer scores. \passthrough{\lstinline!alpha!}
sets the normal confidence level to \(1-\alpha\).

\paragraph{Details}\label{manual-details-23}

The unobserved value \(Y_i^*\) has a Gaussian distribution with mean
\(\beta_0+\sum_j\beta_jX_{ji}^q+\gamma^TZ_i\) and residual standard
deviation \(\sigma\). The observed outcome equals \(Y_i^*\) inside the
limits and the corresponding limit outside them. Here \(X_{ji}^q\) is
exposure \(j\)'s score for row \(i\) and \(Z_i\) holds its adjustment
values. The likelihood uses a normal density for uncensored observations
and normal tail probabilities for values at the limits. The reported
\(\psi=\sum_j\beta_j\) changes the latent mean when all exposure scores
rise by one and the adjustment values stay fixed. The normal interval
and two-sided \(z\) test use the fitted information matrix.

This restricted Python method requires correctly recorded limits and
already-clipped complete numeric outcomes. It has no weights, clusters,
formula, alternative latent distribution or marginal observed-mean
model. The Gaussian distribution and censoring mechanism are model
assumptions; successful fitting does not establish them or a causal
interpretation.

\paragraph{Value}\label{manual-value-30}

\passthrough{\lstinline!TobitResult!} stores read-only coefficients,
coefficient covariance, full covariance including log residual scale,
fitted scale, joint effect and uncertainty, exposure scores, cutpoints
and censor counts.

\paragraph{See Also}\label{manual-see-also-27}

\hyperref[manual-api:quantize]{\passthrough{\lstinline!quantize!}}. Fixed R
comparisons and the independent likelihood check are recorded in the
Tobit validation report.

\subsubsection{\texorpdfstring{\texttt{TobitResult}: Latent-mean effect
and censoring
summary}{TobitResult: Latent-mean effect and censoring summary}}\label{manual-tobitresult-latent-mean-effect-and-censoring-summary}

\paragraph{Description}\label{manual-description-33}

This immutable object stores the restricted Gaussian Tobit fit returned
by \passthrough{\lstinline!fit\_tobit!}.

\paragraph{Value}\label{manual-value-31}

\passthrough{\lstinline!coefficients!} and
\passthrough{\lstinline!covariance!} are read-only arrays for the latent
mean model, in intercept-exposure-adjustment order.
\passthrough{\lstinline!full\_covariance!} also includes the fitted
log-scale parameter. \passthrough{\lstinline!coefficient\_names!} gives
the mean-model order. \passthrough{\lstinline!sigma!} is the latent
residual standard deviation. \passthrough{\lstinline!psi!},
\passthrough{\lstinline!var\_psi!}, \passthrough{\lstinline!se!},
\passthrough{\lstinline!ci!}, \passthrough{\lstinline!test\_statistic!},
\passthrough{\lstinline!p\_value!} and \passthrough{\lstinline!alpha!}
give the joint latent effect and model-based normal inference.
\passthrough{\lstinline!left!}, \passthrough{\lstinline!right!},
\passthrough{\lstinline!left\_count!} and
\passthrough{\lstinline!right\_count!} record the limits and the number
at each limit. \passthrough{\lstinline!iterations!} counts fitting
steps. \passthrough{\lstinline!scores!} and
\passthrough{\lstinline!breaks!} retain exposure grouping.
\passthrough{\lstinline!family!} is
\passthrough{\lstinline!"gaussian\_tobit"!};
\passthrough{\lstinline!reference\_version!} records the R qgcomp
package version used for comparison.

\paragraph{Methods}\label{manual-methods-3}

\passthrough{\lstinline!summary(digits=6)!} returns a text display of
saved values without refitting. No observed-mean prediction method is
provided.

\paragraph{See Also}\label{manual-see-also-28}

\hyperref[manual-api:fit-tobit]{\passthrough{\lstinline!fit\_tobit!}}.

\subsubsection{\texorpdfstring{\texttt{fit\_multinomial}: Conditional
multinomial-outcome
fit}{fit\_multinomial: Conditional multinomial-outcome fit}}\label{manual-api:fit-multinomial}

\paragraph{Description}\label{manual-description-34}

Fit an additive multinomial-logit model for at least three named outcome
categories. The first supplied category is the reference. The result
gives a separate estimated joint log odds ratio \(\widehat\psi_k\) for
each other category \(k\) relative to that reference; there is no single
pooled multinomial effect.

\paragraph{Usage}\label{manual-usage-34}

\begin{lstlisting}[language=Python]
fit_multinomial(y, exposures, covariates=None, *,
                levels, q=4, breaks=None, alpha=0.05)
\end{lstlisting}

\paragraph{Arguments}\label{manual-arguments-27}

\passthrough{\lstinline!y!} is a complete vector of category names.
\passthrough{\lstinline!levels!} lists every category in the intended
order, with the reference first; at least three distinct nonempty
strings are required. \passthrough{\lstinline!exposures!} and optional
\passthrough{\lstinline!covariates!} are complete numeric columns.
\passthrough{\lstinline!q!} and \passthrough{\lstinline!breaks!} follow
\passthrough{\lstinline!quantize!}; \passthrough{\lstinline!q=None!}
requires nonnegative integer exposure scores.
\passthrough{\lstinline!alpha!} sets the normal confidence level to
\(1-\alpha\).

\paragraph{Details}\label{manual-details-24}

For outcome category \(k\) other than the reference, the model's log
odds relative to the reference is
\(\beta_{0k}+\sum_j\beta_{jk}X_{ji}^q+\gamma_k^TZ_i\). Here \(X_{ji}^q\)
is exposure \(j\)'s score for row \(i\) and \(Z_i\) holds its adjustment
values. The category-specific joint effect is
\(\psi_k=\sum_j\beta_{jk}\): every exposure score rises by one while the
adjustment values stay fixed. The fitted category probabilities sum to
one. The full coefficient covariance provides a covariance matrix for
the separate \(\psi_k\) values. Normal intervals and \(z\) tests
describe each category-versus-reference effect. Two chi-square tests
check the global null that all joint effects are zero and the
homogeneity null that they are equal. They do not test a marginal
population-average effect.

This Python subset uses complete numeric model inputs and explicit
category order. It has no case weights, clusters, formula or nonlinear
terms. The separate bootstrap function estimates a marginal effect. A
successful fit does not establish exchangeability, model fit or a causal
interpretation.

\paragraph{Value}\label{manual-value-32}

\passthrough{\lstinline!MultinomialResult!} stores coefficient and
joint-effect matrices, their covariance, conditional fitted
probabilities, exposure scores and cutpoints, named outcome levels,
uncertainty and fitting iterations.

\paragraph{See Also}\label{manual-see-also-29}

\hyperref[manual-api:quantize]{\passthrough{\lstinline!quantize!}}. Fixed R
comparisons under default and tighter R optimizer settings are in the
multinomial validation report.

\subsubsection{\texorpdfstring{\texttt{fit\_multinomial\_bootstrap}:
Marginal multinomial row
bootstrap}{fit\_multinomial\_bootstrap: Marginal multinomial row bootstrap}}\label{manual-fit_multinomial_bootstrap-marginal-multinomial-row-bootstrap}

\paragraph{Description}\label{manual-description-35}

Estimate category-specific marginal log odds slopes after jointly
setting every exposure to each intervention score. This uses a simulated
outcome at each score and a fixed-cut row bootstrap for uncertainty. It
is distinct from the conditional effect returned by
\passthrough{\lstinline!fit\_multinomial!}.

\paragraph{Usage}\label{manual-usage-35}

\begin{lstlisting}[language=Python]
fit_multinomial_bootstrap(y, exposures, covariates=None, *,
                          levels, q=4, breaks=None,
                          alpha=0.05, B=200, seed=None)
\end{lstlisting}

\paragraph{Arguments}\label{manual-arguments-28}

The outcome, exposure, adjustment, quantization and interval inputs
follow \passthrough{\lstinline!fit\_multinomial!}.
\passthrough{\lstinline!B!} is the number of independent-row bootstrap
samples and must be at least two. \passthrough{\lstinline!seed!}
controls a local NumPy random generator. The global random state is
unchanged.

\paragraph{Details}\label{manual-details-25}

The conditional model predicts outcome-category probabilities under each
joint intervention. One category is drawn for every row and score. A
multinomial logit with an intercept and a linear joint-score term is
fitted to those simulated outcomes. Each bootstrap sample refits both
stages on sampled rows, retaining the original exposure cutpoints. A
failed sample raises its position rather than being replaced. The
confidence interval uses a normal critical value and the standard
deviation of bootstrap slopes. It does not correct separately for the
finite number of simulated outcomes. This follows R qgcomp 2.19.6's
method stages, but the random stream and optimizer differ. Installed-R
and repeated-data checks are open.

\paragraph{Value}\label{manual-value-33}

The result stores one marginal slope per nonreference outcome in
\passthrough{\lstinline!psi!}. Its \passthrough{\lstinline!covariance!},
\passthrough{\lstinline!se!}, \passthrough{\lstinline!ci!} and
\passthrough{\lstinline!p\_value!} fields give bootstrap uncertainty.

The \passthrough{\lstinline!intervention\_levels!} field holds joint
scores. The \passthrough{\lstinline!intervention\_probabilities!} field
holds mean conditional category probabilities in outcome-level order.
\passthrough{\lstinline!bootstrap\_effects!} holds one slope row per
resample.

The \passthrough{\lstinline!point\_fit!} field holds the conditional
fit. The \passthrough{\lstinline!odds\_ratios!} property exponentiates
marginal slopes, and \passthrough{\lstinline!summary()!} displays them.

\paragraph{See Also}\label{manual-see-also-30}

\hyperref[manual-api:fit-multinomial]{\passthrough{\lstinline!fit\_multinomial!}}.

\subsubsection{\texorpdfstring{\texttt{MultinomialResult}:
Category-specific joint
effects}{MultinomialResult: Category-specific joint effects}}\label{manual-multinomialresult-category-specific-joint-effects}

\paragraph{Description}\label{manual-description-36}

This immutable object stores the restricted conditional multinomial fit
returned by \passthrough{\lstinline!fit\_multinomial!}.

\paragraph{Value}\label{manual-value-34}

\passthrough{\lstinline!coefficients!} has one row per nonreference
outcome and columns in intercept-exposure-adjustment order.
\passthrough{\lstinline!coefficient\_names!},
\passthrough{\lstinline!levels!},
\passthrough{\lstinline!exposure\_names!} and
\passthrough{\lstinline!covariate\_names!} name these dimensions.
\passthrough{\lstinline!covariance!} is the full read-only coefficient
covariance, ordered by category rows. \passthrough{\lstinline!psi!},
\passthrough{\lstinline!psi\_covariance!}, \passthrough{\lstinline!se!},
\passthrough{\lstinline!ci!} and \passthrough{\lstinline!p\_value!} give
category-specific joint log odds effects and normal inference.
\passthrough{\lstinline!odds\_ratios!} exponentiates the joint effects.
\passthrough{\lstinline!fitted\_probabilities!} contains training
probabilities in \passthrough{\lstinline!levels!} order.
\passthrough{\lstinline!scores!} and \passthrough{\lstinline!breaks!}
save exposure grouping. \passthrough{\lstinline!alpha!} and
\passthrough{\lstinline!iterations!} describe inference and fitting.
\passthrough{\lstinline!family!} is
\passthrough{\lstinline!"multinomial"!};
\passthrough{\lstinline!reference\_version!} records the R qgcomp
package version used for comparison.

\paragraph{Methods}\label{manual-methods-4}

\passthrough{\lstinline!predict\_proba!} returns conditional category
probabilities for new exposure and adjustment rows, reusing the fitted
cutpoints. It warns about exposures outside their training ranges.
\passthrough{\lstinline!summary(digits=6)!} displays the saved
category-specific estimates without refitting.
\passthrough{\lstinline!joint\_test()!} returns the global-null
chi-square statistic, degrees of freedom and p-value.
\passthrough{\lstinline!homogeneity\_test()!} returns the corresponding
test that all nonreference joint effects are equal.

\paragraph{See Also}\label{manual-see-also-31}

\hyperref[manual-api:fit-multinomial]{\passthrough{\lstinline!fit\_multinomial!}}.

\subsection{Zero-inflated and hurdle
counts}\label{manual-zero-inflated-and-hurdle-counts}

These models handle count data with many zeros. A zero-inflated model
allows zeros from two sources: a separate always-zero process and the
count process. A hurdle model first estimates whether the count is
positive, then models its size among positive counts. Their component
effects have different meanings and should not be read as changes in the
observed mean count.

\subsubsection{\texorpdfstring{\texttt{fit\_zero\_inflated\_poisson}:
Conditional zero-inflated Poisson
model}{fit\_zero\_inflated\_poisson: Conditional zero-inflated Poisson model}}\label{manual-api:fit-zero-inflated}

\paragraph{Description}\label{manual-description-37}

Fit a Poisson count model and a separate model for observations that
always have zero counts. The fitted effect \(\widehat\psi_c\) belongs to
the count model; \(\widehat\psi_z\) belongs to the always-zero model.

\paragraph{Usage}\label{manual-usage-36}

\begin{lstlisting}[language=Python]
fit_zero_inflated_poisson(y, exposures, covariates=None, *,
                          zero_covariates="same",
                          exposure_component="both", q=4,
                          breaks=None, alpha=0.05)
\end{lstlisting}

\paragraph{Arguments}\label{manual-arguments-29}

\passthrough{\lstinline!y!} contains complete nonnegative integer
counts, including zeros and positive values.
\passthrough{\lstinline!exposures!} and optional
\passthrough{\lstinline!covariates!} are complete numeric columns.
\passthrough{\lstinline!q=None!} means the exposures are already
nonnegative integer scores. \passthrough{\lstinline!breaks!} supplies
fixed cuts. \passthrough{\lstinline!alpha!} sets the normal confidence
level to \(1-\alpha\). By default,
\passthrough{\lstinline!zero\_covariates="same"!} reuses
\passthrough{\lstinline!covariates!} in the structural-zero component. A
separate complete numeric matrix supplies different adjustment columns;
explicit \passthrough{\lstinline!None!} leaves that component without
adjustment. Prediction requires the separate columns again.
\passthrough{\lstinline!exposure\_component!} places the full exposure
set in \passthrough{\lstinline!"count"!},
\passthrough{\lstinline!"zero"!}, or the default
\passthrough{\lstinline!"both"!} component predictors.

\paragraph{Details}\label{manual-details-26}

Let \(\mu_i=\exp(\eta_{ci})\) be the Poisson count mean and
\(\pi_i=\mathrm{logit}^{-1}(\eta_{zi})\) the chance of a structural
zero. The observed zero chance is \(\pi_i+(1-\pi_i)e^{-\mu_i}\); the
expected observed count is \((1-\pi_i)\mu_i\). The full exposure set
enters the selected component models. They may use different adjustment
covariates. When every exposure score rises by one, \(\psi_c\) changes
\(\log\mu\) in the count model and \(\psi_z\) changes the log odds of an
always-zero observation. Each effect is the sum of its component's
exposure coefficients. Exponentiating an effect gives a ratio for that
component, not for the observed mean count. The inverse observed
information gives normal standard errors and intervals. When an entire
component omits the exposures, its joint effect and uncertainty fields
are \passthrough{\lstinline!None!}; its intercept and adjustment terms
remain in the model.

This subset has no negative-binomial component, weights, clusters,
offsets, formulas or a partial exposure set in one component. A separate
function supports a restricted marginal bootstrap. A fitted association
needs further assumptions for a causal interpretation.

\paragraph{Value}\label{manual-value-35}

\passthrough{\lstinline!ZeroInflatedPoissonResult!} stores component
effects, uncertainty, full coefficient covariance, saved scores and
cutpoints.

\paragraph{See Also}\label{manual-see-also-32}

\hyperref[manual-api:quantize]{\passthrough{\lstinline!quantize!}}. The fixed
installed-R checks and optimizer tolerance appear in the zero-inflated
validation report.

\subsubsection{\texorpdfstring{\texttt{ZeroInflatedPoissonResult}:
Separate count and zero
effects}{ZeroInflatedPoissonResult: Separate count and zero effects}}\label{manual-api:zero-inflated-result}

\paragraph{Description}\label{manual-description-38}

Immutable result from
\passthrough{\lstinline!fit\_zero\_inflated\_poisson!}.

\paragraph{Value}\label{manual-value-36}

\passthrough{\lstinline!psi\_count!},
\passthrough{\lstinline!se\_count!}, \passthrough{\lstinline!ci\_count!}
and \passthrough{\lstinline!p\_value\_count!} summarize the Poisson
count component. The matching \passthrough{\lstinline!psi\_zero!},
\passthrough{\lstinline!se\_zero!}, \passthrough{\lstinline!ci\_zero!}
and \passthrough{\lstinline!p\_value\_zero!} fields summarize
structural-zero log odds. Fields for a component without exposures are
\passthrough{\lstinline!None!}. \passthrough{\lstinline!count\_ratio!}
and \passthrough{\lstinline!structural\_zero\_odds\_ratio!} exponentiate
the joint effects. \passthrough{\lstinline!coefficients!} and
\passthrough{\lstinline!covariance!} preserve the full fit.

\passthrough{\lstinline!coefficient\_names!} and
\passthrough{\lstinline!zero\_coefficient\_names!} name the two
component designs. Their adjustment columns are named by
\passthrough{\lstinline!covariate\_names!} and
\passthrough{\lstinline!zero\_covariate\_names!}.
\passthrough{\lstinline!exposure\_names!} and
\passthrough{\lstinline!scores!} retain the exposure scores;
\passthrough{\lstinline!breaks!}, \passthrough{\lstinline!alpha!},
\passthrough{\lstinline!iterations!}, \passthrough{\lstinline!family!}
and \passthrough{\lstinline!reference\_version!} preserve other details.

\paragraph{Methods}\label{manual-methods-5}

\passthrough{\lstinline!predict\_components!} returns the conditional
Poisson mean and structural-zero probability in that order.
\passthrough{\lstinline!predict!} returns their combined expected
observed count. Both reuse saved exposure cuts. When the zero adjustment
differs, pass \passthrough{\lstinline!zero\_covariates!} again at
prediction. \passthrough{\lstinline!summary(digits=6)!} displays
separate component estimates.

\paragraph{See Also}\label{manual-see-also-33}

\hyperref[manual-api:fit-zero-inflated]{\passthrough{\lstinline!fit\_zero\_inflated\_poisson!}}.

\subsubsection{\texorpdfstring{\texttt{fit\_zero\_inflated\_negative\_binomial}:
Zero-inflated count
fit}{fit\_zero\_inflated\_negative\_binomial: Zero-inflated count fit}}\label{manual-api:fit-zero-inflated-negbin}

\paragraph{Description}\label{manual-description-39}

Fit a negative-binomial count model, which allows more count variation
than a Poisson model, and a separate logistic model for always-zero
observations. The fitted effects \(\widehat\psi_c\) and
\(\widehat\psi_z\) refer to the count and structural-zero components,
respectively.

\paragraph{Usage}\label{manual-usage-37}

\begin{lstlisting}[language=Python]
fit_zero_inflated_negative_binomial(
    y, exposures, covariates=None, *, zero_covariates="same",
    exposure_component="both", q=4, breaks=None, alpha=0.05)
\end{lstlisting}

\paragraph{Arguments}\label{manual-arguments-30}

The complete numeric inputs and exposure controls match
\hyperref[manual-api:fit-zero-inflated]{\passthrough{\lstinline!fit\_zero\_inflated\_poisson!}}.
The full exposure set may enter the count component, structural-zero
component, or both. The components can have different numeric adjustment
columns. Independent rows, full-rank designs and both zero and positive
outcomes are required.

\paragraph{Details}\label{manual-details-27}

Let \(\mu_i=\exp(\eta_{ci})\) be the underlying count mean and
\(\theta>0\) the negative-binomial size. The count variance is
\(\mu_i+\mu_i^2/\theta\), and its zero chance is
\((\theta/(\theta+\mu_i))^\theta\). Let
\(\pi_i=\mathrm{logit}^{-1}(\eta_{zi})\) be the structural-zero chance.
The observed zero chance is
\(\pi_i+(1-\pi_i)(\theta/(\theta+\mu_i))^\theta\); the expected observed
count is \((1-\pi_i)\mu_i\). The joint count effect is a log ratio of
count-model means. The always-zero effect is a log odds ratio. Neither
is a ratio of observed mean counts. The covariance includes uncertainty
from estimating \(\theta\). Normal intervals use the relevant parts of
that covariance.

This conditional subset does not support weights, clusters, offsets,
formulas, partial exposure sets, or a negative-binomial marginal
bootstrap. A fitted association needs further assumptions for causal
interpretation.

\paragraph{Value}\label{manual-value-37}

\passthrough{\lstinline!ZeroInflatedNegativeBinomialResult!} stores both
component effects, their intervals, the dispersion
\passthrough{\lstinline!theta!}, the standard error of
\passthrough{\lstinline!log(theta)!}, the complete coefficient
covariance including dispersion, saved scores and cutpoints. If a
component omits exposures, its effect fields are
\passthrough{\lstinline!None!}.

\paragraph{See Also}\label{manual-see-also-34}

\hyperref[manual-api:fit-zero-inflated]{\passthrough{\lstinline!fit\_zero\_inflated\_poisson!}}
and
\hyperref[manual-api:fit-hurdle-negbin]{\passthrough{\lstinline!fit\_hurdle\_negative\_binomial!}}.
The installed-\passthrough{\lstinline!pscl!} comparison and
installed-qgcomp loading limit are in
\passthrough{\lstinline!validation/ZERO\\\_INFLATED\\\_NEGBIN\\\_RESULTS.md!}.

\subsubsection{\texorpdfstring{\texttt{ZeroInflatedNegativeBinomialResult}:
Count and structural-zero
effects}{ZeroInflatedNegativeBinomialResult: Count and structural-zero effects}}\label{manual-zeroinflatednegativebinomialresult-count-and-structural-zero-effects}

\paragraph{Description}\label{manual-description-40}

Immutable result from
\passthrough{\lstinline!fit\_zero\_inflated\_negative\_binomial!}.

\paragraph{Value and methods}\label{manual-value-and-methods}

The component effect, interval, covariance, prediction and summary
fields follow the \hyperref[manual-api:zero-inflated-result]{Poisson result}.
\passthrough{\lstinline!theta!} is the estimated count size;
\passthrough{\lstinline!se\_logtheta!} is its uncertainty on the log
scale. \passthrough{\lstinline!covariance\_full!} includes both
regression components and log dispersion.
\passthrough{\lstinline!predict\_components!} returns the underlying
count mean and structural-zero probability.
\passthrough{\lstinline!predict!} returns their combined expected
observed count.

\paragraph{See Also}\label{manual-see-also-35}

\hyperref[manual-api:fit-zero-inflated-negbin]{\passthrough{\lstinline!fit\_zero\_inflated\_negative\_binomial!}}.

\subsubsection{\texorpdfstring{\texttt{fit\_zero\_inflated\_poisson\_bootstrap}:
Marginal zero-inflated Poisson row
bootstrap}{fit\_zero\_inflated\_poisson\_bootstrap: Marginal zero-inflated Poisson row bootstrap}}\label{manual-api:fit-zero-inflated-bootstrap}

\paragraph{Description}\label{manual-description-41}

Fit degree-one marginal count and structural-zero slopes after setting
all exposures to each joint score. A row bootstrap estimates uncertainty
while keeping the original exposure cutpoints fixed.

\paragraph{Usage}\label{manual-usage-38}

\begin{lstlisting}[language=Python]
fit_zero_inflated_poisson_bootstrap(
    y, exposures, covariates=None, *, q=4, breaks=None,
    alpha=0.05, B=200, MCsize=10000, seed=None)
\end{lstlisting}

\paragraph{Arguments}\label{manual-arguments-31}

The outcome, exposure, adjustment, quantization and interval inputs
follow \passthrough{\lstinline!fit\_zero\_inflated\_poisson!}.
\passthrough{\lstinline!B!} is the number of independent-row resamples;
it must be at least two. \passthrough{\lstinline!MCsize!} is the number
of sampled adjustment profiles per joint score. The default 10,000
follows R qgcomp 2.19.6. A smaller value can increase simulation error.
\passthrough{\lstinline!seed!} controls a local NumPy random generator
without changing global random state.

\paragraph{Details}\label{manual-details-28}

The conditional model supplies the Poisson mean \(\mu\) and
structural-zero chance \(\pi\) for each sampled adjustment profile under
every joint score. A simulated count is a Poisson draw multiplied by a
Bernoulli draw with success chance \(1-\pi\). The simulated counts are
fitted with a marginal zero-inflated Poisson model containing an
intercept and joint-score term in both components. Each bootstrap sample
refits the conditional and marginal stages on sampled rows. The
confidence interval uses a normal critical value and the standard
deviation of the bootstrap slopes. It does not remove finite-simulation
error separately. Failed samples report their position and are not
replaced.

This follows R's default component-simulation stages. Python uses a
different random stream and optimizer. The two slopes describe model
components; neither alone is the ratio of observed mean counts. A causal
reading needs further assumptions. Installed-R bootstrap and
repeated-data interval checks remain open.

\paragraph{Value}\label{manual-value-38}

The result stores marginal count and structural-zero slopes in
\passthrough{\lstinline!psi\_count!} and
\passthrough{\lstinline!psi\_zero!}. The
\passthrough{\lstinline!covariance!}, \passthrough{\lstinline!se!},
\passthrough{\lstinline!ci!} and \passthrough{\lstinline!p\_value!}
fields give bootstrap uncertainty in that order.
\passthrough{\lstinline!intervention\_levels!} holds joint scores, and
\passthrough{\lstinline!intervention\_means!} holds the mean conditional
expected observed count \(E[(1-\pi)\mu]\) across sampled adjustment rows
at each score. \passthrough{\lstinline!bootstrap\_effects!} and
\passthrough{\lstinline!bootstrap\_means!} hold one row per resample.
\passthrough{\lstinline!point\_fit!} and
\passthrough{\lstinline!marginal\_fit!} retain the two point-fit stages;
\passthrough{\lstinline!summary()!} displays the marginal effects.

\paragraph{See Also}\label{manual-see-also-36}

\hyperref[manual-api:fit-zero-inflated]{\passthrough{\lstinline!fit\_zero\_inflated\_poisson!}}.

\subsubsection{\texorpdfstring{\texttt{fit\_zero\_inflated\_negative\_binomial\_bootstrap}:
Negative-binomial marginal
bootstrap}{fit\_zero\_inflated\_negative\_binomial\_bootstrap: Negative-binomial marginal bootstrap}}\label{manual-api:fit-zero-inflated-negbin-bootstrap}

\paragraph{Description}\label{manual-description-42}

Fit degree-one marginal count and structural-zero slopes after simulated
joint exposure changes, with fixed-cut independent-row bootstrap
intervals. The count process has estimated negative-binomial dispersion.

\paragraph{Usage}\label{manual-usage-39}

\begin{lstlisting}[language=Python]
fit_zero_inflated_negative_binomial_bootstrap(
    y, exposures, covariates=None, *, q=4, breaks=None,
    alpha=0.05, B=200, MCsize=10000, seed=None)
\end{lstlisting}

\paragraph{Arguments}\label{manual-arguments-32}

The complete numeric inputs and quantization controls match
\hyperref[manual-api:fit-zero-inflated-negbin]{the conditional NB fit}.
\passthrough{\lstinline!B!} is the number of independent-row resamples,
at least two. \passthrough{\lstinline!MCsize!} is the number of sampled
adjustment profiles per score, at least two; small values can give
unstable fits. \passthrough{\lstinline!seed!} starts a local random
generator.

\paragraph{Details}\label{manual-details-29}

The conditional model supplies the underlying negative-binomial count
mean, structural-zero chance and dispersion. At each joint score, the
method samples adjustment profiles, draws an NB count and applies a
structural-zero indicator. It then fits a marginal NB mixture with an
intercept and joint-score slope in each component. Both stages,
including dispersion, are refitted after resampling independent rows
with fixed exposure cuts. The marginal slope is a population-average
component parameter; it need not equal the conditional coefficient sum.
The expected observed count at a score is the average of
\((1-\pi_i)\mu_i\) over sampled profiles. Normal intervals use the
bootstrap slope standard deviations.

The R qgcomp source uses NB component draws for this path, but its
random stream and optimizer can differ. The installed-R end-to-end
comparison and repeated-data interval checks remain open. A failed
replicate is named and not replaced. Finite simulation size adds noise.
This subset does not support component-specific adjustment,
one-component exposure placement, weights, clusters, formulas or
alternate simulation methods.

\paragraph{Value}\label{manual-value-39}

\passthrough{\lstinline!ZeroInflatedNegativeBinomialBootstrapResult!}
stores the two marginal effects, bootstrap covariance and intervals,
intervention-level expected counts, all resampled effects, and the
conditional and marginal point fits.
\passthrough{\lstinline!summary(digits=6)!} reports the component
meanings and dispersion.

\paragraph{See Also}\label{manual-see-also-37}

\hyperref[manual-api:fit-zero-inflated-negbin]{the conditional NB fit} and
\hyperref[manual-api:fit-zero-inflated-bootstrap]{the Poisson marginal
bootstrap}.

\subsubsection{\texorpdfstring{\texttt{ZeroInflatedNegativeBinomialBootstrapResult}:
Marginal count and zero
effects}{ZeroInflatedNegativeBinomialBootstrapResult: Marginal count and zero effects}}\label{manual-zeroinflatednegativebinomialbootstrapresult-marginal-count-and-zero-effects}

\paragraph{Description}\label{manual-description-43}

Immutable result from the negative-binomial marginal bootstrap.

\paragraph{Value and methods}\label{manual-value-and-methods-1}

\passthrough{\lstinline!psi\_count!} and
\passthrough{\lstinline!psi\_zero!} are marginal count and
structural-zero slopes. The \passthrough{\lstinline!covariance!},
\passthrough{\lstinline!se!}, \passthrough{\lstinline!ci!} and
\passthrough{\lstinline!p\_value!} fields summarize bootstrap
uncertainty. \passthrough{\lstinline!intervention\_levels!} and
\passthrough{\lstinline!intervention\_means!} retain the standardized
score table; \passthrough{\lstinline!bootstrap\_effects!} and
\passthrough{\lstinline!bootstrap\_means!} retain resample rows.
\passthrough{\lstinline!point\_fit!} and
\passthrough{\lstinline!marginal\_fit!} retain fitted NB dispersions.
\passthrough{\lstinline!summary()!} displays the effect scales and
bootstrap settings.

\paragraph{See Also}\label{manual-see-also-38}

\hyperref[manual-api:fit-zero-inflated-negbin-bootstrap]{\passthrough{\lstinline!fit\_zero\_inflated\_negative\_binomial\_bootstrap!}}.

\subsubsection{\texorpdfstring{\texttt{fit\_hurdle\_poisson}:
Conditional Poisson hurdle
model}{fit\_hurdle\_poisson: Conditional Poisson hurdle model}}\label{manual-api:fit-hurdle}

\paragraph{Description}\label{manual-description-44}

Fit a logistic model for the chance of a positive count and a Poisson
model restricted to positive counts. The second model excludes zero
counts. Its effects differ from the always-zero effect in a
zero-inflated model.

\paragraph{Usage}\label{manual-usage-40}

\begin{lstlisting}[language=Python]
fit_hurdle_poisson(y, exposures, covariates=None, *,
                   zero_covariates="same",
                   exposure_component="both", q=4,
                   breaks=None, alpha=0.05)
\end{lstlisting}

\paragraph{Arguments}\label{manual-arguments-33}

\passthrough{\lstinline!y!} contains complete nonnegative integer
counts, including zeros and positive values.
\passthrough{\lstinline!exposures!} and optional
\passthrough{\lstinline!covariates!} are complete numeric columns.
\passthrough{\lstinline!q=None!} means exposures are already nonnegative
integer scores. \passthrough{\lstinline!breaks!} supplies fixed cuts.
\passthrough{\lstinline!alpha!} sets the normal confidence level to
\(1-\alpha\). By default,
\passthrough{\lstinline!zero\_covariates="same"!} reuses
\passthrough{\lstinline!covariates!} in the nonzero component. A
separate complete numeric matrix supplies different adjustment columns;
explicit \passthrough{\lstinline!None!} leaves that component
unadjusted. Prediction requires the separate columns again.
\passthrough{\lstinline!exposure\_component!} places the full exposure
set in \passthrough{\lstinline!"count"!},
\passthrough{\lstinline!"zero"!}, or the default
\passthrough{\lstinline!"both"!}. Here \passthrough{\lstinline!"zero"!}
refers to the nonzero-probability component, following R's component
label.

\paragraph{Details}\label{manual-details-30}

Let \(\mu_i=\exp(\eta_{ci})\) be the underlying Poisson parameter and
\(p_i=\mathrm{logit}^{-1}(\eta_{zi})\) the chance that \(Y_i>0\). The
mean among positive counts is \(\mu_i/(1-e^{-\mu_i})\), because zero
values are excluded from that Poisson component. The expected observed
count is \(p_i\mu_i/(1-e^{-\mu_i})\). The full exposure set enters the
selected component models; their adjustment covariates may differ. The
joint count effect sums exposure coefficients on the \emph{underlying}
Poisson log-parameter scale. The joint hurdle effect sums them on the
log odds scale for having a positive count. Neither alone is an
observed-count ratio. The positive-count and nonzero likelihood factors
yield separate coefficient covariance blocks and normal intervals. When
an entire component omits the exposures, its joint effect and
uncertainty fields are \passthrough{\lstinline!None!}; its intercept and
adjustment terms remain in the model.

This conditional subset has no negative-binomial component, weights,
clusters, offsets, formulas or a partial exposure set in one component.
A fitted association needs further assumptions for a causal
interpretation.

\paragraph{Value}\label{manual-value-40}

\passthrough{\lstinline!HurdlePoissonResult!} stores both component
effects, uncertainty, full block-diagonal coefficient covariance, saved
scores and cutpoints.

\paragraph{See Also}\label{manual-see-also-39}

\hyperref[manual-api:quantize]{\passthrough{\lstinline!quantize!}} and
\hyperref[manual-api:fit-zero-inflated]{\passthrough{\lstinline!fit\_zero\_inflated\_poisson!}}.
Fixed installed-R checks and optimizer tolerance are in the hurdle
validation report.

\subsubsection{\texorpdfstring{\texttt{HurdlePoissonResult}: Separate
positive-count and nonzero
effects}{HurdlePoissonResult: Separate positive-count and nonzero effects}}\label{manual-api:hurdle-result}

\paragraph{Description}\label{manual-description-45}

Immutable result from \passthrough{\lstinline!fit\_hurdle\_poisson!}.

\paragraph{Value}\label{manual-value-41}

\passthrough{\lstinline!psi\_count!},
\passthrough{\lstinline!se\_count!}, \passthrough{\lstinline!ci\_count!}
and \passthrough{\lstinline!p\_value\_count!} summarize the underlying
positive-count Poisson parameter.
\passthrough{\lstinline!psi\_nonzero!},
\passthrough{\lstinline!se\_nonzero!},
\passthrough{\lstinline!ci\_nonzero!} and
\passthrough{\lstinline!p\_value\_nonzero!} summarize the nonzero log
odds. Fields for a component without exposures are
\passthrough{\lstinline!None!}.
\passthrough{\lstinline!count\_parameter\_ratio!} and
\passthrough{\lstinline!nonzero\_odds\_ratio!} exponentiate these
effects. \passthrough{\lstinline!coefficients!} and
\passthrough{\lstinline!covariance!} preserve the full fit.

\passthrough{\lstinline!coefficient\_names!} and
\passthrough{\lstinline!zero\_coefficient\_names!} name the two
component designs. Their adjustment columns are named by
\passthrough{\lstinline!covariate\_names!} and
\passthrough{\lstinline!zero\_covariate\_names!}.
\passthrough{\lstinline!exposure\_names!} and
\passthrough{\lstinline!scores!} retain the exposure scores;
\passthrough{\lstinline!breaks!}, \passthrough{\lstinline!alpha!},
\passthrough{\lstinline!iterations!}, \passthrough{\lstinline!family!}
and \passthrough{\lstinline!reference\_version!} preserve other details.

\paragraph{Methods}\label{manual-methods-6}

\passthrough{\lstinline!predict\_components!} returns the underlying
Poisson parameter and nonzero probability.
\passthrough{\lstinline!predict!} returns their combined expected
observed count after positive-count truncation. Both reuse saved
exposure cuts. When the nonzero adjustment differs, pass
\passthrough{\lstinline!zero\_covariates!} again at prediction.
\passthrough{\lstinline!summary(digits=6)!} displays separate component
estimates.

\paragraph{See Also}\label{manual-see-also-40}

\hyperref[manual-api:fit-hurdle]{\passthrough{\lstinline!fit\_hurdle\_poisson!}}.

\subsubsection{\texorpdfstring{\texttt{fit\_hurdle\_negative\_binomial}:
Conditional negative-binomial hurdle
model}{fit\_hurdle\_negative\_binomial: Conditional negative-binomial hurdle model}}\label{manual-api:fit-hurdle-negbin}

\paragraph{Description}\label{manual-description-46}

Fit a logistic model for the chance of any positive count and a
zero-truncated negative-binomial model for positive counts. This method
estimates the count dispersion from the positive observations.

\paragraph{Usage}\label{manual-usage-41}

\begin{lstlisting}[language=Python]
fit_hurdle_negative_binomial(
    y, exposures, covariates=None, *, zero_covariates="same",
    exposure_component="both", q=4, breaks=None, alpha=0.05)
\end{lstlisting}

\paragraph{Arguments}\label{manual-arguments-34}

The complete numeric inputs and exposure controls match
\hyperref[manual-api:fit-hurdle]{\passthrough{\lstinline!fit\_hurdle\_poisson!}}.
The full exposure set may enter the count, nonzero, or both components.
The two components can have different numeric adjustment columns.
Independent rows, a full-rank design, and both zero and positive
outcomes are required.

\paragraph{Details}\label{manual-details-31}

Let \(\mu_i=\exp(\eta_{ci})\) be the underlying count mean and
\(\theta>0\) the negative-binomial size. Before truncation, the variance
is \(\mu_i+\mu_i^2/\theta\) and the zero chance is
\((\theta/(\theta+\mu_i))^\theta\). Let
\(p_i=\mathrm{logit}^{-1}(\eta_{zi})\) be the chance of a positive
count. The expected observed count is
\(p_i\mu_i/[1-(\theta/(\theta+\mu_i))^\theta]\). The joint count effect
is a log ratio of underlying means; the joint nonzero effect is a log
odds ratio. Neither is an observed-count ratio. The count covariance
includes uncertainty from estimating \(\theta\). Normal intervals use
the separate component covariance blocks.

This conditional subset does not support weights, clusters, offsets,
formulas, partial exposure sets, or a negative-binomial marginal
bootstrap. A fitted association needs further assumptions for causal
interpretation.

\paragraph{Value}\label{manual-value-42}

\passthrough{\lstinline!HurdleNegativeBinomialResult!} stores both
effects and their intervals, the dispersion
\passthrough{\lstinline!theta!}, the standard error of
\passthrough{\lstinline!log(theta)!}, the full count covariance
including dispersion, and the coefficient covariance for the two
regression components. If a component omits the exposures, its effect
fields are \passthrough{\lstinline!None!}.

\paragraph{See Also}\label{manual-see-also-41}

\hyperref[manual-api:fit-hurdle]{\passthrough{\lstinline!fit\_hurdle\_poisson!}}
and
\hyperref[manual-api:zero-inflated-result]{\passthrough{\lstinline!ZeroInflatedPoissonResult!}}.
The direct installed-\passthrough{\lstinline!pscl!} comparison and
installed-qgcomp loading limit are in
\passthrough{\lstinline!validation/HURDLE\\\_NEGBIN\\\_RESULTS.md!}.

\subsubsection{\texorpdfstring{\texttt{HurdleNegativeBinomialResult}:
Negative-binomial hurdle
effects}{HurdleNegativeBinomialResult: Negative-binomial hurdle effects}}\label{manual-hurdlenegativebinomialresult-negative-binomial-hurdle-effects}

\paragraph{Description}\label{manual-description-47}

Immutable result from
\passthrough{\lstinline!fit\_hurdle\_negative\_binomial!}.

\paragraph{Value and methods}\label{manual-value-and-methods-2}

The component effect, interval, covariance, prediction and summary
fields follow
\hyperref[manual-api:hurdle-result]{\passthrough{\lstinline!HurdlePoissonResult!}}.
\passthrough{\lstinline!theta!} is the estimated count size;
\passthrough{\lstinline!se\_logtheta!} is its uncertainty on the log
scale. \passthrough{\lstinline!count\_covariance\_full!} includes the
count coefficients and log dispersion.
\passthrough{\lstinline!predict\_components!} returns the underlying
count mean and nonzero probability. \passthrough{\lstinline!predict!}
applies negative-binomial zero truncation to return the expected
observed count.

\paragraph{See Also}\label{manual-see-also-42}

\hyperref[manual-api:fit-hurdle-negbin]{\passthrough{\lstinline!fit\_hurdle\_negative\_binomial!}}.

\subsubsection{\texorpdfstring{\texttt{fit\_hurdle\_poisson\_bootstrap}:
Marginal Poisson hurdle row
bootstrap}{fit\_hurdle\_poisson\_bootstrap: Marginal Poisson hurdle row bootstrap}}\label{manual-api:fit-hurdle-bootstrap}

\paragraph{Description}\label{manual-description-48}

Fit a degree-one marginal Poisson hurdle model after simulated joint
exposure interventions, with independent-row bootstrap uncertainty.

\paragraph{Usage}\label{manual-usage-42}

\begin{lstlisting}[language=Python]
fit_hurdle_poisson_bootstrap(
    y, exposures, covariates=None, *, q=4, breaks=None,
    alpha=0.05, B=200, MCsize=10000, seed=None)
\end{lstlisting}

\paragraph{Arguments}\label{manual-arguments-35}

The complete numeric inputs and quantization controls match
\hyperref[manual-api:fit-hurdle]{\passthrough{\lstinline!fit\_hurdle\_poisson!}}.
\passthrough{\lstinline!B!} is the number of independent-row resamples,
at least two. \passthrough{\lstinline!MCsize!} is the number of sampled
adjustment profiles per joint score, at least two.
\passthrough{\lstinline!seed!} starts a local random generator.

\paragraph{Details}\label{manual-details-32}

The conditional hurdle model supplies the chance of a positive count and
the underlying Poisson parameter for each intervention profile. Positive
counts are drawn from a zero-truncated Poisson distribution. A hurdle
model with an intercept and joint-score slope in each component is
fitted to the simulated counts. The observed rows are then resampled,
keeping exposure cuts fixed, and both stages are refitted. Normal
intervals use the sample standard deviation of the bootstrap slopes. A
failed replicate is named and is not replaced. Small
\passthrough{\lstinline!MCsize!} adds simulation noise.

R qgcomp 2.19.6 labels its hurdle bootstrap experimental and defaults to
sampling from a finite fitted count-category table. Python draws from
the full zero-truncated Poisson support. Random streams and optimizer
paths can also differ. Installed-R bootstrap comparison and
repeated-data interval checks remain open. Neither component slope is an
observed-count ratio.

\paragraph{Value}\label{manual-value-43}

\passthrough{\lstinline!HurdlePoissonBootstrapResult!} holds
\passthrough{\lstinline!psi\_count!} and
\passthrough{\lstinline!psi\_nonzero!}, the two marginal component
slopes. Its \passthrough{\lstinline!covariance!},
\passthrough{\lstinline!se!}, \passthrough{\lstinline!ci!} and
\passthrough{\lstinline!p\_value!} fields describe bootstrap
uncertainty. \passthrough{\lstinline!intervention\_levels!} and
\passthrough{\lstinline!intervention\_means!} give joint scores and
standardized expected observed counts.
\passthrough{\lstinline!bootstrap\_effects!} and
\passthrough{\lstinline!bootstrap\_means!} retain the resample results.
\passthrough{\lstinline!point\_fit!},
\passthrough{\lstinline!marginal\_fit!},
\passthrough{\lstinline!alpha!}, \passthrough{\lstinline!B!},
\passthrough{\lstinline!MCsize!} and \passthrough{\lstinline!seed!}
retain fit details. \passthrough{\lstinline!summary()!} displays the
effects and intervals.

\paragraph{See Also}\label{manual-see-also-43}

\hyperref[manual-api:fit-hurdle]{\passthrough{\lstinline!fit\_hurdle\_poisson!}}.

\subsubsection{\texorpdfstring{\texttt{fit\_hurdle\_negative\_binomial\_bootstrap}:
Negative-binomial hurdle
bootstrap}{fit\_hurdle\_negative\_binomial\_bootstrap: Negative-binomial hurdle bootstrap}}\label{manual-api:fit-hurdle-negbin-bootstrap}

\paragraph{Description}\label{manual-description-49}

Fit degree-one marginal positive-count and nonzero slopes after
simulated joint exposure changes, with fixed-cut independent-row
bootstrap intervals. The positive-count model has estimated
negative-binomial dispersion.

\paragraph{Usage}\label{manual-usage-43}

\begin{lstlisting}[language=Python]
fit_hurdle_negative_binomial_bootstrap(
    y, exposures, covariates=None, *, q=4, breaks=None,
    alpha=0.05, B=200, MCsize=10000, seed=None)
\end{lstlisting}

\paragraph{Arguments}\label{manual-arguments-36}

The complete numeric inputs and quantization controls match
\hyperref[manual-api:fit-hurdle-negbin]{the conditional NB hurdle fit}.
\passthrough{\lstinline!B!} is the number of independent-row resamples,
at least two. \passthrough{\lstinline!MCsize!} is the number of
adjustment profiles per score, at least two; small values can give
unstable fits. \passthrough{\lstinline!seed!} starts a local random
generator.

\paragraph{Details}\label{manual-details-33}

At each joint score, the conditional hurdle model supplies an underlying
NB mean \(\mu_i\), nonzero chance \(p_i\) and dispersion \(\theta\). The
method samples adjustment profiles and nonzero status, then draws
positive counts from the NB distribution conditional on exceeding zero.
Their expected observed count is
\(p_i\mu_i/[1-(\theta/(\theta+\mu_i))^\theta]\). A marginal NB hurdle
model with a joint-score slope in each component is fitted to the
simulated outcomes. Both stages, including dispersion, are refitted
after independent-row resampling with fixed exposure cuts. Normal
intervals use bootstrap slope standard deviations. The marginal slopes
need not equal conditional coefficient sums; neither alone is an
observed-count ratio.

R qgcomp 2.19.6 labels this method experimental. Its default component
path can produce zero after the nonzero gate because it draws from an
untruncated NB distribution. Python uses positive-truncated counts to
keep the hurdle interpretation. Random streams and optimizers also
differ; end-to-end numerical equality is not claimed. A failed replicate
is named and not replaced. Finite simulation size adds noise. This
subset uses shared adjustment and exposures in both components; it has
no weights, clusters, formulas or alternate simulation modes.

\paragraph{Value}\label{manual-value-44}

\passthrough{\lstinline!HurdleNegativeBinomialBootstrapResult!} stores
the two marginal slopes, bootstrap covariance and intervals,
standardized expected counts by intervention score, all resampled
effects, and both point fits with their estimated dispersions.
\passthrough{\lstinline!summary(digits=6)!} displays the effect scales
and bootstrap settings.

\paragraph{See Also}\label{manual-see-also-44}

\hyperref[manual-api:fit-hurdle-negbin]{the conditional NB hurdle fit} and
\hyperref[manual-api:fit-hurdle-bootstrap]{the Poisson marginal hurdle
bootstrap}.

\subsubsection{\texorpdfstring{\texttt{HurdleNegativeBinomialBootstrapResult}:
Marginal positive-count and nonzero
effects}{HurdleNegativeBinomialBootstrapResult: Marginal positive-count and nonzero effects}}\label{manual-hurdlenegativebinomialbootstrapresult-marginal-positive-count-and-nonzero-effects}

\paragraph{Description}\label{manual-description-50}

Immutable result from the negative-binomial marginal hurdle bootstrap.

\paragraph{Value and methods}\label{manual-value-and-methods-3}

\passthrough{\lstinline!psi\_count!} and
\passthrough{\lstinline!psi\_nonzero!} are marginal positive-count and
nonzero slopes. \passthrough{\lstinline!covariance!},
\passthrough{\lstinline!se!}, \passthrough{\lstinline!ci!} and
\passthrough{\lstinline!p\_value!} summarize bootstrap uncertainty.
\passthrough{\lstinline!intervention\_levels!} and
\passthrough{\lstinline!intervention\_means!} retain the standardized
score table. \passthrough{\lstinline!bootstrap\_effects!} and
\passthrough{\lstinline!bootstrap\_means!} retain resample rows;
\passthrough{\lstinline!point\_fit!} and
\passthrough{\lstinline!marginal\_fit!} retain the fitted NB
dispersions. \passthrough{\lstinline!summary()!} displays the effect
scales and bootstrap settings.

\paragraph{See Also}\label{manual-see-also-45}

\hyperref[manual-api:fit-hurdle-negbin-bootstrap]{\passthrough{\lstinline!fit\_hurdle\_negative\_binomial\_bootstrap!}}.

\subsection{Summaries and shared
tools}\label{manual-summaries-and-shared-tools}

\subsubsection{\texorpdfstring{\texttt{QGCompWarning}: Warnings about a
fit or
prediction}{QGCompWarning: Warnings about a fit or prediction}}\label{manual-api:warning}

\paragraph{Description}\label{manual-description-51}

A Python warning category for numerical limitations during fitting and
exposure extrapolation during prediction. It subclasses
\passthrough{\lstinline!RuntimeWarning!} and adds no custom methods.

\paragraph{Usage}\label{manual-usage-44}

\begin{lstlisting}[language=Python]
import warnings
from quantilemix import QGCompWarning

warnings.simplefilter("always", QGCompWarning)
\end{lstlisting}

\paragraph{Arguments}\label{manual-arguments-37}

The class uses the standard Python warning machinery. Supply
\passthrough{\lstinline!QGCompWarning!} as the category in
\passthrough{\lstinline!warnings.simplefilter!} or
\passthrough{\lstinline!warnings.filterwarnings!}. It is not a fitting
argument.

\paragraph{Details}\label{manual-details-34}

The implementation emits this category in three circumstances:

\begin{enumerate}
\def\labelenumi{\arabic{enumi}.}
\item
  The design condition number exceeds \(10^8\). The result records
  \passthrough{\lstinline!ill\_conditioned=True!}.
\item
  Residual or contrast variation is numerically zero. The result records
  \passthrough{\lstinline!inference\_degenerate=True!}; the standard
  error is zero, the interval is a point, and the test statistic and
  p-value are \passthrough{\lstinline!NaN!}. Predictions can still be
  calculated.
\item
  Prediction exposures lie outside their observed training ranges. Saved
  boundaries are still used. Call \passthrough{\lstinline!transform!} to
  find the affected row positions and columns.
\end{enumerate}

A warning does not remove observations or change the specified model.
The condition threshold is a numerical diagnostic, not a general
statistical validity test. Neither \passthrough{\lstinline!quantize!}
nor \passthrough{\lstinline!transform!} emits its own exposure-range
warning. Inputs outside stored cut endpoints raise an error rather than
producing an extrapolation result.

\paragraph{Value}\label{manual-value-45}

A warning category, not an estimator result. Standard warning filters
control how it is displayed or captured.

\paragraph{Examples}\label{manual-examples-21}

Using \passthrough{\lstinline!raw\_fit!} from the prediction example:

\begin{lstlisting}[language=Python]
import warnings
from quantilemix import QGCompWarning

with warnings.catch_warnings(record=True) as caught:
    warnings.simplefilter("always", QGCompWarning)
    predicted = raw_fit.predict([-1, 1.75, 1.8, 7, 9])

print(len(caught))                         # 1
print(caught[0].category.__name__)          # QGCompWarning
print(str(caught[0].message))
\end{lstlisting}

\begin{lstlisting}
Some exposures are outside their training ranges; fitted cutpoints are reused. Call transform() for row and column diagnostics.
\end{lstlisting}

The returned means are \passthrough{\lstinline![1, 1, 3, 7, 7]!}.

\paragraph{See Also}\label{manual-see-also-46}

\hyperref[manual-api:fit]{\passthrough{\lstinline!fit!}},
\hyperref[manual-api:result]{\passthrough{\lstinline!QGCompResult!}},
\hyperref[manual-api:predict]{\passthrough{\lstinline!QGCompResult.predict!}}.

\subsubsection{\texorpdfstring{\texttt{se\_comb}: Standard error of a
named coefficient
combination}{se\_comb: Standard error of a named coefficient combination}}\label{manual-api:se-comb}

\paragraph{Description}\label{manual-description-52}

Calculate the standard error of a linear combination from a coefficient
covariance matrix.

\paragraph{Usage}\label{manual-usage-45}

\begin{lstlisting}[language=Python]
se_comb(exposure_names, covariance, *, names, grad=None)
\end{lstlisting}

\paragraph{Arguments}\label{manual-arguments-38}

\passthrough{\lstinline!exposure\_names!} selects distinct coefficient
names to sum. \passthrough{\lstinline!covariance!} is a finite symmetric
positive-semidefinite matrix; \passthrough{\lstinline!names!} gives its
row and column order. By default, each selected coefficient has weight
one. \passthrough{\lstinline!grad!} may be one scalar weight for all
selected coefficients or a full vector in
\passthrough{\lstinline!names!} order.

\paragraph{Details}\label{manual-details-35}

If \(C\) is the covariance matrix and \(g\) the vector of weights, the
standard error is \(\sqrt{g^TCg}\). The calculation includes covariance
between different selected coefficients. A zero variance returns zero.

\paragraph{Value}\label{manual-value-46}

A finite nonnegative Python float.

\paragraph{See Also}\label{manual-see-also-47}

\hyperref[manual-api:vc-comb]{\passthrough{\lstinline!vc\_comb!}} and the named
covariance-helper validation report.

\subsubsection{\texorpdfstring{\texttt{vc\_comb}: Covariance of a
coefficient and a
combination}{vc\_comb: Covariance of a coefficient and a combination}}\label{manual-api:vc-comb}

\paragraph{Description}\label{manual-description-53}

Return the covariance matrix for one named coefficient and a selected
linear combination of coefficients.

\paragraph{Usage}\label{manual-usage-46}

\begin{lstlisting}[language=Python]
vc_comb(exposure_names, covariance, *, names,
        aname="(Intercept)", grad=None)
\end{lstlisting}

\paragraph{Arguments}\label{manual-arguments-39}

\passthrough{\lstinline!exposure\_names!},
\passthrough{\lstinline!covariance!}, \passthrough{\lstinline!names!}
and \passthrough{\lstinline!grad!} have the meanings given for
\passthrough{\lstinline!se\_comb!}. \passthrough{\lstinline!aname!}
selects the coefficient paired with the exposure combination; its
default is the intercept.

\paragraph{Details}\label{manual-details-36}

For a coefficient vector \(\beta\), a unit vector \(e_a\) selecting
\passthrough{\lstinline!aname!} and a contrast vector \(g\), the
returned matrix is the covariance of \((e_a^T\beta,g^T\beta)\):

\[
\begin{pmatrix}e_a^TCe_a&e_a^TCg\\g^TCe_a&g^TCg\end{pmatrix}.
\]

R qgcomp 2.19.6 omits a non-unit \passthrough{\lstinline!grad!} from the
off-diagonal term. This Python function uses the same supplied weights
in all entries. Its default unweighted result agrees with R; the
weighted difference is documented and checked.

\paragraph{Value}\label{manual-value-47}

A read-only 2-by-2 NumPy covariance array.

\paragraph{See Also}\label{manual-see-also-48}

\hyperref[manual-api:se-comb]{\passthrough{\lstinline!se\_comb!}} and the named
covariance-helper validation report.

\subsubsection{\texorpdfstring{\texttt{split\_data}: Split training and
validation
rows}{split\_data: Split training and validation rows}}\label{manual-split_data-split-training-and-validation-rows}

\paragraph{Usage}\label{manual-usage-47}

\begin{lstlisting}[language=Python]
split_data(data, *, cluster=None, prop_train=0.4, seed=None)
\end{lstlisting}

\paragraph{Arguments and value}\label{manual-arguments-and-value-3}

\passthrough{\lstinline!data!} is a named-column mapping or a
two-dimensional table. \passthrough{\lstinline!prop\_train!} is the
fraction of rows, or whole clusters when a
\passthrough{\lstinline!cluster!} column is named, assigned to training.
\passthrough{\lstinline!seed!} controls Python's random generator. The
returned \passthrough{\lstinline!DataSplit!} has zero-based
\passthrough{\lstinline!train\_indices!} and
\passthrough{\lstinline!validation\_indices!}, plus corresponding
\passthrough{\lstinline!train\_data!} and
\passthrough{\lstinline!validation\_data!}. Both sides must be nonempty.
R 2.19.6 samples within clusters despite documenting a
cluster-preserving split. This Python helper keeps every cluster on one
side. Equal integer seeds need not select the same rows in R and Python.

\subsubsection{\texorpdfstring{\texttt{qgcomp\_partials}: Held-out
additive partial
effects}{qgcomp\_partials: Held-out additive partial effects}}\label{manual-qgcomp_partials-held-out-additive-partial-effects}

\paragraph{Usage}\label{manual-usage-48}

\begin{lstlisting}[language=Python]
qgcomp_partials(train_y, train_exposures, validation_y,
                validation_exposures, *, train_covariates=None,
                validation_covariates=None, q=4,
                fixed_breaks=True, alpha=0.05,
                family="gaussian")
\end{lstlisting}

\paragraph{Details}\label{manual-details-37}

The training fit selects positive and negative coefficient groups. Each
nonempty group is then estimated on validation rows, adjusting for all
other exposures. A validation coefficient stays in its training-selected
group even if its sign changes. Raw-exposure cuts come from training by
default; use \passthrough{\lstinline!q=None!} for prequantized scores.
Both halves need matching exposure columns and complete numeric values.
Choose the split without inspecting validation outcomes. Choose
\passthrough{\lstinline!family="gaussian"!},
\passthrough{\lstinline!"binomial"!} or
\passthrough{\lstinline!"poisson"!}. Binary held-out effects are
conditional log odds ratios; Poisson effects are conditional log mean
ratios for equal observation periods. One Gaussian and four binary/count
fixed cases agree with installed R. Repeated-sample interval coverage,
other family variants, weights, offsets, formulas, global cuts and
cross-fitting remain open; see the two sample-split validation reports.

\paragraph{Value}\label{manual-value-48}

The \passthrough{\lstinline!GaussianPartialResult!} or
\passthrough{\lstinline!GLMPartialResult!} retains
\passthrough{\lstinline!training\_fit!} and two name lists:
\passthrough{\lstinline!positive\_exposures!} and
\passthrough{\lstinline!negative\_exposures!}. It also stores
\passthrough{\lstinline!positive\_fit!} and
\passthrough{\lstinline!negative\_fit!}; an empty group has
\passthrough{\lstinline!None!}, not an estimated zero effect. Each fit
provides \passthrough{\lstinline!psi!}, \passthrough{\lstinline!se!},
\passthrough{\lstinline!ci!}, \passthrough{\lstinline!p\_value!} and
\passthrough{\lstinline!summary()!}.

\subsubsection{\texorpdfstring{\texttt{GLMPartialResult}: Binary or
count held-out
summary}{GLMPartialResult: Binary or count held-out summary}}\label{manual-glmpartialresult-binary-or-count-held-out-summary}

\paragraph{Description}\label{manual-description-54}

Returned when \passthrough{\lstinline!qgcomp\_partials!} uses
\passthrough{\lstinline!family="binomial"!} or
\passthrough{\lstinline!family="poisson"!}. Its
\passthrough{\lstinline!family!} names the outcome model.
\passthrough{\lstinline!training\_fit!},
\passthrough{\lstinline!positive\_fit!} and
\passthrough{\lstinline!negative\_fit!} retain the corresponding fitted
models. The selected exposure names remain available even when a
validation coefficient changes sign. An empty training group has no
held-out fit.

\paragraph{Interpretation}\label{manual-interpretation}

The result's \passthrough{\lstinline!summary()!} prints held-out
log-scale effects, model-based standard errors and normal confidence
intervals. A binary effect is a conditional log odds ratio; a Poisson
effect is a conditional log mean ratio. Exponentiating either value
gives its conditional ratio. These are not population-average effects.
See the binary/count partial-effect validation report.

\subsubsection{\texorpdfstring{\texttt{pointwisebound\_noboot}: Compare
conditional joint-score
effects}{pointwisebound\_noboot: Compare conditional joint-score effects}}\label{manual-pointwisebound_noboot-compare-conditional-joint-score-effects}

\paragraph{Usage}\label{manual-usage-49}

\begin{lstlisting}[language=Python]
pointwisebound_noboot(result, *, pointwiseref=1, alpha=0.05)
\end{lstlisting}

\paragraph{Arguments}\label{manual-arguments-40}

\passthrough{\lstinline!result!} is an additive Gaussian, binomial or
Poisson fit with non-null \passthrough{\lstinline!q!}. R's one-based
\passthrough{\lstinline!pointwiseref!} is 1 for score zero and
\passthrough{\lstinline!q!} for the highest score.
\passthrough{\lstinline!alpha!} sets each normal interval's error level.
Poisson observation-time offsets are outside this interface.

\paragraph{Method and interpretation}\label{manual-method-and-interpretation-2}

At joint score \(s\), set every exposure score to \(s\) and adjustment
variables to their encoded reference value. With intercept
\(\widehat\beta_0\) and sum of exposure coefficients \(\widehat\psi\),
the partial linear predictor is \(h(s)=\widehat\beta_0+s\widehat\psi\).
Relative to score \(r\), its contrast is \((s-r)\widehat\psi\) with
standard error \(|s-r|\mathrm{SE}(\widehat\psi)\). Gaussian output is a
conditional mean difference. Exponentiating the contrast and its normal
limits gives a binary odds ratio or an equal-period Poisson mean ratio.
These are neither individual prediction intervals nor population-average
effects. Six fixed cases agree with installed R qgcomp 2.19.6; see the
validation report.

\paragraph{Value}\label{manual-value-49}

\passthrough{\lstinline!PointwiseBoundResult.table!} maps R's column
names to read-only NumPy arrays. Common columns are
\passthrough{\lstinline!quantile!},
\passthrough{\lstinline!quantile.midpoint!} and
\passthrough{\lstinline!linpred!}. Family-specific columns give the
contrast, link-scale standard error and limits:
\passthrough{\lstinline!mean.diff/se.diff/ll.diff/ul.diff!},
\passthrough{\lstinline!or/se.lnor/ll.or/ul.or!} or
\passthrough{\lstinline!rr/se.lnrr/ll.rr/ul.rr!}.
\passthrough{\lstinline!columns!},
\passthrough{\lstinline!pointwiseref!}, \passthrough{\lstinline!alpha!},
\passthrough{\lstinline!family!}, \passthrough{\lstinline!link!} and
\passthrough{\lstinline!measure!} are also stored.

\subsubsection{\texorpdfstring{\texttt{pointwisebound\_boot}: Compare
marginal joint-score
effects}{pointwisebound\_boot: Compare marginal joint-score effects}}\label{manual-pointwisebound_boot-compare-marginal-joint-score-effects}

\paragraph{Usage}\label{manual-usage-50}

\begin{lstlisting}[language=Python]
pointwisebound_boot(result, *, pointwiseref=1, alpha=0.05)
\end{lstlisting}

\paragraph{Arguments}\label{manual-arguments-41}

\passthrough{\lstinline!result!} is an additive Gaussian or Poisson
row-bootstrap fit with a non-null quantile count
\passthrough{\lstinline!q!}. Poisson observation-time offsets are
allowed. \passthrough{\lstinline!pointwiseref!} is one-based: 1 selects
joint score zero. \passthrough{\lstinline!alpha!} sets the error level
of each normal interval. The function does not return a simultaneous
band across all joint scores.

\paragraph{Method and interpretation}\label{manual-method-and-interpretation-3}

For each joint score, the fit averages its predicted outcome across the
observed covariate rows. Each bootstrap draw repeats this calculation on
its resampled rows. The standard error for a comparison is the standard
deviation of its bootstrap contrast with the selected reference.
Gaussian contrasts are differences in marginal means. Poisson contrasts
are ratios of marginal means; their standard errors use differences of
log-scale marginal-model predictions. This matches the covariance used
by installed R qgcomp 2.19.6. The limits are normal Wald limits, not
percentile intervals. They describe uncertainty in a fitted contrast,
not the range of individual outcomes. Six fixed Gaussian, Poisson and
offset-Poisson tables agree with installed R; see
\passthrough{\lstinline!validation/POINTWISE\_BOOT\_RESULTS.md!}.
Repeated-data coverage of these pointwise intervals has not been
checked.

\paragraph{Value}\label{manual-value-50}

The read-only \passthrough{\lstinline!PointwiseBoundResult!} uses R's
names: \passthrough{\lstinline!quantile!},
\passthrough{\lstinline!quantile.midpoint!},
\passthrough{\lstinline!linpred!}, the contrast, its standard error and
limits, plus \passthrough{\lstinline!ll.linpred!} and
\passthrough{\lstinline!ul.linpred!} for display. Gaussian contrast
columns are \passthrough{\lstinline!mean.diff!},
\passthrough{\lstinline!se.diff!}, \passthrough{\lstinline!ll.diff!} and
\passthrough{\lstinline!ul.diff!}. Poisson columns are
\passthrough{\lstinline!rr!}, \passthrough{\lstinline!se.lnrr!},
\passthrough{\lstinline!ll.rr!} and \passthrough{\lstinline!ul.rr!}. The
bootstrap result also saves read-only
\passthrough{\lstinline!bootstrap\_means!}, one row per resample and one
column per joint score.

\subsubsection{\texorpdfstring{\texttt{modelbound\_boot}: Marginal
fitted-curve
bounds}{modelbound\_boot: Marginal fitted-curve bounds}}\label{manual-modelbound_boot-marginal-fitted-curve-bounds}

\paragraph{Usage}\label{manual-usage-51}

\begin{lstlisting}[language=Python]
from quantilemix import modelbound_boot
bounds = modelbound_boot(bootstrap_result, alpha=0.05)
pointwise = modelbound_boot(bootstrap_result, pwonly=True)
\end{lstlisting}

\paragraph{Arguments}\label{manual-arguments-42}

\passthrough{\lstinline!bootstrap\_result!} is an additive Gaussian or
Poisson row-bootstrap fit with a non-null quantile count. Poisson
observation-time offsets are allowed. \passthrough{\lstinline!alpha!} is
the tail area, strictly between zero and one.
\passthrough{\lstinline!pwonly=True!} omits the simultaneous calculation
and leaves those two output columns missing.

\paragraph{Method and interpretation}\label{manual-method-and-interpretation-4}

At each joint score, the pointwise limit uses the standard deviation of
the saved bootstrap marginal predictions. Gaussian predictions stay on
the mean scale. Poisson uncertainty is computed on the log scale, then
its limits are exponentiated to the expected-count scale. The pointwise
limits each address one score and do not jointly cover the curve.

For simultaneous limits, the function follows R qgcomp 2.19.6. It
estimates the covariance of the bootstrap marginal intercept and slope,
retains draws whose coefficient vector lies inside a chi-square
ellipsoid around the original coefficients, and takes the lowest and
highest retained fitted value at each score. An ill-conditioned
covariance or empty retained set raises an error. These bounds are based
on the reference procedure; fixed-data agreement does not show their
repeated-data coverage.

\paragraph{Value}\label{manual-value-51}

\passthrough{\lstinline!ModelBoundResult.table!} maps the R column names
to read-only arrays. Common columns are
\passthrough{\lstinline!quantile!},
\passthrough{\lstinline!quantile.midpoint!},
\passthrough{\lstinline!linpred!}, \passthrough{\lstinline!se.pw!},
\passthrough{\lstinline!ll.pw!}, \passthrough{\lstinline!ul.pw!},
\passthrough{\lstinline!ll.simul!} and
\passthrough{\lstinline!ul.simul!}. The Gaussian mean column is
\passthrough{\lstinline!m!}; the Poisson expected-count column is
\passthrough{\lstinline!r!}. The result also stores
\passthrough{\lstinline!columns!}, \passthrough{\lstinline!alpha!},
\passthrough{\lstinline!family!}, \passthrough{\lstinline!link!} and
\passthrough{\lstinline!pwonly!}. Three fixed Gaussian, Poisson and
offset-Poisson cases give six tables matching installed R; see
\passthrough{\lstinline!validation/MODELBOUND\_BOOT\_RESULTS.md!}.

\subsubsection{\texorpdfstring{\texttt{modelbound\_ee}: Clustered
Gaussian marginal-curve
bounds}{modelbound\_ee: Clustered Gaussian marginal-curve bounds}}\label{manual-modelbound_ee-clustered-gaussian-marginal-curve-bounds}

\paragraph{Usage}\label{manual-usage-52}

\begin{lstlisting}[language=Python]
from quantilemix import modelbound_ee
bounds = modelbound_ee(ee_result, alpha=0.05, seed=418)
pointwise = modelbound_ee(ee_result, pwonly=True, seed=418)
\end{lstlisting}

\paragraph{Arguments}\label{manual-arguments-43}

\passthrough{\lstinline!ee\_result!} is an additive Gaussian fit
returned by \passthrough{\lstinline!fit\_gaussian\_ee!}. It must have
consecutive joint intervention levels and a positive-definite covariance
matrix for its marginal intercept and slope.
\passthrough{\lstinline!alpha!} is the tail area, strictly between zero
and one. \passthrough{\lstinline!pwonly=True!} leaves the simultaneous
columns missing. \passthrough{\lstinline!seed!} is an optional
nonnegative integer for repeatable Python draws.

\paragraph{Method and interpretation}\label{manual-method-and-interpretation-5}

The function samples 5,000 intercept--slope pairs from the fitted normal
approximation using the cluster-robust covariance matrix. The standard
deviation of the resulting fitted means gives each pointwise standard
error. For simultaneous bounds, it keeps coefficient draws inside a
chi-square ellipsoid around the fitted pair and uses the lowest and
highest retained mean at each joint score. R qgcomp 2.19.6 uses the same
calculation. Python and R use different random generators; equal seeds
alone do not match draws. The simultaneous columns follow the R
procedure, not a demonstrated repeated-data coverage guarantee.

\paragraph{Value}\label{manual-value-52}

The read-only \passthrough{\lstinline!ModelBoundResult!} has
\passthrough{\lstinline!quantile!},
\passthrough{\lstinline!quantile.midpoint!},
\passthrough{\lstinline!linpred!}, \passthrough{\lstinline!m!},
\passthrough{\lstinline!se.pw!}, \passthrough{\lstinline!ll.pw!},
\passthrough{\lstinline!ul.pw!}, \passthrough{\lstinline!ll.simul!} and
\passthrough{\lstinline!ul.simul!} columns. Two tables agree with
installed R when supplied the same coefficient draws; see
\passthrough{\lstinline!validation/MODELBOUND\_EE\_RESULTS.md!}. R's
helper errors for prequantized \passthrough{\lstinline!q=NULL!}; Python
accepts a fitted result with consecutive levels. Repeated-data coverage
has not been checked.

\subsubsection{\texorpdfstring{\texttt{msm\_predict}: Marginal-model
outcome
predictions}{msm\_predict: Marginal-model outcome predictions}}\label{manual-msm_predict-marginal-model-outcome-predictions}

\paragraph{Usage}\label{manual-usage-53}

\begin{lstlisting}[language=Python]
from quantilemix import msm_predict
predicted = msm_predict(result, newdata=None)
predicted_new = msm_predict(result, {"psi": [0, 0.5, 1]})
\end{lstlisting}

\paragraph{Arguments}\label{manual-arguments-44}

\passthrough{\lstinline!result!} is an additive Gaussian or Poisson
row-bootstrap fit, with or without an observation-time offset in the
Poisson fit. \passthrough{\lstinline!newdata!} may be omitted, or it may
be a mapping or data frame with a one-dimensional numeric
\passthrough{\lstinline!psi!} column. A one-dimensional numeric score
array is also accepted. Each score is the level assigned jointly to all
exposures. Scores outside the fitted range extrapolate the model; they
are not new exposure measurements to be regrouped into quantiles. Inputs
must be finite and nonempty.

\paragraph{Method and interpretation}\label{manual-method-and-interpretation-6}

Let \(a_0\) be the first fitted joint score and \(\widehat\psi\) the
fitted one-category joint effect. For Gaussian outcomes,
\(\widehat\mu(a_0)\) is the predicted mean after setting every exposure
score to \(a_0\). The prediction at score \(s\) is
\(\widehat\mu(a_0)+(s-a_0)\widehat\psi\). For Poisson outcomes,
\(\widehat M(a_0)\) is the expected count at \(a_0\), and the prediction
at \(s\) is \(\widehat M(a_0)\exp\{(s-a_0)\widehat\psi\}\). These
formulas use the observed distribution of adjustment covariates. A
Poisson fit with unequal observation times also keeps those fitted
times, so its output is an expected count over that time distribution.
Without \passthrough{\lstinline!newdata!}, the function returns the
fitted mean once for every original row at each score, in consecutive
score blocks. R's exported \passthrough{\lstinline!msm.predict!} uses
the same order. This function predicts from the marginal model, not the
conditional outcome model. It gives neither a confidence interval nor an
individual prediction interval. Five fixed default and supplied-score
cases agree with isolated installed R qgcomp 2.19.6; see
\passthrough{\lstinline!validation/MSM\_PREDICT\_RESULTS.md!}.

\paragraph{Value}\label{manual-value-53}

The return value is a read-only one-dimensional NumPy array. With
\passthrough{\lstinline!newdata!}, it has one value for each supplied
score. Without \passthrough{\lstinline!newdata!}, it has \(nq\) values,
where \(n\) is the original row count and \(q\) is the number of fitted
intervention levels.

\subsubsection{\texorpdfstring{\texttt{tidy}: Compact conditional
joint-effect
table}{tidy: Compact conditional joint-effect table}}\label{manual-tidy-compact-conditional-joint-effect-table}

\paragraph{Usage}\label{manual-usage-54}

\begin{lstlisting}[language=Python]
table = tidy(result, conf_level=None, exponentiate=False, quick=False)
\end{lstlisting}

\paragraph{Arguments and Value}\label{manual-arguments-and-value-4}

\passthrough{\lstinline!result!} is an additive conditional Gaussian,
binomial, Poisson or Cox fit. The read-only
\passthrough{\lstinline!TidyResult!} has \passthrough{\lstinline!term!},
\passthrough{\lstinline!estimate!},
\passthrough{\lstinline!std\_error!},
\passthrough{\lstinline!lower\_ci!},
\passthrough{\lstinline!upper\_ci!}, \passthrough{\lstinline!statistic!}
and \passthrough{\lstinline!p\_value!} columns. Cox has only a
\passthrough{\lstinline!psi1!} row; the other families also have an
intercept row. \passthrough{\lstinline!quick=True!} retains only terms
and estimates. \passthrough{\lstinline!conf\_level!} defaults to the
fitted confidence level. \passthrough{\lstinline!exponentiate=True!}
applies to log and logit links; standard errors remain on the link
scale. Gaussian mean differences cannot be exponentiated. The intercept
has its own test result and an exponentiated upper confidence limit uses
the actual upper bound. These correct two R qgcomp 2.19.6
\passthrough{\lstinline!tidy!} display errors. Broader result families
remain open.

\subsubsection{\texorpdfstring{\texttt{glance}: One-row conditional
model fit
table}{glance: One-row conditional model fit table}}\label{manual-glance-one-row-conditional-model-fit-table}

\paragraph{Usage}\label{manual-usage-55}

\begin{lstlisting}[language=Python]
row = glance(result)
print(row["log_likelihood"], row["aic"], row["deviance"])
\end{lstlisting}

\paragraph{Arguments and Value}\label{manual-arguments-and-value-5}

\passthrough{\lstinline!result!} is an additive conditional Gaussian,
binomial or Poisson fit. The read-only
\passthrough{\lstinline!GlanceResult!} holds one row of fit measures,
including \passthrough{\lstinline!nobs!},
\passthrough{\lstinline!df\_model!},
\passthrough{\lstinline!df\_residual!},
\passthrough{\lstinline!log\_likelihood!},
\passthrough{\lstinline!aic!}, \passthrough{\lstinline!bic!},
\passthrough{\lstinline!deviance!},
\passthrough{\lstinline!null\_deviance!} and
\passthrough{\lstinline!df\_null!}. Gaussian fits also give
\passthrough{\lstinline!r\_squared!},
\passthrough{\lstinline!adjusted\_r\_squared!},
\passthrough{\lstinline!sigma!}, \passthrough{\lstinline!f\_statistic!}
and \passthrough{\lstinline!f\_p\_value!}. These describe the underlying
conditional fit, not a mixture intervention or causal effect. Case
weights enter the likelihood and deviance, and the Poisson null model
retains each row's observation time. A Gaussian fit with zero residual
sum of squares returns \passthrough{\lstinline!None!} for its degenerate
likelihood measures. These fields follow standard conditional-model
definitions. Fixed known-answer and weighted-row-replication checks
pass. Seven fixed installed-R cases agree on shared
\passthrough{\lstinline!tidy!} values; six agree on shared
\passthrough{\lstinline!glance!} values. Cox, bootstrap and
estimating-equation results are unsupported because their likelihood
summaries have different meanings.

\subsubsection{\texorpdfstring{\texttt{plot\_weights}: Descriptive
exposure-weight
plot}{plot\_weights: Descriptive exposure-weight plot}}\label{manual-plot_weights-descriptive-exposure-weight-plot}

\paragraph{Usage}\label{manual-usage-56}

\begin{lstlisting}[language=Python]
from quantilemix import plot_weights
ax = plot_weights(fitted_result, ax=None)
\end{lstlisting}

\paragraph{Arguments and Value}\label{manual-arguments-and-value-6}

\passthrough{\lstinline!fitted\_result!} is an additive conditional
Gaussian, binomial or Poisson result. \passthrough{\lstinline!ax!} is an
optional Matplotlib Axes. If omitted, the function creates one; install
the optional \passthrough{\lstinline!quantilemix[plot]!} extra for this
use. The function returns the Axes and does not show or save the figure.

\paragraph{Details}\label{manual-details-38}

The horizontal bars show each exposure's share within the positive or
negative coefficient group. These groups are normalized separately. A
zero exposure coefficient appears in neither group. The plot describes
the fitted coefficients, not causal importance, selection probability or
uncertainty. The coefficient scale is the outcome scale for Gaussian,
log odds for binomial, and log mean for Poisson. No model is refitted.

\subsubsection{\texorpdfstring{\texttt{plot\_response}: Standardized
response
plot}{plot\_response: Standardized response plot}}\label{manual-plot_response-standardized-response-plot}

\paragraph{Usage}\label{manual-usage-57}

\begin{lstlisting}[language=Python]
from quantilemix import plot_response
ax = plot_response(bootstrap_result, ax=None)
\end{lstlisting}

\paragraph{Arguments and Value}\label{manual-arguments-and-value-7}

\passthrough{\lstinline!bootstrap\_result!} is an additive Gaussian or
Poisson marginal bootstrap result, or a binary marginal risk-ratio or
odds-ratio result. \passthrough{\lstinline!ax!} and the optional
plotting extra work as above. The returned Axes show the response values
already saved at each joint exposure score.

\paragraph{Details}\label{manual-details-39}

The vertical axis says standardized mean outcome for Gaussian, expected
count for Poisson, or event risk for binomial. Each score sets every
exposure to that category while retaining the model's adjustment
distribution. The line connects fitted points for display; it adds no
confidence band or new fit. Poisson values include saved observation
times, if present. Other nonlinear and specialist result types have
different response definitions and are not accepted by this helper.

\section{Validation and scope}\label{manual-validation-and-scope}

\subsection{Validation evidence}\label{manual-validation-evidence}

The package has two kinds of checks for selected methods. First, we
compare Python and installed R qgcomp 2.19.6 on fixed datasets to check
that they produce the expected numbers. Second, we fit many datasets
generated under stated conditions to study bias and confidence-interval
coverage. Neither kind of check proves a causal effect or covers every
method in the package.

\begin{itemize}
\item
  \textbf{Gaussian.} Nine fixed cases agree with installed R. Across
  24,000 generated datasets with known exposure scores, the reported
  95\% intervals contained the true effect in 94.55\% to 96.10\% of
  cases, depending on the condition. The study used fixed population
  score boundaries. It did not check boundaries estimated from each
  dataset. See the
  \href{https://github.com/nilson01/quantilemix/blob/main/docs/reference.md}{R
  comparison guide} and
  \href{https://github.com/nilson01/quantilemix/blob/main/validation/GAUSSIAN_RESULTS.md}{Gaussian
  study report}.
\item
  \textbf{Binary and count.} Separate studies cover conditional binary
  log odds and Poisson log rates, and selected marginal binary and count
  methods. The low-count Poisson null condition showed 93.4\% coverage
  and a 6.6\% false-positive rate. See the
  \href{https://github.com/nilson01/quantilemix/blob/main/validation/BINOMIAL_CALIBRATION_RESULTS.md}{binary
  study report},
  \href{https://github.com/nilson01/quantilemix/blob/main/validation/POISSON_CALIBRATION_RESULTS.md}{Poisson
  study report}, and other reports in the
  \href{https://github.com/nilson01/quantilemix/tree/main/validation/}{validation
  directory}.
\item
  \textbf{Survival.} A Cox resampling study with known scores returned
  intervals for 7,994 of 8,000 datasets. Six fits failed. Coverage among
  successful fits does not describe those six attempts. See the
  \href{https://github.com/nilson01/quantilemix/blob/main/validation/COX_BOOTSTRAP_STUDY_RESULTS.md}{survival
  study report}.
\item
  \textbf{Other methods.} Reports in the
  \href{https://github.com/nilson01/quantilemix/tree/main/validation/}{validation
  directory} state their fixed comparisons, study settings, failures and
  limits. The
  \href{https://github.com/nilson01/quantilemix/blob/main/docs/r-api-parity.md}{coverage
  map} identifies checks that remain open.
\end{itemize}

The linked reports give the study designs, exact counts, numerical
differences and steps to repeat the checks. Use them to assess the
validation results. This manual explains the methods and how to call
them.

\subsection{Method source}\label{manual-method-source}

Keil, A. P., Buckley, J. P., O'Brien, K. M., Ferguson, K. K., Zhao, S.,
and White, A. J. (2020). ``A Quantile-Based g-Computation Approach to
Addressing the Effects of Exposure Mixtures.'' \emph{Environmental
Health Perspectives}, 128(4), 047004.
\href{https://doi.org/10.1289/EHP5838}{doi:10.1289/EHP5838}.

\end{document}